\documentclass[twocolumn,aps,rmp]{revtex4-1}
\usepackage[T1]{fontenc}
\usepackage[latin9]{inputenc}
\usepackage{verbatim}
\usepackage{amsthm}
\usepackage{amsmath}
\usepackage{amssymb}
\usepackage{revsymb}
\usepackage{graphicx}
\usepackage{esint}
\usepackage{epstopdf}
\usepackage{color}
\usepackage{mathrsfs}
\usepackage{mathtools}

\inputencoding{utf8}
\usepackage[
unicode=true,
bookmarks=true,
bookmarksnumbered=true,
bookmarksopen=false,
breaklinks=true,
pdfborder={0 0 1},
backref=false,
colorlinks=false
]
 {hyperref}
\hypersetup{pdftitle={Quantum limits in optical interferometry},
 pdfauthor={R.Demkowicz-Dobrzanski, M. Jarzyna, J. Kolodynski},
 pdfkeywords={quantum optics, quantum metrology, quantum information}}
\inputencoding{latin9}
\makeatletter

\newcommand{\eqnref}[1]{Eq.~(\ref{#1})}
\newcommand{\eqnsref}[2]{Eqs.~(\ref{#1}) and (\ref{#2})}
\newcommand{\eref}[1]{(\ref{#1})}
\newcommand{\figref}[1]{Fig.~\ref{#1}}

\newcommand{\secref}[1]{Sec.~\ref{#1}}

\newcommand{\bra}[1]{\langle #1|}
\newcommand{\ket}[1]{| #1 \rangle}

\newcommand{\ketE}[1]{\left| #1 \right\rangle}

\newcommand{\braket}[2]{\langle #1 | #2 \rangle }
\newcommand{\e}{\mathrm{e}}
\newcommand{\ii}{\mathrm{i}}
\newcommand{\Tr}{\textrm{Tr}}
\renewcommand{\t}[1]{\textrm{#1}}
\newcommand{\mat}[2]{\left(\begin{array}{#1}#2\end{array}\right)}
\newcommand{\mN}{\langle N \rangle} %mean number of photons
\newcommand{\mean}[1]{\left\langle #1 \right\rangle } % mean
\newcommand{\noon}{NOON }% NOON statemean

\makeatother
\begin{document}

\title{Quantum limits in optical interferometry}

\author{R. Demkowicz-Dobrza{\'n}ski}
%\ead{demko@fuw.edu.pl}
\author{M. Jarzyna}
%\ead{mjarz@fuw.edu.pl}
\author{J. Ko{\l}ody{\'n}ski}
%\ead{jankolo@fuw.edu.pl}
\affiliation{Faculty of Physics, University of Warsaw, ul. Ho\.{z}a 69, PL-00-681 Warszawa, Poland},


\begin{abstract}
Non-classical states of light find applications in enhancing the performance of optical interferometric experiments, with notable example
 of gravitational wave-detectors. Still, the presence of decoherence hinders significantly the performance
 of quantum-enhanced protocols. In this review, we summarize the developments of quantum metrology with particular focus on optical interferometry and derive fundamental bounds on achievable quantum-enhanced precision
 in optical interferometry taking into account the most relevant decoherence processes including: phase diffusion, losses and imperfect interferometric visibility. We introduce all the necessary tools of quantum optics as well as quantum estimation theory required to derive the bounds.
We also discuss the practical attainability of the bounds derived and stress in particular that the techniques of
quantum-enhanced interferometry which are being implemented in modern gravitational wave detectors are close to the optimal ones.
\end{abstract}
% keywords: quantum metrology, quantum interferometry, decoherence, phase estimation, squeezed states of light, gravitational wave detection
\maketitle

\tableofcontents
%\noindent\rule[0.5ex]{1\columnwidth}{1pt}
%\newpage

%\input{introduction}  %Rafal
%\input{states} % Marcin
%\input{interferometers} % Rafal
%\input{estimation} % Janek
%\input{ideal} % Marcin
%\input{decoherence} % Janek
%\input{conclusions} % Rafal
\section{Introduction}
Without much exaggeration one may say that optics is basically the science of light interference.
Light interference effects were behind the final acceptance of classical wave optics and abandoning the Newtonian corpuscular theory of light.
Conceptual insight into the process of interfering light waves prompted development of a number of
measurement techniques involving well controlled interference effects and gave birth to the
field of optical interferometry \citep{Hariharan2003}. At the fundamental level, classical interferometry
is all about observing light intensity variations (intensity fringes) resulting from a change in relative phases between
two (or more) overlapping light waves, e.g.~when a single light beam is split into a number of paths
with tunable optical path-length differences and made to interfere on the screen. The number of applications
is stunning and ranges from basic length measurements via spectroscopic interferometric techniques to
the most spectacular examples involving stellar interferometry and gravitational-wave detectors \citep{Pitkin2011}.

Coherent properties of light as well as the degree of overlap between the interfering beams determine
the visibility of the observed intensity fringes and are crucial for the quality of any interferometric measurement.
Still, when asking for fundamental limitations on precision of estimating e.g.~a phase difference between the arms
of a Mach-Zehnder interferometer, there is no particular answer within a purely classical theory, where both
the light itself as well as the detection process are treated classically. In classical theory intensity of light can in
principle be measured with arbitrary precision and as such allows to detect in principle arbitrary small phase shifts
in an interferometric experiment.

This, however, is no longer true when semi-classical theory is considered in which the
light is still treated classically but the detection process is quantized, so that instead of a continuous intensity parameter
the number of energy quanta (photons) absorbed is being measured. The absorption process within the semi-classical theory
has a stochastic character and the number of photons detected obeys Poissonian or super-Poissonian statistics \citep{Fox2006}.
If light intensity fluctuations can be neglected, the number of photons detected, $N$, follows the Poissonian statistics with
photon number standard deviation $\Delta N=\sqrt{\mN}$, where $\mN$ denotes the mean number of photons detected.
This implies that the determination of the relative phase difference $\varphi$ between the arms of the interferometer,
based on the number of photons detected at the output ports, will be affected by the
relative uncertainty $\Delta \varphi \propto \Delta N/ \mN = 1/\sqrt{\mN}$ referred to as the \emph{shot noise}.
The shot noise plays a fundamental role in the semi-classical theory and in many cases it is indeed the
factor limiting achievable interferometric sensitivities. In modern gravitational wave detectors, in particular, the shot noise
is the dominant noise term in the noise spectral density for frequencies above few hundred \t{Hz} \citep{LIGO2011,LIGO2013,Pitkin2011}.

Yet, the shot noise should not be regarded as a fundamental bound whenever non-classical states of light are considered---we
review the essential aspects of non-classical light relevant from the point of view of interferometry in \secref{sec:states}.
The \emph{sub-Poissonian} statistics characteristic for the so-called squeezed states of light
may offer a precision enhancement in interferometric scenarios by reducing the photon number
fluctuations at the output ports. First proposals demonstrated that sending coherent light together
with the squeezed vacuum state into the two separate input ports of a Mach-Zehnder interferometer
offers estimation precision beating the shot noise
and attaining the $1/\mN^{2/3}$ scaling of the phase estimation precision \citep{Caves1981}, while consideration of more
general two-mode squeezed states showed that even $1/\mN$ scaling is possible \citep{Yurke1986,Bondurant1984}.
A number of papers followed, studying in more detail the phase-measurement probability distribution and
proposing various strategies leading to $1/\mN$ precision scaling \citep{Braunstein1992,Holland1993,Sanders1995,Dowling1998}.
Similar observations have been made in the context of precision spectroscopy, where spin-squeezed states have been shown
to offer a $1/N$ scaling of atomic transition frequency estimation precision, where $N$ denotes the number of atoms employed \citep{Wineland1992,Wineland1994}.
\secref{sec:interferometers} provides a detailed framework for deriving the above results.
However, already at this point a basic intuition should be conveyed that
one \emph{cannot} go beyond the shot noise limit whenever an interferometric
experiment may be regarded in a spirit that each photon
interferes only with itself. In fact, the only possibility of surpassing
this bound is to use light sources that exhibit correlations
in between the constituent photons, e.g.~squeezed light,
so that the interference process may
benefit from the properties of inter-photonic entanglement.

These early results provided a great physical insight into the possibilities of quantum enhanced interferometry and
the class of states that might be of practical interest for this purpose. The papers lacked
generality, however, by considering specific measurement-estimation strategies, in which the error in the estimated phase was
related via a simple error-propagation formula to the variance of some experimentally accessible observable, e.g.~photon number
difference at the two output ports of the interferometer, or by studying the width of the peaks in the shape of the
phase-measurement probability distribution.

Given a particular state of light fed into the interferometer, it is a priori not clear what
is the best measurement and estimation strategy yielding the optimal estimation precision.
Luckily, the tools designed to answer these kinds of questions had already been present
in the literature under the name of \emph{quantum estimation theory} \citep{Helstrom1976,Holevo1982}.
The \emph{Quantum Fisher Information} (QFI) as well as the \emph{cost of Bayesian inference} provide a systematic way
to quantify the ultimate limits on performance of phase-estimation strategies for a given quantum state,
which are already optimized over all theoretically admissible quantum measurements and estimators.
The concept of the QFI and the Bayesian approach to quantum estimation are reviewed in \secref{sec:estimation}.
As a side remark, we should note, that by treating the phase as an evolution parameter to be estimated
and separating explicitly the measurement operators from the estimator function, quantum estimation theory
circumvents some of the mathematical difficulties
that arise if one insists on the standard approach to quantum measurements and attempts to define
the \emph{quantum phase operator} representing the phase observable being measured \citep{Lynch1995,Barnett1992,Noh1992, Summy1990}.

The growth of popularity of the QFI in the field of quantum metrology was triggered by the seminal paper of \citet{Braunstein1994} advocating
the use of QFI as a natural measure of distance in the space of quantum states.
The QFI allows to pin down the optimal probe states that are the most sensitive to small variations of the estimated parameter
by establishing the fundamental bound on the corresponding parameter sensitivity valid for arbitrary measurements and estimators.
Following these lines of reasoning the $1/N$ bound,
referred to as the \emph{Heisenberg limit}, on the phase estimation precision using $N$-photon states has been claimed fundamental
and the \noon states were formally proven to saturate it \citep{Bollinger1996}.
Due to close mathematical analogies between optical and atomic interferometry \citep{Bollinger1996, Lee2002}
similar bounds hold for the problem of atomic transition-frequency estimation and
more generally for any arbitrary unitary parameter estimation problem, i.e.~the one in which an
$N$-particle state evolves under a unitary $U_\varphi^{\otimes N}$, $U_\varphi = \exp(- \mathrm{i} \hat{H} \varphi)$,
with $\hat{H}$ being a general single-particle evolution generator and $\varphi$ the parameter to be estimated
 \citep{Giovannetti2004,Giovannetti2006}.

A complementary framework allowing to determine the fundamental bounds in interferometry is the Bayesian approach,
in which one assumes the estimated parameter to be a random variable itself and explicitly defines its \emph{prior} distribution
to account for the initial knowledge about $\varphi$ before performing the estimation.
In the case of interferometry
the typical choice is the flat prior $p(\varphi)=1/2\pi$ which reflects the initial ignorance of the phase.
The search for the optimal estimation strategies within the Bayesian approach is possible thanks to the general theorem on the optimality
of the \emph{covariant measurements} in estimation problems satisfying certain group symmetry \citep{Holevo1982}.
In the case of interferometry, a flat prior guarantees the phase-shift, U(1), symmetry and as a result the optimal
measurement operators can be given explicitly and they coincide with the eigenstates of the Pegg-Barnett ``phase operator''
\citep{Barnett1992}.
This makes it possible to optimize the strategy over the input states and for simple cost functions allows
to find the optimal probe states \citep{Luis1996, Buzek1999,Berry2000}. In particular,
for the $4 \sin^2(\delta \varphi/2)$ cost function which approximates
the variance for small phase deviations $\delta\varphi$,
the corresponding minimal estimation uncertainty has been found to read $\Delta \varphi \approx \pi/N$
%the optimal precision scaling has been found to be $\approx \pi/N$
for large $N$  providing again a proof of the possibility of achieving the Heisenberg scaling, yet with an additional $\pi$ coefficient. It should be noted
that the optimal states in the above approach that have been found independently in \cite{Summy1990, Luis1996, Berry2000} have completely different structure to the \noon states which are optimal when QFI is considered as the figure of merit. This is not that surprising taking into account that the \noon states
suffer from the $2\pi/N$ ambiguity in retrieving the estimated phase, and hence are designed only to work in the local estimation approach when
phase fluctuations can be considered small.
Derivations of the Heisenberg bounds for phase interferometry using both the QFI and Bayesian approaches
are reviewed in \secref{sec:ideal}. We also
discuss the problem of deriving the bounds for states with indefinite photon number in which case replacing
$N$ in the derived bounds with the mean number of photons $\mN$ is not always legitimate, so that
in some cases the ``naive'' Heisenberg bound $1/\mN$ may in principle be beaten \citep{Anisimov2010,Hofmann2009,Giovanetti2012}.

Further progress in theoretical quantum metrology stemmed from the need to incorporate
realistic decoherence processes in the analysis of the optimal estimation strategies. While deteriorating effects of
noise on precision in quantum-enhanced metrological protocols have been realized by many authors working in the field
\citep{Caves1981,Xiao1987,Huelga1997,Sarovar2006,Shaji2007,Rubin2007,Huver2008,Gilbert2008, Datta2011}, it has long remained an open question to what extent decoherence effects may be circumvented by employing either more sophisticated states of light or more advanced measurements strategies including e.g.~adaptive techniques.

With respect to the most relevant decoherence process in optical applications, i.e.~the photonic losses, strong numerical evidence
based on the QFI \citep{Dorner2009,Demkowicz2009} indicated that in the asymptotic limit of large number of photons
the precision of the optimal quantum protocols approaches $\t{const}/\sqrt{N}$, and hence the gain over classical strategies
is bound to a constant factor. This fact has been first rigorously proven within the Bayesian approach \citep{Kolodynski2010}
and then independently using the QFI \citep{Knysh2011}. Both approaches yielded the same
 fundamental bound on precision in the lossy optical interferometry: $\Delta \varphi  \geq \sqrt{(1-\eta)/(\eta N)}$,
 where $\eta$ is the overall power transmission of an interferometric setup. This bound is also valid after
 replacing $N$ with $\mN$ when dealing with states of indefinite photon number, and moreover can be easily saturated
 using the most popular scheme involving a coherent and a squeezed-vacuum state impinged onto two input ports of the Mach-Zehnder
 interferometer \citep{Caves1981}. This fact also implies that the presently implemented quantum enhanced schemes in gravitational
 wave detection, based on interfering the squeezed vacuum with coherent light, operate
 close to the fundamental bound \citep{Demkowicz2013}, i.e.~they make almost optimal use of non-classical features of light
 for enhanced sensing given light power and loss levels present in the setup.
Based on the mathematical analysis of the geometry of quantum channels \citep{Fujiwara2008,Matsumoto2010}
general frameworks have been developed allowing to find fundamental bounds on quantum precision enhancement
for general decoherence models \citep{Escher2011,Demkowicz2012}.
% including the phase diffusion which may be relevant
%to interferometry when e.g.~mirror-position fluctuations cannot be neglected \citep{Genoni2011,Escher2012}.
These tools allow to investigate optimality of estimation strategies for basically any decoherence model
and typically provide the maximum allowable constant factor improvements forbidding better than $1/\sqrt{N}$ asymptotic scaling of precision.
Detailed presentation of the above mentioned results is given in \secref{sec:decoherence}.

Other approaches to derivation of fundamental metrological bounds have  been advocated recently.
 Making use of the calculus of variations it was shown in \citep{Knysh2014} how to obtain exact formulas for the achievable
asymptotic precision for some decoherence models, while in  \citep{Alipour2014, Alipour2014b} a variants of QFI have been considered
in order to obtain easier to calculate, yet weaker, bounds on precision. While detailed discussion of these approaches is beyond the scope of the present review, in \secref{sub:phasediff} we make use of the result from \citep{Knysh2014} to benchmark the precision bounds derived
in the case of phase diffusion noise model.

The paper concludes with \secref{sec:conclusions} with a summary and an outlook on challenging problems
in the theory of quantum enhanced metrology.

\section{Quantum states of light \label{sec:states}}

The advent of the laser, light-squeezing and single-photon light sources triggered developments in interferometry that could benefit from the non-classical features of light \citep{Buzek1995, Chekhova2011, Torres2011}. In this section, we focus on the quantum-light description of relevance to quantum optical interferometry. We discuss the mode description of light and the most commonly used states in quantum optics---the Gaussian states. In the end, we consider states of definite photon-number and study their particle-description, in particular, investigating their relevant entanglement properties.

%%%%%%%%%%%%%%%%%%%%%%%%%%%%%%%%%%

\subsection{Mode description}

Classically, electromagnetic field can be divided into orthogonal modes distinguished by their characteristic spatial, temporal and polarization properties. This feature survives in the quantum description of light, where formally we may associate a separate quantum subsystem with each of these modes. Each subsystem is described by its own Hilbert space and, because photons are bosons, can be occupied by an arbitrary number of particles \citep{Mandel1995, Walls1995}.
%Usually it is not specified which photon is in which state, one rather consider states in an occupation basis, where it is specified only how many photons are in each of the modes.
The most general $M$-mode state of light may be then written as:
\begin{equation}
\label{eq:rhogeneral}
\rho=\sum_{\mathbf{n},\mathbf{n}^\prime}\rho_{\mathbf{n},\mathbf{n}^{\prime}}\ket{\mathbf{n}} \bra{\mathbf{n}^\prime}, \quad \t{Tr}(\rho)=1, \quad \rho \geq 0
\end{equation}
with $\mathbf{n} = \{n_1,\dots,n_M\}$ %$\sum_{\mathbf{n}} |c_{\mathbf{n}}|^2 = 1$,
and $\ket{\mathbf{n}}= \ket{n_1} \otimes \dots \otimes \ket{n_M}$  representing a Fock state with \emph{exactly} $n_i$ photons occupying the $i$-th mode.
States $\ket{n_i}$ may be further expressed in terms of the respective creation and annihilation operators
${\hat{a}_i}^\dagger$, $\hat{a}_i$ obeying $[\hat{a}_i,\hat{a}_j^\dagger]=\delta_{ij}$:
\begin{equation}
\ket{n_i}=\frac{\hat{a}_i^{\dagger n}}{\sqrt{n!}}\ket{0}, \  \hat{a}\ket{n_i}= \sqrt{n_i} \ket{n_i-1}, \  \hat{a}_i^\dagger \ket{n_i} = \sqrt{n_i+1}\ket{n_i+1},
\end{equation}
where $\ket{0}$ is the vacuum state with no photons at all.

In the context of optical interferometry, modes are typically taken to be distinguishable by their spatial separation, corresponding to different arms of an interferometer, whereas
the various optical devices such as mirrors, beam-splitters or phase-delay elements transform the state on its way through the
interferometer. Eventually, photon numbers are detected in the output modes allowing to infer the value of the relative phase difference
between the arms of the interferometer.

In many applications, the above standard state representation may not be convenient and phase-space description is used instead---in particular, the Wigner function representation \citep{Wigner1932,Schleich2001}. Adopting the convention in which the quadrature operators read $\hat{x}_i=\hat{a}^\dagger_i+\hat{a}_i$ and $\hat{p}_i=\mathrm{i}(\hat{a}_i^\dagger-\hat{a}_i)$,
the Wigner function may be regarded as a quasi-probability distribution on the quadrature phase space:
\begin{equation}
W(\textbf{x},\textbf{p}) =
\frac{1}{(2\pi^{2})^M} \int\!\! \t{d}^{M}\textbf{x}^\prime\,\t{d}^{M}\textbf{p}^\prime  \;
\Tr\!\left( \rho \, \e^{\mathrm{i}[\textbf{p}^\prime(\hat{\textbf{x}}-\textbf{x}) - \textbf{x}^\prime(\hat{\textbf{p}} -\textbf{p}) ]} \right),
\end{equation}
where $\textbf{x}=\{x_1,\dots,x_M\}$, $\hat{\textbf{x}}=\{\hat{x}_1,\dots,\hat{x}_M\}$
and similarly for $\mathbf{p}$ and $\hat{\mathbf{p}}$.
As a consequence, the Wigner function is real, integrates to $1$ over the whole phase space and its marginals yield the correct
probability densities of each of the phase space variables.
Yet, since it may take negative values it cannot be regarded as a proper probability distribution. Most importantly, it may be reconstructed from experimental data either by tomographic methods \citep{Schleich2001} or by direct probing of the phase space \citep{Banaszek1999}, and hence is an extremely useful representation both for theoretical and experimental purposes.

%In general, different modes of light do not interact with each other if there is no kind of coupling device involved in the process, which is %reflected in a fact that annihilation operators for different modes commutes with each other. We may however make them to interact by some devices %which act on two or more modes, for example beam-splitters or some nonlinear crystals. In mathematical description, those devices mix annihilation and %creation operators of different modes effectively acting as some Bogoliubov transformation. This enables one to create various entangled states %between modes of light which is of great interest in current physics and plays a great role especially in quantum-enhanced interferometry.

%%%%%%%%%%%%%%%%%%%%%%%%%%%%%%%%%%

\subsection{Gaussian states}
\label{sec:gaussianstates}

From the practical point of view, the most interesting class of states are the Gaussian states \citep{Paris1995,Olivares2007,Pinel2012,Pinel2013}.
The great advantage of using them is that they are relatively easy to produce in the laboratory with the help of standard laser-sources
and non-linear optical elements, which allow to introduce non-classical features such as squeezing or entanglement.
Gaussian states have found numerous application in various fields of quantum information processing \citep{Adesso2006}
and are also extensively employed in quantum metrological protocols.

Gaussian states of $M$ modes are fully characterized by their first and second quadrature moments
 and are most conveniently represented using the Wigner function which is then just a
multidimensional Gaussian distribution
\begin{equation}
W(\textbf{z})=\frac{1}{(2\pi)^M\sqrt{\det \sigma}}\e^{-\frac{1}{2}
({\mathbf z}-\mean{\hat{\mathbf{z}}})^{T}\sigma^{-1}(\mathbf{z}-\mean{\hat{\mathbf{z}}})},
\end{equation}
where for a more compact notation we have introduced: the phase space variable $\mathbf{z}=\{x_1,p_1,\dots,x_M,p_M\}$,
the vector containing mean quadrature values $\mean{\hat{\mathbf{z}}}=\{\mean{\hat{x}_1},\mean{\hat{p}_1},\dots,\mean{\hat{x}_M},\mean{\hat{p}_M}\}$,
$\mean{\hat{z}_i}= \t{Tr}(\hat{z}_i \rho) = \int \t{d}^{2M}\mathbf{z} \, W(\mathbf{z})z_i$,
and the $2M$ dimensional {\it covariance matrix} $\sigma$:
\begin{equation}\label{eq:wigner}
\sigma_{i,j}=\frac{1}{2}\langle{ \hat{z}}_i{ \hat{z}}_j+{ \hat{z}}_j{ \hat{z}}_i\rangle-\langle{ \hat{z}}_i\rangle\langle{ \hat{z}}_j\rangle.
\end{equation}
Gaussian states remain Gaussian under arbitrary evolution involving Hamiltonians at most quadratic in the quadrature operators, what
includes all passive devices such as beam-splitters and phase-shifters as well as single- and multi-mode squeezing operations.
Below we focus on a few classes of Gaussian states highly relevant to interferometry.
%The point which mostly distinguish Gaussian states from Fock states or states of the form of eq.~(\ref{eq:genstate}) is that they do not posses an %exact photon number, they rather have average photon number $\mN$, defined analogously as $\mN=\bra{\Psi}\hat{N}\ket{\Psi}$. In the Fock state %basis this corresponds to the fact that Gaussian states are superpositions or mixtures of the Fock states with different total photon numbers. \

\subsubsection {Coherent states\label{sub:coherent}}
Coherent states are the Gaussian states with identity covariance matrix $\sigma = \openone$,
so that the uncertainties are equal for all quadratures saturating the Heisenberg uncertainty relations $\Delta^2 x_i \Delta^2 p_i = 1$ and
there are no correlations between the modes. Mean values of quadratures may be arbitrary and correspond to the
coherent state complex amplitude $\boldsymbol{\alpha} = (\mean{\hat{\textbf{x}}} + \mathrm{i} \mean{\hat{\textbf{p}}})/2.$
These are the states produced by any phase-stabilized laser, what makes them almost a fundamental tool in the theoretical description of many quantum optical experiments.
Moreover, coherent sates have properties that resemble features of classical light, and thus enable to establish a bridge between the quantum and classical descriptions of light.

In the standard representation, an $M$-mode coherent state $ \ket{\boldsymbol{\alpha}} = \ket{\alpha_1} \otimes \dots \otimes \ket{\alpha_M}$
is a tensor product of single-mode coherent states, whereas a single-mode coherent state
is an eigenstate of the respective annihilation operator:
\begin{equation}
\hat{a}\ket{\alpha}=\alpha\ket{\alpha},
\end{equation}
where $\alpha=|\alpha|\e^{i\theta}$ and $|\alpha|,\,\theta$ are respectively the amplitude and the phase of a coherent state.
Equivalently,  we may write
\begin{equation}
\ket{\alpha}=\hat{D}(\alpha)\ket{0},
\end{equation}
where $\hat{D}(\alpha)=\e^{\alpha\hat{a}^\dagger-\alpha^*\hat{a}}$ is the so-called {\it displacement operator}
or write the coherent state explicitly as a superposition of consecutive Fock states:
\begin{equation}
\ket{\alpha}=\e^{-|\alpha|^2/2}\sum_{n=0}^{\infty}\frac{\alpha^n}{\sqrt{n!}}\ket{n}.
\label{eq:coherentFock}
\end{equation}
From the formula above, it is clear that coherent states do not have a definite photon number
and if a photon number $n$ is measured its distribution follows the Poissonian
statistics $P(n)=\e^{-|\alpha|^2}\frac{|\alpha|^{2n}}{n!}$
with average $\langle n\rangle=|\alpha|^2$ and standard deviation $\Delta n = |\alpha|$.
Thus, the relative uncertainty $\Delta n/\langle n\rangle$ in the measured photon number scales like
$1/\sqrt{\langle n\rangle}$ and hence
in the classical limit of large $\langle n\rangle$
the beam power may be determined up to arbitrary precision.
Moreover, the evolution of the coherent state amplitude is identical to the evolution of
a classical-wave amplitude. In particular, an optical phase delay $\varphi$ transforms the state $\ket{\alpha}$ into
$\ket{\alpha\,\e^{\mathrm i \varphi} }$.
These facts justify a common jargon of calling coherent states the classical states of light, even though
for relatively small amplitudes different coherent states may be hard to distinguish due to their non-orthogonality
$|\braket{\alpha}{\beta}|^2=e^{|\alpha - \beta|^2}$.
More generally, we call $\rho_{\t{cl}}$ a classical state of light if and only if it can be written as a mixture of coherent states:
\begin{equation}
\label{eq:class}
\rho_{\t{cl}} = \int\!\! \t{d}^{2M} \boldsymbol{\alpha}\; P(\boldsymbol{\alpha}) \ket{\boldsymbol{\alpha}}\bra{\boldsymbol{\alpha}}
\end{equation}
with $P(\boldsymbol\alpha)\geq 0$, which is equivalent to the statement that $\rho_{\t{cl}}$ admits a non-negative Glauber $P$-representation
\citep{Walls1995,Glauber1963}. Classical states are often used as a benchmark to test the degree of possible quantum
enhancement which may be obtained by using more general states outside this class.

\subsubsection{Single-mode squeezed states}

\begin{figure}[t!]
  \includegraphics[width=0.99\columnwidth]{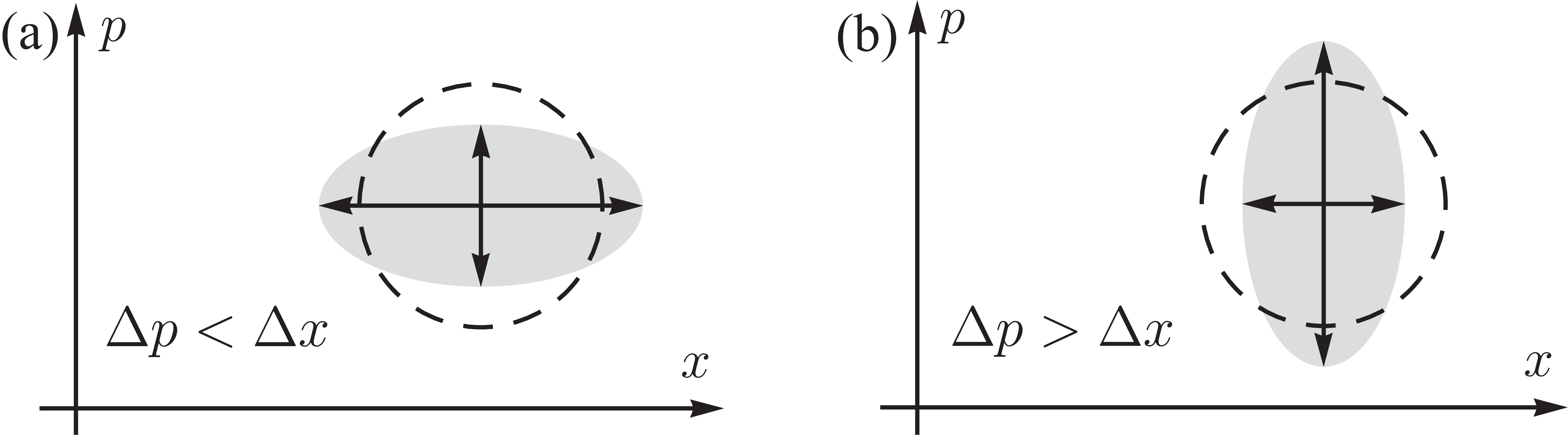}
\caption{Phase-space diagrams denoting uncertainties in different quadratures
for momentum squeezed states (a) and for position squeezed states (b).
Dashed circles denote corresponding uncertainties for coherent states $\Delta x=\Delta p$.
\label{fig:squeezed}}
\end{figure}

Heisenberg uncertainty principle imposes that $\Delta x\Delta p\geq 1$ for all possible quantum states.
Single-mode states that saturate this inequality are called the \emph{single-mode squeezed states} \citep{Walls1995}.
As mentioned above, coherent states fall into such a category serving as a special example for which $\Delta x=\Delta p=1$.
Yet, as for general squeezed states $\Delta x\neq\Delta p$, the noise
in one of the quadratures can be made smaller than in the other.
Formally, a single-mode squeezed state may always be expressed as
\begin{equation}\label{eq:squeezed}
\ket{\alpha,r}=\hat{D}(\alpha)\hat{S}(r)\ket{0},
\end{equation}
where $\hat{S}(r)=\exp(\frac{1}{2}r^*\hat{a}^2-\frac{1}{2}r\hat{a}^{\dagger 2})$ is the
\emph{squeezing operator}, $r=|r|\e^{\ii\theta}$ is a complex number and $|r|$ and $\theta$
are the \emph{squeezing factor} and the \emph{squeezing angle} respectively.
In fact, any pure Gaussian one-mode state may be written in the above form.
For $\theta=0$, uncertainties in the quadratures $x$ and $p$ read $\Delta x=\e^{-r}$ and $\Delta p=\e^{r}$---reduction
of noise in one quadrature is accompanied by an added noise in the other one.
This may be conveniently visualized in the phase-space picture by error disks representing
uncertainty in quadratures in different directions, see \figref{fig:squeezed}.
In such a representation squeezed states correspond to ellipses while coherent states are represented by circles.
Importantly, the fact that the uncertainty of one of the quadratures can be less relatively to the other makes it possible
to design an interferometric scheme where the measured photon number fluctuations are below
that of a coherent state and allows for a sub-shot noise phase estimation precision, see \secref{sec:interferometers}.
%\begin{figure}[h!]\label{fig:squeezed}
%\begin{centering}
%\epsfxsize=200pt
%\epsfxsize=200pt\epsfbox{squeezedp1.eps}\epsfxsize=200pt\epsfbox{squeezed3.eps}
%\caption{Phase-space diagrams denoting uncertainties in different quadratures for momentum squeezed states (left) and position squeezed state for (right). Dashed circles denote %corresponding uncertainties for coherent states. Squeezed state have different uncertainties $\Delta x\neq\Delta p$ while for coherent state $\Delta x=\Delta p$.}
%\end{centering}
%\end{figure}%
As squeezed states in general cannot be described as mixtures of coherent states,
they are non-classical and their features cannot be fully described by classical electrodynamics.
Nevertheless, they can be relatively easily prepared using non-linear optical elements in the process
of parametric down conversion \citep{Bachor2004}.

The special type of squeezed states which is most relevant from the metrological perspective is
the class of the \emph{squeezed vacuum states} that possess vanishing mean
values of their quadratures, i.e.~$\mean{\hat{z}}=0$:
\begin{equation}
\ket{r}=\hat{S}(r)\ket{0}.
\end{equation}
In the Fock basis a squeezed vacuum state reads
\begin{equation}\label{eq:SV}
\ket{r}=\frac{1}{\sqrt{\cosh r}}\sum_{n=0}^{\infty}\frac{H_n(0)}{\sqrt{n!}}\Big(\frac{\tanh r}{2}\Big)^{\frac{n}{2}}\e^{\ii \frac{n\,\theta}{2}}\ket{n},
\end{equation}
where $H_n(0)$ denotes values of $n$-th Hermite polynomial at $x=0$.
As for odd $n$ $H_n(x)$ is antisymmetric and thus $H_n(0)=0$,
it follows that squeezed vacuum states are superpositions of Fock states with only \emph{even} photon numbers.
The average number of photons in a squeezed vacuum state is given by $\langle n\rangle=\sinh^2r$, what means that,
despite their name, a squeezed vacuum states contain photons, possibly a lot of them.
%Because in quantum interferometry there are rather used SV states than that of eq.~(\ref{eq:squeezed}) we will further refer to them as squeezed %states whereas states of the general form of eq.~(\ref{eq:squeezed}) would be called general Gaussian one mode states.

\subsubsection{Two-mode squeezed states}

The simplest non-classical two-mode Gaussian state is the so-called two-mode squeezed vacuum state or the \emph{twin-beam
 state} \citep{Walls1995}. Mathematically, such a state is generated from the vacuum by a two-mode squeezing operation, so that
\begin{equation}
\label{eq:TBS}
\ket{\xi}_2=\hat{S}_2(\xi)\ket{0,0},
\end{equation}
where $\hat{S}_2(\xi)=\exp(\xi^*\hat{a}\hat{b}-\xi\hat{a}^{\dagger}\hat{b}^{\dagger})$ and $\xi=|\xi|\e^{\ii\theta}$,
whereas in the Fock basis it reads
\begin{equation}\label{eq:TBSFock}
\ket{\xi}_2=\frac{1}{\cosh{\xi}}\sum_{n=0}^{\infty}(-1)^n \e^{\ii\theta}\tanh^n \xi\,\ket{n,n}.
\end{equation}
A notable feature of the twin-beam state, which may be clearly seen from \eqnref{eq:TBSFock}, is that it is not a product of squeezed states in modes $a$ and $b$, but rather it is correlated in between them being a superposition of terms with the same number of photons in both modes.
Its first moments of all quadratures are zero, $\mean{\hat{z}}=0$, whereas in the case of $\xi=|\xi|$ its covariance matrix has a particularly simple form:
\begin{equation}
\sigma=\left(
\begin{array}{cccc}
\cosh(2\xi) & 0 & \sinh(2\xi) & 0\\
0 & \cosh(2\xi) & 0 & -\sinh(2\xi)\\
\sinh(2\xi) & 0 & \cosh(2\xi) & 0\\
0 & -\sinh(2\xi) & 0 & \cosh(2\xi)
\end{array}\right).
\end{equation}
Such a covariance matrix clearly indicates the presence of correlations between the modes and since the state \eref{eq:TBS} is pure this implies
immediately the presence of the \emph{mode-entanglement}. In fact, in the limit of large squeezing coefficient $|\xi|\to\infty$ such twin-beam state
becomes the original famous Einstein-Podolsky-Rosen state \citep{Banaszek1998,Adesso2006} that violates assumptions of any realistic local hidden variable theory.

Twin-beam states may be generated in a laboratory by various non-linear processes such as
four- and three-wave mixing \citep{Reid1988,Bachor2004}. Alternatively, they may be produced by mixing
two single-mode squeezed vacuum states with opposite squeezing angles on a fifty-fifty beam-splitter.

\subsubsection{General two-mode Gaussian states}

General two-mode Gaussian state is rather difficult to write in the Fock basis, so  it is best characterized by its $4 \times 4$ real symmetric covariance matrix
\begin{equation}
\label{eq:cov}
\sigma=\left(
\begin{array}{cc}
{[}\sigma_{11}{]} & {[}\sigma_{12}{]}\\
{[}\sigma_{21}{]} & {[}\sigma_{22}{]}
\end{array}
\right),
\end{equation}
where $\sigma_{ij}$ represent blocks with $2 \times 2$ matrices describing correlations between the $i$-th and the $j$-th mode,
and the vector of the first moments $\mean{\hat{\mathbf{z}}}=\{\mean{\hat{x}_1},\,\mean{\hat{p}_1},\,\mean{\hat{x}_2},\,\mean{\hat{p}_2}\}$. This in total gives up to fourteen real parameters describing the state:
ten covariances, two displacement amplitudes and two phases of displacement.
General Gaussian states are in principle feasible within current technological state of art, as any pure Gaussian state can theoretically be generated from the vacuum by utilizing only a combination of one-, two-mode squeezing and displacement operations with help of beam-splitters and one-mode rotations \citep{Adesso2006}. Furthermore, mixed Gaussian states are obtained as a result of tracing out some of the system degrees of freedom, which
is effectively the case in the presence of light losses, or by adding a Gaussian noise to the state.

%%%%%%%%%%%%%%%%%%%%%%%%%%%%%%%%%%

\subsection{Definite photon number states \label{sub:def_ph_N_states}}

Gaussian states are important from the practical point of view due to the relative ease with which their may be prepared.
From a conceptual point of view, however, when asking fundamental questions on limits to quantum enhancement in interferometry,
states with a definite photon number prove to be a better choice. The main reason is that photons are typically regarded as a resource in interferometry and when benchmarking different interferometric schemes it is natural to restrict the class of states with the same number of photons, i.e.~the same resources consumed.
A general $M$-mode state consisting of $N$ photons is given by:
\begin{equation}
\rho_N = \sum_{|\mathbf{n}| = |\mathbf{n^\prime}| = N} \rho_{\mathbf{n},\mathbf{n}^\prime} \ket{\mathbf{n}}\bra{\mathbf{n}^\prime},
\end{equation}
where  $|\mathbf{n}| =\sum_i n_i $, so that the summation is restricted only to terms with exactly $N$ photons in all the modes.
Apart from the vacuum state $\ket{0}$ no Gaussian state falls into this category.

States with an exact photon number are extensively used in other fields of quantum information processing, including quantum communication and quantum computing \citep{Kok2007,Pan2012}. Most of the quantum computation and communication schemes are
designed with such states in mind, as they provide the most intuitive and clear picture of the role the quantum features
play in these tasks. For large $N$, however, states with a definite photon numbers are notoriously hard to prepare
and states with $N$ of the order of $10$ can only be produced
with the present technology pushed to its limits \citep{Hofheinz2008,Sayrin,Torres2011}.
When considering states with definite $N$, it is also possible to easily switch between the mode- and  particle-description of the states
of light, which is a feature that we discuss in the following section.

\subsection{Particle description \label{sub:par_descr}}

When dealing with states of definite photon number, instead of thinking about modes as quantum subsystems that possess some
number of excitations (photons), we may equivalently consider the ``first quantization'' formalism and regard photons themselves as elementary subsystems.
Fundamentally, photons are indistinguishable particles and since they are bosons their wave function should always be permutation-symmetric.
Still, it is common in the literature to use a description in which photons are regarded as distinguishable particles and adopt
a notation such that
\begin{equation}
\ket{\textbf{m}} = \ket{m_1}_1 \otimes \ket{m_2}_2 \otimes \dots \otimes \ket{m_N}_N
\label{eq:|m>}
\end{equation}
denotes a product state of $N$ photons, where the $i$-th photon occupies the mode $m_i$.
We explicitly add subscripts to the kets above labeling each constituent photon, in order to distinguish this notation from the mode description
of \eqnref{eq:rhogeneral}, where kets denoted various modes and not the distinct particles.
The description \eref{eq:|m>} is legitimate
provided there are some degrees of freedom that ascribe a meaning to the statement \emph{``the $i$-th photon''}.
For example, in the case when photons are prepared in non-overlapping time-bins, the time-bin degree of freedom
plays the role of the label indicating a \emph{particular} photon, whereas the spatial characteristics determine the state of a \emph{given} photon.
Nevertheless, if we assume the overall wave function describing also the temporal degrees of freedom of the complete state
to be fully symmetric, the notion of the ``$i$-th photon'' becomes meaningless.

A general pure state of $N$ ``distinguishable'' photons has the form:
\begin{equation}
\ket{\psi_N} = \sum_{\textbf{m}} c_{\textbf{m}}\, \ket{\textbf{m}}
\end{equation}
where $\sum_{\textbf{m}} |c_{\textbf{m}}|^2=1$. If indeed there is no additional degree of freedom that makes the notion
of ``the $i$-th photon'' meaningful, the above state should posses the symmetry property such that
$c_{\textbf{m}} = c_{\Pi(\textbf{m})}$, where $\Pi$ is an arbitrary permutation of the $N$ indices.

Consider for example a Fock state $\ket{\textbf{n}}=\ket{n_1}\dots\ket{n_M}$ of $N=n_1+\dots+n_M$ indistinguishable photons in $M$ modes.
In the particle description the state has the form:
\begin{multline}\label{eq:sym}
\ket{\textbf{n}}=\sqrt{\frac{n_1!\dots n_M!}{N!}} \times \\ \sum_{\Pi}\ket{\Pi(\{\underbrace{1,\dots,1}_{n_1},\underbrace{2 \dots, 2}_{n_2}, \dots,
\underbrace{M,\dots,M}_{n_M}\})},
\end{multline}
where the sum is performed over all non-trivial permutations $\Pi$ of the indices inside the curly brackets \citep{Shankar1994}.
Since all quantum states may be written in the Fock basis representation, by the above
construction one can always translate any quantum state to the particle description.

\subsection{Mode vs particle entanglement \label{sub:mode_vs_part_ent}}

One of the most important features which makes the quantum theory different from the classical one is the notion of
entanglement \citep{Horodecki2009}.
This phenomenon plays also an important role in quantum metrology and is often claimed to be the crucial resource for the enhancement
 of the measurement precision \citep{Giovannetti2006, Pezze2009}. Conflicting statements can be found in the literature, however,
as some of the authors claim that entanglement is not indispensable to get a quantum precision enhancement \citep{Benatti2013}.
This confusion stems simply from the fact that entanglement is a relative concept dependent on the way we divide the relevant Hilbert space into
particular subsystems. In order to clarify these issues, it is necessary to explicitly study relation between mode- and particle-entanglement,
i.e.~entanglement with respect to different tensor product structures used in the two descriptions.

Firstly, let us go through basic definitions and notions of entanglement. The state $\rho_{AB}$ of two parties $A$ and $B$ is called
separable if and only if one can write it as a mixture of product of states of individual subsystems:
\begin{equation}
\rho^{(AB)}=\sum_{i}p_i\;\rho_i^{(A)}\otimes\rho_i^{(B)}, \quad p_i \geq 0.
\end{equation}
Entangled states are defined as all states that are not separable. A crucial feature of entanglement is that it depends on the {\it division} into subsystems. For example, consider three qubits $A$, $B$ and $C$ and their joint quantum state $\rho^{(ABC)}=\sum_{i,j=0}^{1}\frac{1}{2}\ket{i}\bra{j} \otimes \ket{i}\bra{j}\otimes\ket{0}\bra{0}$. This state is separable with respect to the $AB|C$ cut but is entangled with respect
to the $A|BC$ cut.

As a first example, consider a two-mode Fock state $\ket{1}_a\ket{1}_b$
which represents  one photon in mode $a$ and one photon in mode $b$. This state written
in the mode formalism of \eqnref{eq:rhogeneral} is clearly separable. On the other hand, photons are indistinguishable bosons and if we would like to write their state in the particle formalism of \eqnref{eq:|m>} we have to symmetrize over all possible permutations of particles, thus obtaining the state
\begin{equation}
\ket{1}_a\ket{1}_b=\frac{1}{\sqrt{2}}(\ket{a}_1\ket{b}_2+\ket{b}_1\ket{a}_2).
\end{equation}
In this representation the state is clearly entangled.
We may thus say that the state contains \emph{particle} entanglement but \emph{not} the \emph{mode} entanglement.

If, however, we perform the Hong-Ou-Mandel experiment and send the $\ket{1}_a\ket{1}_b$  state through a balanced beam-splitter
which transforms mode annihilation operators as $\hat{a} \rightarrow (\hat{a} + \hat{b})/\sqrt{2}$,
$\hat{b} \rightarrow (\hat{a} - \hat{b})/\sqrt{2}$, the resulting state reads:
\begin{equation}\label{eq:11BS}
\ket{1}_a\ket{1}_b\to\frac{1}{\sqrt{2}}(\ket{2}_a\ket{0}_b - \ket{0}_a\ket{2}_b)=\frac{1}{\sqrt{2}}(\ket{a}_1\ket{a}_2 - \ket{b}_1\ket{b}_2),
\end{equation}
which is \emph{both} mode- and particle-entangled. Mode entanglement emerges because the beam-splitter is a joint operation over two modes
that introduces correlations between them. On the other hand, it is a local operation with respect to the particles, i.e.~it can be written as
$U \otimes U$ in the particle representation, and does not couple photons with each other. Thus, using a beam-splitter one may change mode entanglement but not the content of particle entanglement.

As a second example, consider two modes of light, $a$ and $b$, each of them in coherent state with the same amplitude $\alpha$, $\ket{\alpha}_a\ket{\alpha}_b$. This state clearly has \emph{no} mode entanglement.
Since this state does not have a definite photon number, in order to ask questions about the particle entanglement
we first need to consider its projection on one of the $N$-photon subspaces---one can think of a non-demolition
total photon-number measurement yielding result $N$. After normalizing the projected state we obtain:
\begin{equation}
\label{eq:sector}
%\ket{\psi_N}=[\ket{\alpha}_a\ket{\alpha}_b]^{(N)}=e^{-|\alpha|^2}\sum_{n=0}^{N}\frac{\alpha^N}{\sqrt{n!(N-n)!}}\ket{n}_a\ket{N-n}_b
\ket{\psi_N}=[\ket{\alpha}_a\ket{\alpha}_b]^{(N)}=\frac{1}{\sqrt{2^N}}\sum_{n=0}^{N}\sqrt{N \choose n}\ket{n}_a\ket{N-n}_b,
\end{equation}
which in the particle representation reads:
\begin{eqnarray}
\ket{\psi_N}=\frac{1}{\sqrt{2^N}}\bigotimes_{i=1}^N \left(\ket{a}_i+\ket{b}_i\right)
\end{eqnarray}
and is clearly a separable state. The fact that products of coherent states contain no particle entanglement
is in agreement with our definition of classical state given in \eqnref{eq:class} being a mixture of products of coherent states.
A \emph{classical state} according to this definition will contain \emph{neither} mode nor particle entanglement.

As a last example, consider the case of particular interest for quantum interferometry, i.e.
a coherent state of mode $a$ and a squeezed vacuum state of mode $b$: $\ket{\alpha}_a\ket{r}_b$.
Again this state has \emph{no} mode entanglement. On the other hand it \emph{is} particle entangled. To see this, consider e.g.~the two-photon sector,
which up to irrelevant  normalization factor reads:
\begin{multline}\label{eq:squeezedcoherent}
[\ket{\alpha}\ket{r}]^{(N=2)} \propto  \alpha^2\ket{2}_a\ket{0}_b+\tanh r\ket{0}_a\ket{2}_b = \\
 = \alpha^2 \ket{a}_1\ket{a}_2 + \tanh r \ket{b}_1\ket{b}_2
\end{multline}
and contains particle entanglement provided both $\alpha$ and $r$ are non-zero.
We argue and give detailed arguments in \secref{sec:ideal} that it is indeed the particle entanglement
and \emph{not} the mode entanglement that is relevant in quantum-enhanced interferometry. See also \citet{Plenio2014}
for more insight into the relation between mode and particle entanglement.
%This also resolves the issue of squeezing mentioned before: although a single-mode squeezed state is not entangled, in practice it has to be always %accompanied by some other, reference coherent beam which eventually provides a state of the form of \eqnref{eq:squeezedcoherent} which contains %particle entanglement.

\section{Mach-Zehnder interferometry \label{sec:interferometers}}

%Mach-Zehnder interferometer, Michelson, Fabry-Perrot  [mozna chyba skorzystac z Yurke PRA 33 4033 (1986)],
%introduce angular momentum formalism (to sie chyba nazywa Schwinger representation), and analyze precision via simple
%error propagation formulas - in particular recover $1/N^{3/2}$ result by Caves etc. Reference beam

We begin the discussion of quantum-enhancement effects in optical interferometry by discussing the
paradigmatic model of the Mach-Zehnder (MZ) interferometer. We analyze the most popular interferometric schemes
involving the use of coherent and squeezed states of light accompanied by a basic measurement-estimation procedure,
in which the phase is estimated
%via simple formula
based on the value of the photon-number difference between the two output ports of the interferometer.
Such a protocol provides us with a benchmark that we may use in the following sections
when discussing the optimality of the interferometry schemes both with respect to the states of light used as well as the measurements and the estimation
procedures employed.

\subsection{Phase-sensing uncertainty \label{sub:phase_sens_uncer}}

\begin{figure}
  \centering
  \includegraphics[width=0.9\columnwidth]{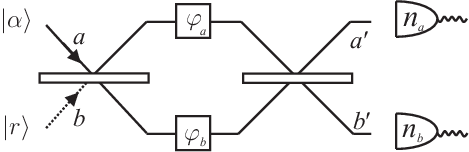}
  \caption{The Mach-Zehnder interferometer, with two input light modes $a$, $b$ and two output modes $a^\prime, b^\prime$. In a standard configuration a coherent state of light $\ket{\alpha}$ is sent into mode $a$.
In order to obtain quantum enhacement, one needs to make use of the  $b$ input port also, sending e.g.~the squeezed vacuum state $\ket{r}$.
}
\label{fig:machzehnder}
\end{figure}

In the standard MZ configuration, depicted in \figref{fig:machzehnder}, a coherent state of light is split on a balanced beam-splitter, the two beams acquire
phases $\varphi_a$, $\varphi_b$ respectively, interfere on the second beam-splitter and finally the photon numbers $n_a$, $n_b$ are measured at the output ports. Let $\hat{a}, \hat{b}$ and $\hat{a}^\prime, \hat{b}^\prime$ be the annihilation operators corresponding to the two input and the two output modes respectively. The combined action of the beam-splitters and the phase delays results in the effective transformation of the annihilation operators:
\begin{multline}
\label{eq:inputoutputMZ}
\mat{c}{\hat{a}^\prime \\ \hat{b}^\prime} = \frac{1}{2}\mat{cc}{1 & \mathrm{i} \\ \mathrm{i} & 1} \mat{cc}{\e^{i\varphi_a} & 0 \\ 0 & \e^{\mathrm{i} \varphi_b}} \mat{cc}{1 & -\mathrm{i} \\ -\mathrm{i} & 1} \mat{c}{\hat{a} \\ \hat{b}}= \\
=\e^{\mathrm{i}(\varphi_a+\varphi_b)/2}\mat{cc}{ \cos(\varphi/2) & -\sin(\varphi/2) \\ \sin(\varphi/2) & \cos(\varphi/2)} \mat{c}{\hat{a} \\ \hat{b}},
\end{multline}
where $\varphi=\varphi_b  - \varphi_a$ is the relative phase delay
and for convenience we assume that the beams acquire a $-\pi/2$ or $\pi/2$ phase when transmitted through the first or the second beam-splitter respectively. The common phase factor $\e^{\mathrm{i}(\varphi_a + \varphi_b)/2}$ is irrelevant for further discussion in this section and will be omitted.
%dropped.

In order to get a better insight into the quantum-enhancement effects in the operation of the MZ interferometer, it is
useful to make use of the so-called Jordan-Schwinger map \citep{Schwinger1965} and analyse the action of the MZ interferometer in terms of
the algebra of the angular momentum operators \citep{Yurke1986}. Let us define the operators:
\begin{equation}
\label{eq:joperators}
\hat{J}_x = \frac{1}{2}(\hat{a}^\dagger \hat{b} + \hat{b}^\dagger \hat{a}), \quad
\hat{J}_y = \frac{\ii}{2}(\hat{b}^\dagger \hat{a} - \hat{a}^\dagger \hat{b}), \quad
\hat{J}_z = \frac{1}{2}(\hat{a}^\dagger \hat{a} - \hat{b}^\dagger \hat{b} ),
\end{equation}
which fulfill the angular momentum commutation relations $[\hat{J}_i,\hat{J}_j]= \mathrm{i} \epsilon_{ijk} \hat{J}_k$
while the corresponding square of the total angular momentum reads:
\begin{equation}
\hat{J}^2 = \frac{\hat{N}}{2}\left( \frac{\hat{N}}{2} +1\right), \quad \hat{N} = \hat{a}^\dagger \hat{a} + \hat{b}^\dagger \hat{b},
\label{eq:J2_op}
\end{equation}
where $\hat{N}$ is the total photon number operator. The action of linear optical elements appearing in the MZ interferometer
can now be described as rotations in the abstract spin space:
$\hat{a}^\prime = U \hat{a} U^\dagger$, $\hat{b}^\prime = U \hat{b} U^\dagger$, $U = \exp(- \mathrm{i} \alpha \hat{\mathbf{J}} \cdot \mathbf{s})$,
where $\hat{\mathbf{J}}=\{\hat{J}_x,\hat{J}_y,\hat{J}_z\}$ and $\alpha$, $\mathbf{s}$ are the angle and the axis of the rotation respectively.
In particular, the balanced beam splitter is
a rotation around the $x$ axis by an angle $\pi/2$: $U=\exp(- \mathrm{i} \frac{\pi}{2}  \hat{J}_x)$, while the phase delay is a $\varphi$ rotation
around the $z$ axis: $U=\exp( -\mathrm{i} \varphi  \hat{J}_z)$. Instead of analysing the transformation of the annihilation operators,
it is more convenient to look at the corresponding transformation of the $J_i$ operators themselves:
\begin{multline}
\label{eq:jtransformMZ}
\mat{c}{\hat{J}^\prime_x \\ \hat{J}^\prime_y\\\hat{J}^\prime_z}  \!\! =  \!\!\mat{ccc}{1 & 0 & 0 \\ 0  & 0  & 1 \\ 0 & -1 & 0} \!\mat{ccc}{ \cos \varphi  &- \sin\varphi &0 \\ \sin\varphi & \cos\varphi & 0 \\0 & 0 & 1 \\ } \times  \\ \times\!\mat{ccc}{1 & 0 & 0 \\ 0  & 0  & -1 \\ 0 & 1 & 0}\!\mat{c}{\hat{J}_x \\ \hat{J}_y\\\hat{J}_z}
  =  \!\!\mat{ccc}{\cos\varphi & 0 & \sin\varphi \\ 0  & 1  & 0 \\ -\sin \varphi & 0 & \cos\varphi}\!\mat{c}{\hat{J}_x \\ \hat{J}_y\\\hat{J}_z},
\end{multline}%
%\begin{equation}
%\label{eq:jtransformMZ}
%\begin{split}
%\mat{c}{J^\prime_x \\ J^\prime_y\\J^\prime_z} =
%\mat{ccc}{1 & 0 & 0 \\ 0  & 0  & 1 \\ 0 & -1 & 0} \!\mat{ccc}{ \cos \varphi  &- \sin\varphi &0 \\ \sin\varphi & \cos\varphi & 0 \\0 & 0 & 1 \\ }\!\mat{ccc}{1 & 0 & 0 \\ 0  & 0  & -1 \\ 0 & 1 & 0}\!\mat{c}{J_x \\ J_y\\J_z}
%= \\
%= \mat{ccc}{\cos\varphi & 0 & \sin\varphi \\ 0  & 1  & 0 \\ -\sin \varphi & 0 & \cos\varphi}  \mat{c}{J_x \\ J_y\\J_z}
%\end{split}
%\end{equation}
which makes it clear that the sequence of  $\pi/2$, $\varphi$ and $-\pi/2$ rotations around axes $x$, $z$ and $x$ respectively,
results in an effective $\varphi$ rotation around the $y$ axis.

Using the above formalism,  let us now derive a simple formula for uncertainty of phase-sensing based on the
measurement of the photon-number difference at the output. Note that $\hat{n}_a - \hat{n}_b = 2 \hat{J}_z$, so the photon-number
difference measurement is equivalent to the $\hat{J}_z$ measurement.
Utilizing \eqnref{eq:jtransformMZ} in the Heisenberg picture, the average $J_z$ evaluated on the interferometer output state
may be related to the average of $J_z^\prime$ of the input state $\ket{\psi}_{\t{in}}$ as
\begin{equation}
\label{eq:Jzmean}
\langle \hat{J}_z \rangle =  \cos \varphi \langle \hat{J}_z \rangle_{\t{in}} - \sin \varphi \langle \hat{J}_x \rangle_{\t{in}}.
\end{equation}
In order to assess the precision of $\varphi$-estimation, we also calculate the variance of the $\hat{J}_z$ operator
of the output state of the interferometer:
\begin{multline}
\label{eq:Jzvariance}
\Delta^2 J_z =  %\langle J_z^{\prime 2} \rangle_{\t{in}} - \langle J^\prime_z\rangle_{\t{in}}^2 =
\cos^2\varphi\, \Delta^2J_z|_{\t{in}}+ \sin^2\varphi\, \Delta^2 J_x|_{\t{in}} + \\ -2 \sin\varphi\cos\varphi\,\t{cov}(J_x,J_z)|_{\t{in}},
\end{multline}
where $\t{cov}(J_x,J_z) = \frac{1}{2}\langle \hat{J}_x \hat{J}_z + \hat{J}_z \hat{J}_x \rangle - \langle \hat{J}_x \rangle  \langle \hat{J}_z \rangle$ is the covariance
 of the two observables. The precision of estimating $\varphi$ can now be quantified via a simple error-propagation formula:
\begin{equation}
\label{eq:deltaphi}
\Delta \varphi = \frac{\Delta J_z}{\left|\frac{\t{d} \mean{\hat{J}_z}}{\t{d} \varphi}\right|}.
\end{equation}

\subsection{Coherent-state interferometry \label{sub:coh_state_inter}}

Let us now analyze the precision given by \eqnref{eq:deltaphi} for the standard optical interferometry with the input state
 $\ket{\psi}_{\t{in}}=\ket{\alpha}\ket{0}$,
representing a coherent state and no light at all being sent into the input modes $a$ and $b$ of the interferometer in \figref{fig:machzehnder}.
The relevant quantities required for calculating the precision given in \eqnref{eq:deltaphi} read:
\begin{multline}
\langle \hat{J}_z \rangle_{\t{in}} = \frac{1}{2}|\alpha|^2, \ \langle \hat{J}_x \rangle_{\t{in}}=0, \ \Delta^2J_z|_{\t{in}} =  \Delta^2J_x|_{\t{in}} = \frac{1}{4} |\alpha|^2, \\ \t{cov}(J_x,J_z)|_{\t{in}}=0
\end{multline}
yielding the precision:
\begin{equation}
\label{eq:deltaphicoh}
\Delta \varphi^{\ket{\alpha}\ket{0}} =\frac{\frac{1}{2}|\alpha|}{\frac{1}{2} |\alpha|^2 |\sin \varphi|} = \frac{1}{|\alpha \sin\varphi|} = \frac{1}{\sqrt{\mean{{N}}}|\sin \varphi|},
\end{equation}
where the average photon number $\mN = \mean{\hat{N}} = |\alpha|^2$.
The above formula represents  $1/\sqrt{\mN}$ shot noise scaling of precision characteristic for
the classical interferometry. The shot noise is a consequence of the $\Delta J_z$ effectively
representing the Poissonian fluctuations of the photon-number difference measurements at the output ports.
Yet, although such fluctuations are $\varphi$-independent, the average photon-number difference $\langle J_z \rangle$ changes with $\varphi$
with speed proportional to $|\sin \varphi|$ appearing in \eqnref{eq:deltaphicoh}, so
that the optimal operating points are at $\varphi=\pi/2,3\pi/2$.

%The above formula represents  $1/\sqrt{\mN}$ shot noise scaling of precision characteristic for
%the classical interferometry.
%The additional dependence on $\varphi$ stems from the fact that the $\Delta J_z$ is $\varphi$ independent
%and is simply the Poissonian fluctuation of the photon-number difference measurements at the output ports,
%while the average photon-number difference $\langle J_z \rangle$ changes with $\varphi$ with speed proportional to $|\sin \varphi|$
%implying that the optimal operating point is around $\varphi=\{\pi/2,3\pi/2\}$.

\subsection{Fock state interferometry}
\label{sec:fockinterferometry}
We can attempt to reduce the estimation uncertainty using more general states of light at the input. For example, we can replace the coherent state
with an  $N$-photon Fock state, so that  $\ket{\psi}_{\t{in}} = \ket{N} \ket{0}$. This is an eigenstate of $\hat{J}_z$ and
hence $\Delta^2 J_z|_{\t{in}} = 0$, and only the $\Delta^2 J_x|_{\t{in}}$ contributes to the $\hat{J}_z$ variance at the output:
$\Delta^2 J_z  = \frac{1}{4} N \sin^2 \varphi$. Since $\langle \hat{J}_z \rangle = \frac{1}{2}N \cos\varphi$, the corresponding estimation uncertainty reads:
\begin{equation}
\label{eq:fockmzprecision}
\Delta \varphi^{\ket{N}\ket{0}} = \frac{1}{\sqrt{N}},
\end{equation}
being again shot-noise limited. The sole benefit of using the Fock state is the lack of $\varphi$-dependence of the estimation precision.
This, however, is scarcely of any use in practice, since one may always perform rough interferometric measurements and bring the setup close
to the optimal operating points before performing more precise measurements there. Moreover, the $\varphi$-dependence in \eqnref{eq:deltaphicoh}
can be easily removed by taking into account not only the photon-number difference observable $\hat{J}_z$ but also the total photon number measured.
By using their ratio as an effective observable in the r.h.s.~of \eqnref{eq:deltaphi}, or in other words by
considering the ``visibility observable'', the formula \eref{eq:deltaphicoh} is replaced by \ref{eq:fockmzprecision}.

\subsection{Coherent + squeezed-vacuum interferometry \label{sub:coh_sq_vac_inter}}

We thus need to use more general input states in order to surpass the shot noise limit. Firstly, let us note that sending the light solely to one
of the input ports will not provide us with the desired benefit.
As in the end only a photon-number measurement is assumed,
which is not sensitive to any relative phase differences between various Fock terms of the output state,
any scenario involving a single-beam input may always be translated to the the situation in which
an incoherent mixture of Fock states is sent onto the input port.
Since the variance is a convex function with respect to state density matrices,
and thus increasing under mixing, such strategies are of no use for our purposes.

Let us now consider a scheme were apart from the coherent light
we additionally send a squeezed-vacuum state into the other input port \citep{Caves1981}:
$\ket{\psi}_{\t{in}}= \ket{\alpha} \otimes \ket{r}$, see \figref{fig:machzehnder}.
This kind of strategy is being implemented in current most advanced interferometers designed to detect gravitational waves like LIGO or GEO600 \citep{Pitkin2011,LIGO2011,LIGO2013}.
Assuming for simplicity that $r$ is real, the relevant quantities required to calculate the estimation precision read:
\begin{equation}
\label{eq:coh+sq_vac_Jrels}
\begin{split}
&\langle\hat N\rangle = |\alpha|^2 + \sinh^2 r, \ \langle J_z \rangle_{\t{in}} = \frac{1}{2}(|\alpha|^2 - \sinh^2 r), \ \langle J_x \rangle_{\t{in}} = 0 \\
&\Delta^2 J_z|_\t{in} = \frac{1}{4}\left(|\alpha|^2 + \tfrac{1}{2} \sinh^2 2r \right), \quad \t{cov}(J_x,J_z)|_{\t{in}} = 0, \\
& \Delta^2 J_x|_{\t{in}} = \frac{1}{4} \left(|\alpha|^2 \cosh 2r  -  \t{Re}(\alpha^2)  \sinh2r + \sinh^2r\right).
\end{split}
\end{equation}
Hence, the usage of squeezed-vacuum as a second input allows to reduce the variance $\Delta^2 J_x|_{\t{in}}$ thanks to the $\t{Re}(\alpha^2) \sinh 2r$ term above,
which is then maximized by choosing the phase of the coherent state such that $\alpha=\t{Re}(\alpha)$.
%With squeezing we are therefore capable of reducing the $\Delta^2 J_x|_{\t{in}}$ term thanks to the $\t{Re}(\alpha^2) \sinh 2r$ term above.
%Optimally, we should choose the phase of a coherent state such that $\alpha=\t{Re}(\alpha)$, so that the term subtracts the most from the variance.
This corresponds to the situation, when the coherent state is displaced in phase space in the direction in which the squeezed vacuum possesses its lowest variance.
With such an optimal choice of phase, substituting the above formulas into \eqnref{eq:deltaphi}, we obtain the final expression for the phase-estimation precision:
\begin{equation}
\label{eq:coh+sq_vac_prec}
\Delta \varphi^{\ket{\alpha}\ket{r}} = \frac{\sqrt{\cot^2 \varphi \,(|\alpha|^2 + \tfrac{1}{2}\sinh^22r)+|\alpha|^2 \e^{-2r} + \sinh^2 r}}{
\left||\alpha|^2  - \sinh^2r\right|}.
\end{equation}
The optimal operation point are again clearly $\varphi\!=\!\pi/2, 3\pi/2$, since at them the first term under the square root, which is non-negative, vanishes. For a fair comparison with other strategies we should fix the total average number of photons $\mN$, which is regarded as a resource,
and optimize the split of energy between the coherent and the squeezed vacuum beams in order to minimize $\Delta \varphi$.
This optimization can only be done numerically, but the solution can be well approximated analytically in the regime of $\mN \gg 1$.
In this regime the squeezed vacuum should carry approximately $\sqrt{\mN}/2$ of photons, so the squeezing factor obeys $\sinh^2 r \approx \frac{1}{4} \e^{2r} \approx \sqrt{\mN}/2$  while the majority of photons belongs to the coherent beam. The resulting precision reads
\begin{equation}
\Delta \varphi^{\ket{\alpha}\ket{r}} \overset{\mN \gg 1}{\approx}  \frac{\sqrt{\mN/(2 \sqrt{\mN}) + \sqrt{\mN}/2} }{\mN - \sqrt{\mN}/2} \overset{\mN \gg 1}{\approx} \frac{1}{\mN^{3/4}}
\label{eq:precAs_coh_sq_noDec}
\end{equation}
and proves that indeed this strategy offers better than shot noise scaling of precision.

While the above example shows that indeed quantum states of light may lead to an improved sensitivity, the issue of optimality of
the proposed scheme has not been addressed.
%yet.
In fact, keeping the measurement-estimation scheme unchanged, it is possible
to further reduce the estimation uncertainty by sending a more general two-mode
Gaussian states of light with squeezing present in both input ports and reach the $\propto 1/\mN$
scaling of precision \citep{Yurke1986, Olivares2007}. We skip the details here, as the optimization of the  general Gaussian two-mode input state
minimizing the estimation uncertainty of the scheme considered is cumbersome.
More importantly, using the tools of estimation theory introduced in \secref{sec:estimation}, we will later show  in \secref{sec:ideal}
that even with $\ket{\alpha} \otimes \ket{r}$ class of input states it is
possible to reach the $\propto 1/\mN$ scaling of precision, but this requires a significantly modified measurement-estimation scheme  \citep{Pezze2008} and different energy partition between the two input modes.

\subsection{Definite photon-number state interferometry \label{sec:definite}}

Even though definite photon-number states are technically difficult to prepare, they are conceptually appealing
and we make use of them to demonstrate explicitly the possibility of achieving the $1/N$ Heisenberg scaling
of estimation precision in Mach-Zehnder interferometry.  We have already shown that sending an $N$-photon
state into a single input port of the interferometer does not lead to an improved precision compared with a coherent-state--based strategy.
Therefore, we need to consider states with photons being simultaneously sent into both input ports. A general $N$-photon
two-mode input state can be written down using the angular momentum notation as
\begin{equation}
\ket{\psi}_{\t{in}} = \sum_{m=-j}^j c_{m} \ket{j,m},%, \quad J^2 \ket{j,m} = j(j+1) \ket{j,m}, \quad J_z \ket{j,m} \ket{m}
\end{equation}
where $j\!=\!N/2$ and $\ket{j,m}\!=\!\ket{j+m}\ket{j-m}$ in the standard mode-occupation notation. In particular, $\ket{j,j} = \ket{N}\ket{0}$ corresponds to a state with all the photons being sent into the upper input port. One can easily check that angular momentum operators introduced in \eqnref{eq:joperators} act in a standard way on the $\ket{j,m}$ states.
For concreteness assume $N$ is even and consider the state \citep{Yurke1986}:
\begin{multline}
\ketE{\psi}_\t{in} = \frac{1}{\sqrt{2}} \left(\ketE{j,0} + \ketE{j,1}  \right) = \\
 = \frac{1}{\sqrt{2}}\left(\ketE{\tfrac{N}{2}}\ketE{\tfrac{N}{2}} + \ketE{\tfrac{N}{2}+1}\ketE{\tfrac{N}{2}-1}\right),
\end{multline}%
%\begin{equation}
%\ket{\psi} = \frac{1}{\sqrt{2}} \left(\ket{j,0} + \ket{j,1}  \right) = \frac{1}{\sqrt{2}}(\ket{n/2}\ket{n/2} + \ket{n/2+1}\ket{n/2-1})
%\end{equation}
for which:
\begin{equation}
\begin{split}
&\langle \hat{J}_z \rangle_{\t{in}} = \frac{1}{2}, \quad  \langle \hat{J}_x \rangle_{\t{in}} = \frac{1}{2}\sqrt{j(j+1)}, \quad \Delta^2 J_z|_\t{in} = \frac{1}{2}, \\  & \Delta^2 J_x|_{\t{in}} = \frac{1}{2}\,j(j+1)-\frac{1}{4}, \quad \t{cov}(J_x,J_z)|_{\t{in}} = 0.
\end{split}
\end{equation}
Plugging the above expressions into \eqnref{eq:deltaphi}, we get:
\begin{equation}
\Delta \varphi = \frac{\sqrt{ \cos^2 \varphi +  \sin^2 \varphi[j(j+1)-1]}}{| \sin \varphi  + \cos \varphi \sqrt{j(j+1)}|}.
\end{equation}
The optimal operation point corresponds to $\sin\varphi=0$, where we benefit from large $\langle \hat{J}_x \rangle_{\t{in}}$ and low $\Delta^2J_z|_{\t{in}}$
making the state very sensitive to rotations around the $y$ axis. The resulting precision reads:
\begin{equation}
 \Delta \varphi = \frac{1}{\sqrt{j(j+1)}} \overset{N \gg 1}{\approx} \frac{2}{N},
\end{equation}
indicating the possibility of achieving the Heisenberg scaling of precision. We should note here, that
a simpler state $\ket{\psi}_{\t{in}}=\ket{j,0}= \ket{N/2}\ket{N/2}$ called the \emph{twin-Fock state} where $N/2$ photons are simultaneously sent into each of the input ports, is also capable of providing the Heisenberg scaling of precision \citep{Holland1993}, but requires a
different measurement-estimation scheme which goes beyond the analysis of
the average photon-number difference at the output, see \secref{sub:ideal_gaussian}.

One should bear in mind that the expressions for attainable precision
presented in this section base on a simple error propagation formula calculated at a particular
operating point. Therefore, in order to approach any of the precisions claimed, one needs to
first lock the interferometer to operate close to an optimal
point, which
also requires some of the resources to be consumed. Rigorous quantification of the total resources
needed to attain a given estimation precision starting with a completely unknown phase may be difficult in general.
We
return to this issue in \secref{sec:ideal}, where
we are able to resolve this problem by
approaching it with the language of the Bayesian inference.

\subsection{Other interferometers}
\label{sec:otherinterferometers}

Even though we have focused on the Mach-Zehnder interferometer setup, analogous results could be obtained for
other optical interferometric configurations, such as the Michelson interferometer, the Fabry-P{\'e}rrot interferometer, as well
as the atomic interferometry setups utilized in: atomic clocks operation \citep{Diddams2004}, spectroscopy \citep{Leibfried2004}, magnetometry \citep{Budker2007} or the BEC interferometry \citep{Cronin2009}.
We briefly show below that despite physical differences the mathematical framework is common to all these cases and as such the results presented in this review, even though derived with the simple optical interferometry in mind, have a much broader scope of applicability.
\begin{figure}[!t]
  \centering
  \includegraphics[width=0.9\columnwidth]{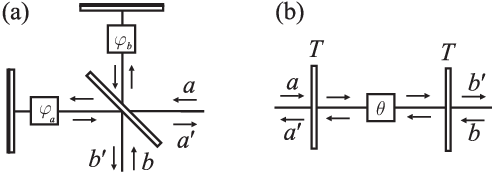}
  \caption{Other popular two-input/two-output mode interferometers: (a) Michelson interferometer with
  one-way phase delays $\varphi_a$, $\varphi_b$ in the respective arms,  (b) Fabry-P{\'e}rrot interferometer with one-way phase delay $\theta$ and
  power transmission $T$ of the mirrors.}
\label{fig:michelson}
\end{figure}

The Michelson interferometer is depicted in \figref{fig:michelson}a. Provided
the output modes $a^\prime$, $b^\prime$ can be separated from the input ones $a$, $b$
via an optical circulator, the Michelson interferometer is formally equivalent to the
Mach-Zehnder interferometer. The input output relations are identical as in \eqnref{eq:inputoutputMZ} with
 $\varphi = 2(\varphi_b  - \varphi_a)$, as the light acquires the relative phase twice---traveling both to
and from the end mirrors.

Consider now the Fabry-P{\'e}rrot interferometer depicted in  \figref{fig:michelson}b. We assume for simplicity that both mirrors have the same
power transmission coefficient $T$ %=\cos^2 \beta$
and the phase $\theta$ is acquired while the light travels from one mirror to the other inside the interferometer.
The resulting input-output relation reads:
\begin{multline}
\mat{c}{\hat{a}^\prime \\ \hat{b}^\prime} = \frac{1}{(2-T)\mathrm{i}\sin \theta - T \cos\theta}  \times \\  \times
\mat{cc}{2\mathrm{i} \sqrt{1-T}\sin\theta & T\\T & 2\mathrm{i} \sqrt{1-T}\sin\theta }\!\!\mat{c}{\hat{a} \\ \hat{b}}
\label{eq:inputoutputFP}
\end{multline}
and up to an irrelevant global phase may be rewritten as:
\begin{equation}
\begin{split}
\mat{c}{\hat{a}^\prime \\ \hat{b}^\prime}  &= \mat{cc}{\cos\varphi/2 & \mathrm{i} \sin \varphi/2   \\  \mathrm{i} \sin \varphi/2 & \cos \varphi/2 }\mat{c}{\hat{a} \\ \hat{b}}, \\
 \varphi & =
2 \arcsin\left(\tfrac{T}{\sqrt{T^2 + 4(1-T)\sin^2 \theta}} \right).
\end{split}
\end{equation}
In terms of the angular momentum operators the above transformation is simply a $\varphi$-rotation around the $x$ axis.
Thus, up to a change of the rotation axis, the action of the Fabry-P{\'e}rrot interferometer with phase delay $\theta$ is equivalent to the one of the MZ interferometer with phase delay $\varphi$. Knowing the formulas for the estimation precision of $\varphi$ in the MZ interferometer, we can easily calculate the corresponding estimation precision of $\theta$ via the error propagation formula obtaining:
\begin{equation}
\Delta \theta = \frac{\Delta \varphi}{|\frac{\partial \varphi}{\partial \theta}|} = \Delta \varphi \frac{T^2 + 4(1-T)\sin^2 \theta}{4T \sqrt{1-T} \cos \theta}.
\end{equation}
As a consequence, the above expression allows to translate
all the results derived for the MZ interferometer to the Fabry-P{\'e}rrot case.

Ramsey interferometry is a popular technique for performing precise spectroscopic measurements of atoms.
It is widely used in atomic clock setups,
where it allows to lock the frequency of an external source of radiation, $\omega$, to a selected atomic transition frequency, $\omega_0$, between the single-atom excited and ground states, $\ket{e}$ and $\ket{g}$ \citep{Diddams2004}.
% were it allows to lock the frequency of an external source of radiation $\omega$ to a selected atomic transition frequency $\omega_0$ between the excited $\ket{e}$ and the ground state $\ket{g}$ \citep{Diddams2004}.
In a typical Ramsey interferometric experiment $N$ atoms
are
%ale
initially prepared in the ground state
%. They
and are subsequently subjected to a $\pi/2$ Rabi pulse,
which transforms each of them
%the atoms
into an equally weighted superposition of ground and excited states.
%$[(\ket{g}+\ket{e})/\sqrt{2}]^{\otimes n}$.
Afterwards, they
%Atoms
evolve freely for time $t$, before finally being subjected to
a second $\pi/2$ pulse, which in the ideal case of $\omega_0=\omega$ would put all the atoms in the exited state.
In case of any frequency mismatch,
%atoms
the probability for an atom to be measured in an excited state is $\cos^2(\varphi/2)$,
where $\varphi= (\omega_0 -\omega)t$.
Hence, by measuring the number of atoms in the excited state, one can estimate $\varphi$ and
consequently knowing $t$ the frequency difference $\omega_0 -\omega$.
\begin{figure}[!t]
  \centering
  \includegraphics[width=0.9 \columnwidth]{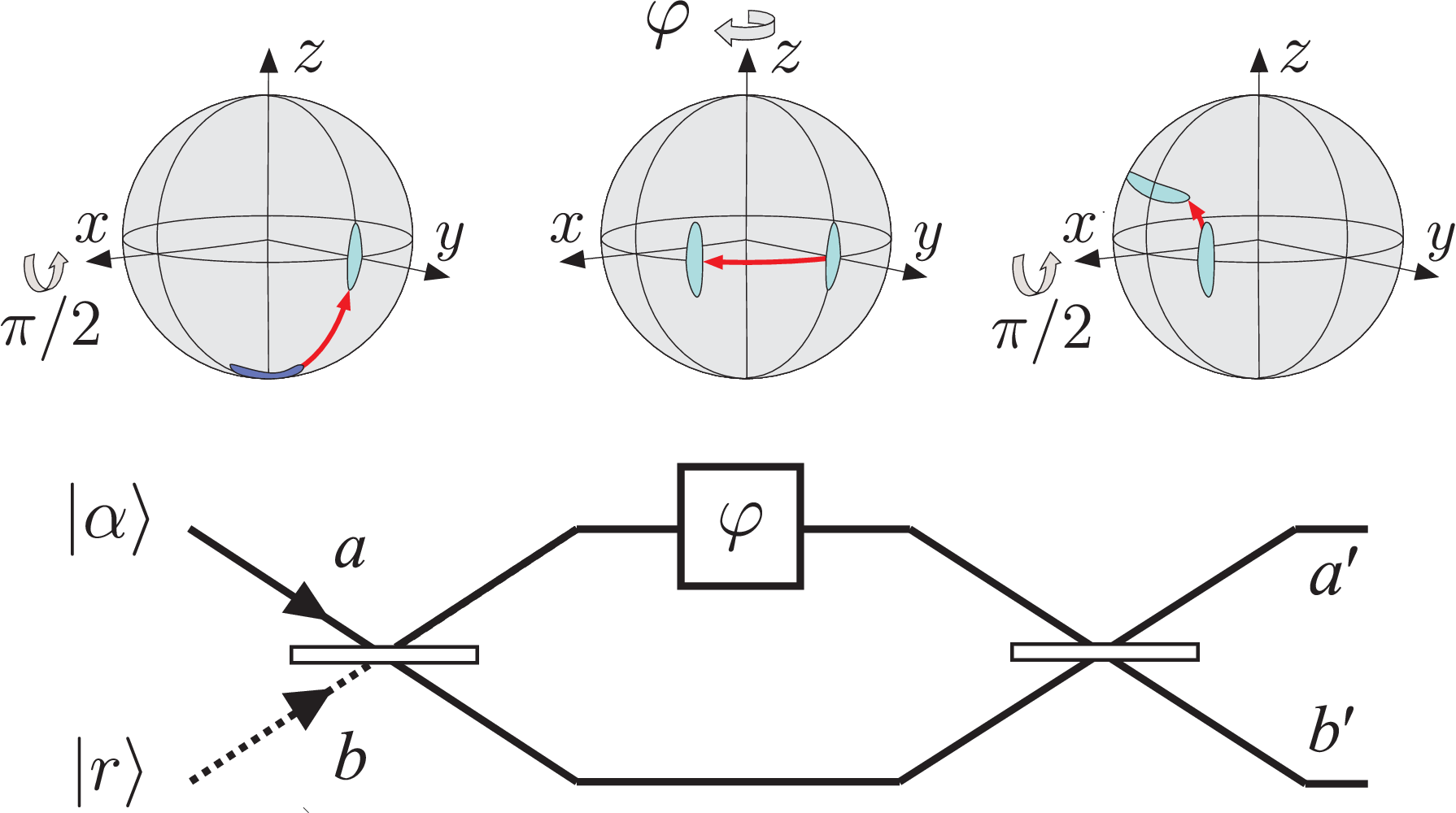}
  \caption{Formal equivalence of Ramsey and MZ interferometry. The two  atomic levels which are used in the Ramsey interferometry play
  analogous roles as the two arms of the interferometer, while the $\pi/2$ pulses equivalent from a mathematical point of view to the action of the beam-splitters. Quantum-enhancement in Ramsey interferometry may be obtained by preparing the atoms in a spin-squeezed input state
  reducing the variances of the relevant total angular momentum operators, in a similar fashion as using squeezed states of light leads
   to an improved sensitivity in the MZ interferometry.
  }
\label{fig:ramsey}
\end{figure}
Treating two-level atoms as spin-$1/2$ particles with their two levels corresponding to up and down projections of the spin $z$ component,
we may introduce the total spin operators $\hat{J}_i = \frac{1}{2} \sum_{k=0}^N \hat{\sigma}^{(k)}_i$, $i=x,y,z$,
where $\hat{\sigma}_i^{(k)}$ are standard Pauli sigma matrices acting on the $k$-th particle.
Evolution of a general input state can be written as:
\begin{equation}
\ket{\psi_{\varphi}} =\e^{-\mathrm{i} \hat{J}_x \pi/2} \e^{\mathrm{i} \hat{J}_z \varphi} \e^{-\mathrm{i} \hat{J}_x \pi/2}\ket{\psi}_{\t{in}}.
\end{equation}
which is completely analogous to the MZ transformation \eref{eq:jtransformMZ} with $\pi/2$ pulses playing the role of the beam-splitters,
see \figref{fig:ramsey}.
The total spin $z$ operator can be written as $\hat{J}_z = 2(\hat{n}_g - \hat{n}_e)$, where  $\hat{n}_g$, $\hat{n}_e$ denote the ground and excited state atom number operators respectively. Therefore, measurement of $\hat{J}_z$ is equivalent to the measurement of the difference of excited and non-excited atoms analogously to the optical case where it corresponded to the measurement of photon-number difference at the two output ports of the interferometer.
Fluctuation of the number of atoms measured limits the estimation precision
and in case of uncorrelated atoms is referred to as the \emph{projection noise}, which may be regarded as an analog of the optical shot noise.
An important difference from the optical case, though, is that when dealing with atoms we are restricted to consider states of definite particle-number. Thus, there is no exact analogue of coherent or squeezed states that we consider in the photonic case.
We can therefore regard atomic Ramsey interferometry as a special case
of the MZ interferometry with inputs restricted to states of definite particle-number, discussed
in \secref{sec:definite}, and further relate the precision of estimating the frequency difference to the precision of phase estimation via
\begin{equation}
\Delta \omega = \frac{\Delta \varphi}{t}.
\end{equation}
Beating the projection noise requires
the input state of the atoms to share some particle entanglement.
From an experimental point of view the most promising class of states are the so-called \emph{one-axis or two-axis spin-squeezed states},
which may be realized in BEC and atomic systems
interacting with light
%respectively
\citep{Kitagawa1993, Ma2011}. In fact, these states may
be regarded as a definite particle-number analogues of optical squeezed states. In particular, starting with atoms in a ground state the two-axis spin-squeezed states may be obtained via:
\begin{equation}
\ket{\psi_\chi} = \e^{-\frac{\chi}{2}(\hat{J}_+^2 - \hat{J}_-^2)} \ket{g}^{\otimes N},
\end{equation}
where $\hat{J}_{\pm} = \hat{J}_x \pm i \hat{J}_y$. The above formula resembles the definition of an optical squeezed state given in \eqnref{eq:squeezed}, where $\hat a$, ${\hat a}^{\dagger}$ operators are replaced by $\hat{J}_{-}$, $\hat{J}_{+}$. With an appropriate choice of squeezing strength $\chi$ as a function of $N$ it is possible to achieve the Heisenberg scaling of precision $\Delta \omega \propto 1/N$ \citep{Wineland1994, Ma2011}.

Analogous schemes may also be implemented in BEC \citep{Cronin2009, Gross2010}. In particular, BEC opens a way of realizing a specially appealing
matter-wave interferometry, in which, similarly to the optical interferometers, the matter-wave is split into two spatial modes,
evolves and finally interferes resulting in a spatial fringe pattern that may be used to estimate the relative phase acquired by
the atoms \citep{Shin2004}. Such a scheme may potentially find applications in precise measurements of the gravitational field \citep{Anderson1998}.
Yet, in this case, the detection involves measurements of positions of the atoms forming the interference fringes,
what makes the estimation procedure more involved than in the simple MZ scheme \citep{Chwedenczuk2011,Chwedenczuk2012},
but eventually the precisions for the optimal estimation schemes should coincide with the ones obtained for the MZ interferometry.

Finally, we should also mention that atomic ensembles interacting with light are excellent candidates for ultra-precise magnetometers
\citep{Budker2007}. Collective magnetic moment of atoms rotates in the presence of magnetic field to be measured, what again can be seen
as an analogue of the MZ transformation on $\hat{J}_i$ in \eqnref{eq:jtransformMZ}. The angle of the atomic magnetic-moment rotation is determined by sending polarized light which due to the Faraday effect is rotated proportionally to the atomic magnetic-moment component
in the direction of the light propagation. In standard scenarios, the ultimate precision will be affected by both the atomic projection noise,
due to characteristic uncertainties of the collective magnetic-moment operator for uncorrelated atoms, and the light shot noise.
The quantum enhancement of precision may again be achieved by squeezing the atomic states \citep{Wasilewski2010, Sewell2012} as well as by using the non-classical states of light \citep{Horrom2012}, what in both cases allows to go beyond the projection and shot-noise limits.

\section{Estimation Theory \label{sec:estimation}}

In this section we review the basics of both classical and quantum estimation theory.
We present Fisher Information and Bayesian approaches to determining the optimal estimation strategies
and discuss tools particularly useful for analysis of optical interferometric setups.

%%%%%%%%%%%%%%%%%%%%%%%%%%%%%%%

\subsection{Classical parameter estimation\label{sec:ClEst}}

The essential question
that has been addressed by statisticians long before the invention
of quantum mechanics is how to most efficiently extract information
from a given data set, which is determined by some non-deterministic
process \citep{Kay1993,Lehmann1998}.

In a typical scenario we are given
an $N$-point data set $\mathbf{x}=\left\{ x_{1},x_{2},\dots,x_{N}\right\} $
which is a realization of $N$ independent identically distributed
random variables, $X^{N}$, each distributed according
to a common Probability Density Function (PDF), $p_{\varphi}(X)$,
that depends on an unknown parameter $\varphi$ we wish to determine.
Our goal is to construct an \emph{estimator $\tilde{\varphi}_{N}(\mathbf{x})$
}that should be interpreted as a function which outputs the most accurate
estimate of the parameter $\varphi$ based on a given data set. Importantly,
as the estimator $\tilde{\varphi}_{N}$ is build on a sample of random
data, it is a random variable itself and the smaller are its fluctuations around the true value $\varphi$
the better it is.
% and its statistical properties,
%such as the mean or the variance, may be deduced from the collective,
%PDF $p_{\varphi}(X)$.

Typically, two approaches to the problem of the choice of the optimal estimator are undertaken.
%, depending on the choice of the nature of the estimated parameter.
 In the so-called frequentist or classical approach, $\varphi$ is assumed to be a
deterministic variable with an unknown value that, if known, could
in principle be stated to any precision. In this case, one of the basic tools in studying optimal estimation strategies
is the Fisher information, and hence we will refer to this approach as the \emph{Fisher information} approach.
In contrast, when following the \emph{Bayesian} paradigm, the estimated parameter is a random
variable itself that introduces some intrinsic error that accounts
for the lack of knowledge about $\varphi$ we possess \emph{prior}
to performing the estimation. We describe both approaches in detail
below.

\subsubsection{Fisher Information approach \label{sub:ClEstF}}

In this approach $p_{\varphi}(X)$
is regarded as a family of PDFs parametrized by $\varphi$---the parameter to be estimated based on the registered data $\mathbf{x}$.
The performance of a given estimator $\tilde{\varphi}_{N}(\mathbf{x})$ is quantified
by the Mean Square Error (MSE) deviation from the true value $\varphi$:
\begin{equation}
\Delta^{2}{\tilde{\varphi}}_{N}=\left\langle \left(\tilde{\varphi}_{N}(\mathbf{x})-\varphi \right)^{2}\right\rangle =
\int\!\!\textrm{d}^{N}\! \mathbf{x}\; p_{\varphi}(\mathbf{x})\left(\tilde{\varphi}_{N}(\mathbf{x})- \varphi \right)^{2}.
\label{eq:EstVar}
\end{equation}
A desired property for an estimator is that it is
\emph{unbiased}:
\begin{equation}
\left\langle \tilde{\varphi}_{N}\right\rangle =\int\!\!\textrm{d}^{N}\! \mathbf{x}\; p_{\varphi}
(\mathbf{x})\,\tilde{\varphi}_{N}(\mathbf{x})=\varphi\,,
\label{eq:GlobUnbiCond}
\end{equation}
so that on average it yields the true parameter value.
The optimal unbiased estimator
is the one that minimizes $\Delta^{2}{\varphi}_{N}$ for all $\varphi$.
Looking for the optimal estimator may be difficult and it may even be the case that there
is no single estimator that minimizes the MSE for all $\varphi$.

Still, one may always construct the so-called \emph{Cramer-Rao Bound} (CRB) that lower-bounds the MSE
of any unbiased estimator $\tilde{\varphi}_{N}$ (see e.g. \citep{Kay1993} for a review):
\begin{equation}
\Delta^{2}{\tilde{\varphi}}_{N} \ge\frac{1}{N\, F\!\left[p_{\varphi}\right]}\,,
\label{eq:CRB}
\end{equation}
where $F$ is the \emph{Fisher Information} (FI), and can be expressed using one of the formulas below:
\begin{multline}
\label{eq:FI}
F\left[p_{\varphi}\right]=\int\!\!\textrm{d}x\,\frac{1}{p_{\varphi}(x)}\left[\frac{\partial\, p_{\varphi}(x)}{\partial\varphi}\right]^{2}= \\
=\left\langle \left(\frac{\partial}{\partial\varphi}\ln p_{\varphi}\right)^{2}\right\rangle =-\left\langle \frac{\partial^{2}}{\partial\varphi^{2}}\ln p_{\varphi}\right\rangle ,
\end{multline}
The basic intuition is that the bigger the FI is the higher estimation precision may be expected.
 The FI is non-negative and  additive for uncorrelated events, so that
$F\!\left[p_{\varphi}^{(1,2)}\right]=F\!\left[p_{\varphi}^{(1)}\right]+F\!\left[p_{\varphi}^{(2)}\right]$,
for $p_{\varphi}^{(1,2)}(x_{1},x_{2})\!=\! p_{\varphi}^{(1)}(x_{1})\, p_{\varphi}^{(2)}(x_{2})$ and in particular:
$F\!\left[p_{\varphi}^{N}\right]=N\, F\!\left[p_{\varphi}\right]$,
which can be easily verified using the last expression in definition \eref{eq:FI}.
The FI is straightforward to calculate and once an estimator is found that saturates the CRB it is guaranteed to be optimal.
In general estimators saturating the CRB are called \emph{efficient}. The sufficient
and necessary condition for efficiency is the following condition on the PDF and the estimator \citep{Kay1993}:
\begin{equation}
\frac{\partial}{\partial\varphi}\ln\, p_{\varphi}(\mathbf{x})=
N\, F \!\left[p_{\varphi}\right]\,\left(\tilde{\varphi}_{k}(\mathbf{x})-\varphi\right).
\label{eq:CRBSatCondLU}
\end{equation}
An estimator $\tilde{\varphi}$ satisfying the above equality exist only for a special class of PDFs belonging to the so called
\emph{exponential family} of PDFs, for which:
\begin{equation}
\label{eq:expfamily}
\ln p_\varphi(x)=a(\varphi)+b(x)+c(\varphi)d(x), \quad \frac{a^\prime(\varphi)}{c^\prime(\varphi)}=-\varphi,
\end{equation}
where $a(\varphi),\,c(\varphi)$ and $b(x),\,d(x)$ are arbitrary functions and primes denote differentiation over $\varphi$. In general, however, the saturability condition cannot be met.

Note that in general FI is a function of $\varphi$, so that depending on the true value of the parameter, the CRB puts weaker or stronger
constraints on the minimal MSE. Actually, one is not always interested in
the optimal estimation strategy that is valid \emph{globally}---for
any potential value of $\varphi$---but may want to design a protocol
that works optimally for $\varphi$ confined to some small parameter range.
In this case one can take a \emph{local approach} and
analyze the CRB at a given point $\varphi=\varphi_0$. Formally derivation of the CRB at a given point requires
only a weaker \emph{local unbiasedness} condition:
\begin{equation}
\left.\frac{\partial}{\partial\varphi}\left\langle \tilde{\varphi}_{N}\right\rangle \right|_{\varphi=\varphi_{0}}=\;1\label{eq:LocUnbiCond}
\end{equation}
at a given parameter value $\varphi_{0}$.
FI at $\varphi_0$ is a \emph{local} quantity that depends only on
$\left.p_{\varphi}(X)\right|_{\varphi\!=\!\varphi_{0}}$ and $\left.\frac{\partial\, p_{\varphi}(X)}{\partial\varphi}\right|_{\varphi\!=\!\varphi_{0}}$,
as explicitly stated in \eqnref{eq:FI}. As a result, the FI is
sensitive to changes of the PDF of the first order in $\delta\varphi=\varphi-\varphi_0$.
Looking for the optimal locally unbiased estimator at a given point $\varphi_0$
makes sense provided one has a substantial prior knowledge that the true value of $\varphi$ is close to $\varphi_0$.
This may be the case if the data is obtained from a well controlled physical system subjected to small external fluctuations
or if some part of the data had been used for preliminary estimation narrowing the range of compatible $\varphi$ to a small
region around $\varphi_0$. In this case, even if the condition for saturability of the CRB cannot be met, it still may
be possible to find a locally unbiased estimator which will saturate the CRB at least at a given point $\varphi_0$. The explicit
form of the estimator can easily be derived from \eqnref{eq:CRBSatCondLU} by substituting $\varphi=\varphi_0$:
\begin{equation}
\label{eq:luoptimal}
\tilde{\varphi}_N^{\varphi_0}(\mathbf{x}) = \varphi_0 + \left.\frac{1}{N F[p_{\varphi}]}
\frac{\partial \ln p_\varphi(\mathbf{x})}{\partial \varphi}\right|_{\varphi=\varphi_0}.
\end{equation}

%Still, finding an estimator that saturates CRB is again possible only for
%PDFs of the form given in \eqnref{eq:expfamily}, but this time with a bit relaxed
% condition on functions $a(\varphi)$, $c(\varphi)$: $ \left.
% \frac{\t{d} (a^{\prime}/c^{\prime})}{\t{d} \varphi}\right|_{\varphi=\varphi_0}=-1$
% which can always be met

Fortunately, difficulties in saturating the CRB are only present in the finite-$N$ regime.
In the asymptotic limit of infinitely many repetitions of an experiment, or equivalently,
for an infinitely large sample, $N\!\rightarrow\!\infty$,
a particular estimator called the \emph{Maximum Likelihood} (ML) estimator saturates the CRB \citep{Kay1993,Lehmann1998}.
%---we say that MLE is \emph{asymptotically efficient}.
The ML estimator formally defined as
\begin{equation}
\tilde{\varphi}_{N}^{\textrm{ML}}\left(\mathbf{x}\right)\,=\,\underset{\varphi}{\textrm{argmax}}\,
p_{\varphi}(\mathbf{x})%\,=\,\underset{\varphi}{\textrm{argmax}}\,\ln p_{\varphi}(\mathbf{x}),
\label{eq:MLE}
\end{equation}
is a function that for a given
instance of outcomes, $\mathbf{x}$, outputs the value of parameter
for which this data sample is the most probable. For finite $N$ the ML estimator is in general biased, but becomes
unbiased asymptotically $\lim_{N\rightarrow \infty} \langle \varphi^{\textrm{ML}}_N(\mathbf{x})\rangle  = \varphi$ and
saturates the CR bound $\lim_{N\rightarrow \infty} N \Delta^2 {\tilde{\varphi}}^{\textrm{ML}}(\mathbf{x}) = F[p_\varphi]$.

\subsubsection{\label{sub:ClEstB}Bayesian approach}

In this approach, the parameter to be estimated, $\varphi$, is assumed
to be a random variable that is distributed according to a \emph{prior}
PDF, $p(\varphi)$, representing the knowledge about $\varphi$ one
possesses before performing the estimation, while $p(x|\varphi)$ denotes the conditional probability
of obtaining result $x$ for parameter value $\varphi$. Notice, a subtle change in notation from $p_\varphi(x)$ in the FI approach
to $p(x|\varphi)$ in the Bayesian approach reflecting the change in the role of $\varphi$ which is a parameter in the FI approach and a random variable
in the Bayesian approach.
If we stick to the MSE as a cost function,
we say that the estimator $\tilde{\varphi}_{N}(\mathbf{x})$ is optimal if it minimizes the average MSE
\begin{equation}
\mean{\Delta^2 \tilde{\varphi}_N} =\iint\!\! d\varphi\, d\mathbf{x}\;
p(\mathbf{x}|\varphi)p(\varphi)\left(\tilde{\varphi}_{N}(\mathbf{x})-\varphi\right)^{2},
\label{eq:BMSE}
\end{equation}
which, in contrast to \eqnref{eq:EstVar}, is also averaged over all
the values of the parameter with the Bayesian prior $p(\varphi)$.
Making use of the Bayes theorem we can rewrite the above expression in the form
\begin{equation}
\mean{\Delta^2 \tilde{\varphi}_N}=\int\!\! d\mathbf{x}\, p(\mathbf{x})\left[\int\!\! d\varphi\, p(\varphi|\mathbf{x})\left(\tilde{\varphi}_{N}(\mathbf{x})-\varphi\right)^{2}\right].
\end{equation}
From the above formula it is clear that the optimal estimator
is the one that minimizes terms in square bracket for each $\mathbf{x}$.
Hence, we can explicitly derive the form of the \emph{Minimum Mean Squared Error} (MMSE)  estimator
\begin{multline}
\frac{\partial}{\partial\tilde{\varphi}_{N}
%(\mathbf{x})
}\!\!\int\!\! d\varphi\, p(\varphi|\mathbf{x})\left(\tilde{\varphi}_{N}(\mathbf{x})\!-\!\varphi\right)^{2}\!=0\;\implies\; \\
\tilde{\varphi}_{N}^{\textrm{MMSE}}\!(\mathbf{x})\!=\!\!\!\int\!\! d\varphi\, p(\varphi|\mathbf{x})\,\varphi\!=\!\left\langle
\varphi\right\rangle _{p(\varphi|\mathbf{x})}\label{eq:BEstMSE}
\end{multline}
which simply corresponds to the \emph{average} value of
the parameter computed with respect to the \emph{posterior} PDF, $p(\varphi|\mathbf{x})$.
The posterior PDF represents the knowledge we possess about the parameter
after inferring the information about it from the sampled data $\mathbf{x}$.
%that again due to the Bayes' theorem \eref{eq:BayesTh} may be defined
%in terms of the available quantities, i.e.
%\begin{equation}
%p(\varphi|\mathbf{x})=\frac{p(\mathbf{x}|\varphi)\, p(\varphi)}{\int\!\! d\varphi\, p(\mathbf{x},\varphi)}=\frac{p(\mathbf{x}|\varphi)\, p(\varphi)}{\int\!\! d\varphi\, p(\mathbf{x}|\varphi)\, p(\varphi)}.\label{eq:PostPDF}
%\end{equation}
The corresponding MMSE reads:
\begin{eqnarray}
\mean{\Delta^2\tilde{\varphi}_N}&=&\int\!\! d\mathbf{x}\, p(\mathbf{x})\left[\int\!\! d\varphi\, p(\varphi|\mathbf{x})\left(\varphi-\left\langle \varphi\right\rangle _{p(\varphi|\mathbf{x})}\right)^{2}\right] \nonumber\\
&=&\int\!\! d\mathbf{x}\, p(\mathbf{x})\left.\Delta^{2}\varphi\right|_{p(\varphi|\mathbf{x})},\label{eq:BMSEmin}
\end{eqnarray}
so that it may be interpreted as the variance of the parameter $\varphi$
computed again with respect to the posterior PDF $p(\varphi|\mathbf{x})$
that is averaged over all the possible outcomes.

The optimal estimation strategy
within the Bayesian approach depends explicitly on the prior PDF assumed.
If either the
prior PDF will be very sensitive to variations of $\varphi$ or the
physical model will predict the data to be weakly affected by any
parameter changes, so that $p(\mathbf{x}|\varphi)p(\varphi)\!\approx\! p(\varphi)$,
the minimal $\Delta^2\tilde{\varphi}_N$ will
be predominantly determined by the prior distribution $p(\varphi)$ and
the sampled data will have limited effect on the estimation process.
Therefore, it is really important in the Bayesian
approach to choose an appropriate prior PDF that, on one hand, should
adequately represent our knowledge about the parameter before the
estimation, but, on the other, its choice should not dominate
the information obtained from the data.

In principle, nothing prevents us to consider more general \emph{cost functions, $C(\tilde{\varphi},\varphi)$,}
that in some situations may be more suitable than the squared error.
The corresponding optimal estimator will be the one that minimizes the average cost function
\begin{eqnarray}
\langle C\rangle &=&\iint\!\! d\varphi\  d\mathbf{x}\ p(\varphi)\,p(\mathbf{x}|\varphi)\,
C\left(\tilde{\varphi}_{N}(\mathbf{x}),\varphi\right).
\label{eq:BRisk}
\end{eqnarray}
In the context of optical interferometry the estimated quantity of interest $\varphi$ is the phase
which is a circular parameter, i.e. may be identified with a point on a circle or more formally as an element
of the circle group $U(1)$ and in particular $\varphi \equiv \varphi + 2 n \pi$.
Following \citep{Holevo1982} the cost function should respect the parameter \emph{topology} and the squared error is clearly not
the proper choice. We require the cost function to be symmetric, $C(\tilde{\varphi},\varphi)\!=\! C(\varphi,\tilde{\varphi})$,
\emph{group invariant}, i.e. $\forall_{\phi\in U(1)}\!:\, C(\tilde{\varphi}\!+\!\phi,\varphi\!+\!\phi)\!=\! C(\tilde{\varphi},\varphi)$,
and \emph{periodic}, $C(\varphi+2n\pi,\tilde{\varphi})= C(\varphi,\tilde{\varphi})$.
This restricts the class of cost functions to:
\begin{equation}
C(\tilde{\varphi},\varphi)=C(\delta \varphi)= \sum_{n=0}^{\infty}c_{n}\cos\!\left[n\,\delta\varphi\right]\;\;\textrm{with}\;\;\delta\varphi = \tilde{\varphi}-\varphi\,.
\end{equation}
Furthermore, we require $C(\delta \varphi)$ must rise monotonically from $C(0)\!=\!0$
at $\delta \varphi\!=\!0$ to some $C(\pi)\!=\! C_{\textrm{max}}$ at
$\delta \varphi\!=\!\pi$, so that $C'(\delta \varphi)\!\ge\!0$, so that the coefficients
$c_{n}$ must fulfill following constraints:
\begin{eqnarray}
\nonumber\sum_{n=0}^{\infty}c_{n}=0,\quad\sum_{n=0}^{\infty}(-1)^{n}c_{n}=C_{\max},\\
\sum_{n=1}^{\infty}n^{2}c_{n}\le0,\quad\sum_{n=1}^{\infty}n^{2}(-1)^{n}c_{n}\ge0,
\end{eqnarray}
which may be satisfied by imposing $\forall_{n>0}\!:\, c_{n}\le0$
and taking $c_{n}$ to decay at least quadratically with $n$.
Lastly, for the sake of compatibility we would like the cost function to
approach the standard variance for small $\delta\varphi$
so that $C(\delta \varphi)\!=\!\delta\varphi^{2}\!+\! O\!\left(\delta\varphi^{4}\right)$,
which is equivalent to $\sum_{n=1}^{\infty}n^{2}c_{n}\!=\!-2$.

In all the Bayesian estimation problems considered in this work, we will consider the simplest
cost function that satisfies all above-mentioned conditions with $c_{0}\!=\!-c_{1}\!=\!2$,
$\forall_{n>1}\!:\, c_{n}=0$ which reads explicitly:
\begin{equation}
C(\tilde{\varphi},\varphi)=4\,\sin^{2}\!\left(\frac{\tilde{\varphi}-\varphi}{2}\right).
\label{eq:CostFunH}
\end{equation}
Following the same argumentation as described in \eqnref{eq:BEstMSE}
when minimizing the averaged MSE, one may prove
that for the above chosen cost function the average cost $\langle C \rangle$ is minimized if an estimator,
$\tilde{\varphi}^C_{N}(\mathbf{x})$,
can be found that for any possible data sample $\mathbf{x}$ collected
satisfies the condition
\begin{equation}
\int\!\! d\varphi\;\; p(\varphi|\mathbf{x})\,\sin\!\left(\tilde{\varphi}^C_{N}(\mathbf{x})-\varphi\right)=0\label{eq:BEstH}.
\end{equation}

\subsubsection{Example: Transmission coefficient estimation \label{sub:ClEst_BinDistr_CoinToss}}

In order to illustrate the introduced concepts let us consider a simple model of parameter
estimation, where a single photon impinges on the beamsplitter with power transmission and reflectivity
equal respectively  $\mathsf{p}$ and $\mathsf{q}\!=\!1\!-\! \mathsf{p}$. The experiment is repeated $N$ times
and based on the data obtained---number of photons transmitted and the number of photons reflected---the
goal is to estimate the transmission coefficient $\mathsf{p}$.
This problem is equivalent to a coin-tossing experiment, where we assume
an unfair coin, which ``heads'' and ``tails'' occurring with probabilities $\mathsf{p}$
and $\mathsf{q}\!=\!1\!-\! \mathsf{p}$ respectively.

Probability that $n$ out of $N$ photons get transmitted is governed by the binomial distribution
\begin{equation}
\label{eq:binomial}
p_{\mathsf{p}}^{N}(n)=\binom{N}{n}\, \mathsf{p}^{n}\, (1-\mathsf{p})^{N-n}.
\end{equation}
The FI equals $F[p^N_{\mathsf{p}}]=N/[\mathsf{p}(1-\mathsf{p})]$ and hence the CRB imposes a lower bound on the achievable
estimation variance:
\begin{equation}
\Delta^{2}\tilde{\mathsf{p}}_{N}\ge\frac{\mathsf{p}(1-\mathsf{p})}{N}
\label{eq:CRBp}.
\end{equation}
Luckily the binomial probability distribution belongs to the exponential family of PDFs specified in
\eqnref{eq:expfamily}, and by inspecting saturability condition \eqnref{eq:CRBSatCondLU} it may be easily checked,
that the simple estimator $\tilde{\mathsf{p}}_N(n)=n/N$ saturates the CRB. It is also worth mentioning that the
optimal estimator also coincides with the ML estimator, hence in this case the ML estimator is optimal also for finite $N$ and not only in the
asymptotic regime.

In the context of optical interferometry, we will deal with an analogous situation, where photons are sent into
one input port of an interferometer and $\mathsf{p}$, $\mathsf{q}$ correspond to the probabilities of
detecting a photon in one of the two output ports.
For a Mach-Zehnder interferometer, see \secref{sec:interferometers}, the probabilities
depend on the relative phase delay difference between the arms of an interferometer $\varphi$:
$\mathsf{p}=\sin^2(\varphi/2)$, $\mathsf{q}=\cos^2(\varphi/2)$, which is the actual parameter of interest.
Probability distribution as a function of $\varphi$ then reads:
\begin{equation}
\label{eq:binomvarphi}
p^N_\varphi(n) =\binom{N}{n}\, \sin(\varphi/2)^{2n}\, \cos(\varphi/2)^{2(N-n)}
\end{equation}
and the corresponding FI and the CRB take the form
\begin{equation}
F\!\left[p_{\varphi}^{N}\right]=N,
\quad \Delta^2 \tilde{\varphi}_N \geq \frac{1}{N}\label{eq:FIBinPhi}.
\end{equation}
Interestingly, FI does not depend on the actual value $\varphi$, what suggests that
the achievable estimation precision may be independent of the actual parameter value.
However, for such a parametrization, the CRB saturability condition \eqnref{eq:CRBSatCondLU}
does not hold and there is no unbiased estimator saturating the bound.
Nevertheless, using \eqnref{eq:luoptimal}, we may still write a locally unbiased estimator
saturating CRB at $\varphi_0$:  $\tilde{\varphi}_N(n) =\varphi_0 -\tan(\varphi_0/2) + 2n/(N \sin \varphi_0)$,
which is possible provided $\sin\varphi_0 \neq 0$.

Since the CRB given in \eqref{eq:FIBinPhi} can only be saturated locally it is worth
looking at the MLE which we know will perform optimally in the asymptotic regime $N\rightarrow \infty$.
Solving \eqnref{eq:MLE} we obtain
\begin{equation}
\tilde{\varphi}_N(n)=\underset{\varphi}{\textrm{argmax}}\,\ln p_{\varphi}^{N}(n)=
 \pm\,2\, \textrm{arctan}\sqrt{\frac{n}{N-n}}\label{eq:MLEBinPhiMaxima}
\end{equation}
and in general we notice that there are two equivalent maxima. This ambiguity is simply the result of invariance of the PDF
$p_\varphi(n)$ with respect to the change $\varphi \rightarrow -\varphi$ and might have been expected.
Hence in practice we would need some additional information, possibly coming from prior knowledge or other observations,
in order to distinguish between this two cases and be entitled to claim that the MLE saturates the CRB asymptotically for all $\varphi$.

We now analyze the same estimation problem employing the Bayesian approach.
Let us first consider the case where $\mathsf{p}$ is the parameter to be estimated and
the relevant conditional probability $p^{N}(n|\mathsf{p})=p^{N}_\mathsf{p}(n)$ is given by \eqnref{eq:binomial}.
Choosing a flat prior distribution $p(\mathsf{p})=1$ and the mean square as a cost function,
we find using \eqnref{eq:BEstMSE} and \eqnref{eq:BMSEmin} that the MMSE estimator and the corresponding minimal averaged MSE equal:
\begin{equation}
\tilde{\mathsf{p}}_N(n) =\dfrac{n+1}{N+2}, \quad \mean{\Delta^2 \tilde{\mathsf{p}}_{N}} = \dfrac{1}{6\left(N+2\right)}\,,
\end{equation}
which may be compared with the previously discussed FI approach where the optimal estimator was
$\tilde{\mathsf{p}}(n)= n/N$ and the resulting variance
when averaged over all $\mathsf{p}$ would yield $\langle \Delta^2\tilde{\mathsf{p}}_N \rangle = \int_0^1 \t{d} p \frac{p(p-1)}{N} = 1/(6N)$.
Hence, in the limit $N \rightarrow \infty$ these results converge to the ones obtained previously in the FI approach. This is a typical situation
that in the case of large amount of data the two approaches yield equivalent results \citep{Vaart2000}.

We now switch to $\varphi$ parametrization and consider $p^{N}(n|\varphi)=p^{N}_{\varphi}(n)$ as in \eqnref{eq:binomvarphi}.
Assuming flat prior distribution $p(\varphi)=1/(2\pi)$ and the previously introduced natural cost function in the case of circular parameter
 $C(\tilde{\varphi},\varphi)= 4 \sin^2[(\tilde{\varphi}-\varphi)/2]$, due to $\pm \varphi$ estimation ambiguity we
 realize that the condition for the optimal estimator given in
\eqnref{eq:BEstH} is satisfied for a trivial estimator $\tilde{\varphi}^C(n) = 0$ which does not take into account the measurement results at all.
This can be understood once we realize that the ambiguity in the sign  of estimated phase $\varphi$ and the possibility of
estimating the phase with the wrong sign is worse than not taking into account the measured data at all.
In order to obtain a more interesting result we need to consider a subset of possible values of $\varphi$ over which
the reconstruction is not ambiguous. If we choose $\varphi \in [0,\pi)$, and the corresponding prior $p(\varphi)=1/\pi$,
\eqnref{eq:BEstH} yields the optimal form of the estimator and the corresponding minimal cost calculated according to \eqnref{eq:BRisk} are as follows:
 \begin{eqnarray}
 \label{eq:transmissionbayesian}
 \tilde{\varphi}^C_N(n) & =& \arctan\left(\tfrac{2}{f(N,n)} \right) \\
 \langle C \rangle & = &2\left(1-\tfrac{1}{\pi(N+1)}\sum_{n=0}^N \sqrt{4 + f(N,n)^2}  \right),
 \end{eqnarray}
where $f(N,n)=(N-2n)(n-1/2)!(N-n-1/2)!/n!(N-n)!$.
Despite its complicated form the above formulas simplify in the limit $N\!\rightarrow\!\infty$.  The optimal estimator
approaches the ML estimator $\tilde{\varphi}^C_N(n) \approx 2\*\arctan\sqrt{n/(N-n)}$
 while the average cost function approaches $\langle C \rangle \approx 1/N$ indicating saturation of the CRB and confirming again that in
 the regime of many experimental repetitions the two discussed approaches coincide.
\subsection{Quantum parameter estimation \label{sec:estimationquantum}}

In a quantum estimation scenario the parameter $\varphi$ is encoded
in a \emph{quantum state} $\rho_{\varphi}$ which is subject to a quantum measurement  $M_{\mathbf{x}}$ yielding measurement result
$\mathbf{x}$ with probability $p_{\varphi}(\mathbf{x})\!=\!\textrm{Tr}\left\{ \rho_{\varphi}M_{\mathbf{x}}\right\}$. The estimation
strategy is complete once an estimator function $\tilde{\varphi}(\mathbf{x})$ is given ascribing estimated
parameters to particular measurement results.
The quantum measurement may be a standard projective von-Neumann measurement,
$M_{\mathbf{x}}M_{\mathbf{x^\prime}} = M_{\mathbf{x}}\delta_{\mathbf{x},\mathbf{x^\prime}}$,
 or a \emph{generalized measurement} where
the measurement operators form a \emph{Positive Operator
Valued Measure} (POVM) with the only constraints being $M_{\mathbf{x}}\geq0$,
$\int\! d\! \mathbf{x}\, M_{\mathbf{x}}\!=\!\openone$ \citep{Nielsen2000,Bengtsson2006}.
Establishing the optimal estimation strategy corresponds then not
only to the most accurate inference of the parameter value from the
data, but also to a non-trivial optimization over the class of all
POVMs to find the measurement scheme maximizing the precision. However,
as soon as we decide on a particular measurement scheme, we obtain
a model that determines probabilities describing the sampled data
$p_{\varphi}(\mathbf{x})$ and their quantum mechanical origin becomes
irrelevant. Hence, the estimation problem becomes then fully classical
and all the techniques developed in \secref{sec:ClEst}
apply.

\subsubsection{Quantum Fisher Information approach \label{sub:EstTh_QFIapproach}}

The problem of determining the optimal measurement scheme for a particular estimation scenario is non-trivial.
Fortunately analogously as in the classical estimation it is relatively easy to obtain
useful lower bounds on the minimal MSE.
The \emph{Quantum Cram\'{e}r-Rao Bound} (QCRB) \citep{Helstrom1976,Holevo1982,Braunstein1994}
is a generalization of the classical CRBs \eref{eq:CRB},
which lower bounds the variance of estimation for all possible locally unbiased estimators and the most general POVM
measurements \footnote{For clarity of notation, in what follows we drop the tilde symbol when writing estimator variance.}:
\begin{equation}
\Delta^{2}{\varphi}_{N} \ge\frac{1}{N\, F_{\textrm{Q}}\!\left[\rho_{\varphi}\right]}\textrm{ \qquad with\qquad}F_{\textrm{Q}}\!\left[\rho_{\varphi}\right]=\textrm{Tr}\!\left\{ \rho_{\varphi}\,L\!\left[\rho_{\varphi}\right]^{2}\right\} \label{eq:qCRB}
\end{equation}
where $F_Q$ is the \emph{Quantum Fisher Information} (QFI) while the Hermitian operator $L[\rho_{\varphi}]$
is the so called \emph{Symmetric Logarithmic Derivative} (SLD), which
can be unambiguously defined for any state $\rho_{\varphi}$ via the
relation $\dot{\rho}_{\varphi}\!=\!\frac{1}{2}\left(\rho_{\varphi}L[\rho_{\varphi}]\!+\! L[\rho_{\varphi}]\rho_{\varphi}\right)$.
Crucially, QFI is solely determined by the dependence of $\rho_\varphi$ on the estimated parameter, and
hence allows to analyze parameter sensitivity of given probe states without considering any particular measurements nor estimators.
Explicitly, the SLD when written in the eigenbasis of
$\rho_{\varphi}\!=\!\sum_{i}\lambda_{i}(\varphi)\left|e_{i}(\varphi)\right\rangle \!\left\langle e_{i}(\varphi)\right|$
reads:
\begin{equation}
L[\rho_{\varphi}]=\sum_{i,j}\frac{2\left<e_{i}\!\left(\varphi\right)\right|\dot{\rho}_{\varphi}\left|e_{j}\!\left(\varphi\right)\right>}{\lambda_{i}\!\left(\varphi\right)+\lambda_{j}\!\left(\varphi\right)}\left|e_{i}\!\left(\varphi\right)\right>\!\left<e_{j}\!\left(\varphi\right)\right|,\label{eq:SLD}
\end{equation}
where the sum is taken over the terms with non-vanishing denominator.
Analogously to the FI \eref{eq:FI}, the QFI is an additive quantity when calculated on product states and in particular
 $F_{\textrm{Q}}\!\left[\rho_{\varphi}^{\otimes N}\right]\!=\! N\, F_{\textrm{Q}}\!\left[\rho_{\varphi}\right]$.
Thus the $N$ term in the denominator of \eqnref{eq:qCRB} may be
equivalently interpreted as the number of independent repetitions
of an experiment with a state $\rho_{\varphi}$ to form the data sample
$\mathbf{x}$ of size $N$, or a single shot experiment with a multi-party
state $\rho_{\varphi}^{\otimes N}$.

Crucially, as proven in \citep{Braunstein1994,Nagaoka2005},
there always exist a measurement strategy---a projection measurement
in the eigenbasis of the SLD---for which the FI calculated for the resulting probability distribution equals the QFI, and
consequently the bounds \eref{eq:CRB} and \eref{eq:qCRB} coincide.
Hence the issue of saturability of the QCRB amounts to the problem of saturability of the corresponding classical CRB.
As discussed in detail in \secref{sub:ClEstF}, the bound is therefore globally saturable for
a special class of probability distributions belonging to the so called exponential family, and if this is not the case
the saturability is achievable either in the asymptotic limit of many independent experiments $N \rightarrow \infty$ or
in the local approach when one estimates small fluctuation of the parameter in the vicinity of a known value $\varphi_0$.

For pure states, $\rho_{\varphi}\!=\!\left|\psi_{\varphi}\right\rangle \!\left\langle \psi_{\varphi}\right|$,
the QFI in \eqnref{eq:qCRB} simplifies to
\begin{equation}
\label{eq:qfipure}
F_{\textrm{Q}}\!\left[\left|\psi_{\varphi}\right\rangle \right]\!=
\!4\left(\left\langle \dot{\psi}_{\varphi}|\dot{\psi}_{\varphi}\right\rangle \!-\!\left|\left\langle
\dot{\psi}_{\varphi}|\psi_{\varphi}\right\rangle \right|^{2}\right), \ \ket{\dot{\psi}_{\varphi}} = \frac{\t{d} \ket{\psi_{\varphi}}}{\t{d} \varphi}.
\end{equation}
Yet, for general mixed states, calculation of the QFI
involves diagonalization of the quantum state $\rho_{\varphi}$, in order to calculate the SLD, and
becomes tedious for probe states living in highly dimensional Hilbert spaces.

Interestingly, the QFI may be alternatively calculated by considering purifications $\ket{\Psi_\varphi}$ of a given family of mixed states
on an extended Hilbert space
$\rho_{\varphi}\!=\!\textrm{Tr}_{\textrm{E}}\!\left\{ \left|\Psi_\varphi\right\rangle \!\left\langle \Psi_\varphi\right|\right\}$,
where by $E$ we denote an ancillary space needed for the purification.
It has been proven by \citet{Escher2011} that the QFI of any $\rho_{\varphi}$
is equal to the smallest QFI of its purifications $\left|\Psi_\varphi\right\rangle $
\begin{equation}
F_{\textrm{Q}}\!\left[\rho_{\varphi}\right]=\min_{\Psi_\varphi}F_{\textrm{Q}}\!\left[\left|\Psi_\varphi\right\rangle \right]=4\min_{\Psi_\varphi}\left\{ \left\langle \dot{\Psi}_\varphi|\dot{\Psi}_\varphi\right\rangle -\left|\left\langle \dot{\Psi}_\varphi|\Psi_\varphi\right\rangle \right|^{2}\right\} \!.
\label{eq:FqPurifEscher}
\end{equation}
Even though minimization over all purification may still be challenging, the above formulation may easily be employed in deriving upper bounds
on QFI by considering some class of purifications. Since upper bounds on QFI translate to lower bounds on estimation uncertainty this
approach proved useful in deriving bounds in quantum metrology in presence of decoherence \citep{Escher2011, Escher2012}.

Independently, in \citep{Fujiwara2008} another purification-based
QFI definition has been constructed:
\begin{equation}
F_{\textrm{Q}}\!\left[\rho_{\varphi}\right]=4\min_{\Psi_\varphi}\left\langle \dot{\Psi}_\varphi|\dot{\Psi}_\varphi\right\rangle \!.\label{eq:FqPurifFuji}
\end{equation}
Despite apparent difference, \eqnsref{eq:FqPurifEscher}{eq:FqPurifFuji}
are equivalent and one can prove that any purification minimizing
one of them is likewise optimal for the other causing the second term of \eqnref{eq:FqPurifEscher} to vanish.
Although for any suboptimal $\ket{\Psi_\varphi}$ \eqnref{eq:FqPurifEscher} must
provide a strictly tighter bound on QFI than \eqnref{eq:FqPurifFuji},
the latter definition, owing to its elegant form allows for a direct and efficient procedure
for derivation of the precision bounds in quantum metrology \citep{Demkowicz2012,Kolodynski2013}.
Derivations of the precision bounds using the above two techniques in the context of optical interferometry are discussed in
\secref{sec:decoherence}

For completeness, we list below some other important properties of the QFI.
QFI does not increase under a parameter independent quantum channel
\begin{equation}
\label{eq:QFI_contract}
F_Q(\rho_\varphi) \geq F_Q[\Lambda(\rho_\varphi)],
\end{equation}
where $\Lambda$ is an arbitrary completely positive (CP) map.
QFI appears in the lowest order expansion of the measure of fidelity $\mathcal{F}$  between two quantum states
\begin{align}
\mathcal{F}(\rho_1,\rho_2) & = \left(\t{Tr}\sqrt{\sqrt{\rho_1} \rho_2 \sqrt{\rho_1}}\right)^2, \\
 \mathcal{F}(\rho_\varphi, \rho_{\varphi+\t{d}\varphi}) & = 1 - \frac{1}{4} F_Q(\rho_\varphi)\t{d} \varphi^2 + O(\t{d} \varphi^4).
\end{align}
QFI is convex
\begin{equation}
\label{QFI_convex}
F_Q\left(\sum_i p_i \rho_\varphi^{(i)}\right) \leq \sum_i p_i F_Q(\rho_\varphi^{(i)}), \quad \sum_i p_i=1,\ p_i\geq 0,
\end{equation}
which reflects the fact that mixing quantum states can only reduce achievable estimation sensitivity.
In a commonly encountered case, specifically in the context of optical interferometry,
when the estimated parameter is encoded on the state by a unitary
\begin{equation}
\label{eq:unitaryencoding}
\rho_{\varphi}\!=\! U_{\varphi}\rho U_\varphi^{\dagger}, \quad U_{\varphi}\!=\!\textrm{e}^{-\textrm{i}\hat{H}\varphi},
\end{equation}
where $\hat{H}$ is the generating ``Hamiltonian'', the general formula for QFI reads:
\begin{equation}
F_Q(\rho_\varphi) = \sum_{i,j} \frac{2 |\bra{e_i}\hat{H} \ket{e_j}|^2(\lambda_i - \lambda_j)^2}{\lambda_i + \lambda_j}
\end{equation}
where $\ket{e_i}$, $\lambda_i$ form the eigendecomposition of $\rho$. Note that in this case the QFI does not depend
on the actual value of $\varphi$. For the pure state estimation case, $\rho = \ket{\psi}\bra{\psi}$,  the QFI is proportional to the variance
of $\hat{H}$:
\begin{equation}
F_Q(\ket{\psi_\varphi}) = 4 \Delta^2 H = 4 (\bra{\psi} \hat{H}^2 \ket{\psi} - \bra{\psi} \hat{H}  \ket{\psi})
\end{equation}
and the QCRB \eqref{eq:qCRB} takes a particular appealing form resembling the form of the energy-time uncertainty relation:
\begin{equation}
\Delta^2 {\varphi } \Delta^2 H \geq \frac{1}{4N}.
\end{equation}
We conclude the discussion of QFI properties by mentioning a very recent and elegant result proving that for unitary parameter encodings,
QFI is proportional to the convex roof of the variance of $\hat{H}$ \citep{Toth2013,Yu2013}:
\begin{equation}
F_Q(\rho_\varphi) = 4 \min_{\{p_i,\ket{\psi^{(i)}}\}} \sum_i p_i \Delta^2 H^{(i)},
\end{equation}
where the minimum is performed over all decompositions of $\rho= \sum_i p_i \* \ket{\psi^{(i)}} \bra{\psi^{(i)}}$,
and $\Delta^2 H^{(i)}$ denotes the variance of $\hat{H}$ calculated on $\ket{\psi^{(i)}}$.

\subsubsection{Bayesian approach \label{sec:quantumbayesian}}

As the quantum mechanical estimation scenario, with a particular
measurement scheme chosen, represents nothing but a probabilistic model
with outcome probabilities $p_{\varphi}(\mathbf{x})$, we may also
apply the Bayesian techniques described in \secref{sub:ClEstB}.
The quantum element of the problem, however, i.e. minimization of the average cost function over the choice of
measurements is in general highly non-trivial. Fortunately, provided the problem
possesses a particular kind of symmetry it  may be solved using the concept of covariant measurements \citep{Holevo1982}.

In the context of optical interferometry, it is sufficient to consider
the unitary parameter encoding case as defined in \eqref{eq:unitaryencoding}, where the estimated parameter $\varphi$
will correspond to the phase difference in an interferometer, see \figref{fig:interchannel} in \secref{sec:ideal}.
Let us denote a general POVM measurement as $M_{\mathbf{x}}$ and the corresponding estimator as $\tilde{\varphi}$.
Since we need to minimize the average cost over both the measurements and the estimators it is
convenient to combine these two elements into one by labeling the POVM elements with the estimated values themselves:
$M_{\tilde{\varphi}} = \int{\t{d} \mathbf{x}} M_{\mathbf{x}} \delta[\tilde{\varphi}- \tilde{\varphi}(\mathbf{x})]$.
The expression for the average cost function, \eqref{eq:BRisk}, takes the form:
\begin{equation}
\langle C \rangle = \iint\!\! \frac{d\varphi}{2\pi} \frac{d\tilde{\varphi}}{2\pi} \, p(\varphi) \t{Tr}(U_\varphi \rho U^\dagger_\varphi M_{\tilde{\varphi}}) \,
C\left(\tilde{\varphi},\varphi\right).
\label{eq:BCquantum}
\end{equation}
and leaves us with the problem of minimization $\langle C \rangle$ over a general POVM $M_{\tilde{\varphi}}$ with standard constraints $M_{\tilde{\varphi}}\geq 0$,
$\int \frac{\t{d} \tilde{\varphi}}{2\pi} M_{\tilde{\varphi}} = \openone$. Note that for clarity we use the normalized measure $\frac{\t{d}\varphi}{2\pi}$.

%The continuously parametrised
%measure $\left\{ M_{\tilde{\varphi}}\right\} $ employed in \eqnref{eq:BRiskHQ}
%will definitely not be realizable in a potential real-life experiment
%that has only a discrete set of settings. However, once the optimal
%continuously parametrised POVM is found, one may seek for its discrete
%version that leads to equal precision of estimation, what is normally
%possible for quantum systems consisting of finite number of particles
%\citep{Derka1998}.

Provided the problem enjoys the $\varphi$ shift symmetry, so that $p(\varphi+\varphi_0) = p(\varphi)$,
$C(\tilde{\varphi}+\varphi_0,\varphi+\varphi_0)=C(\tilde{\varphi},\varphi)$, it may be shown
that one does not loose optimality by restricting the class of POVM measurements
to
\begin{equation}
M_{\tilde{\varphi}}=U_{\tilde{\varphi}}\,\Xi\, U_{\tilde{\varphi}}^{\dagger},\label{eq:CovPOVM}
\end{equation}
which is a special case of the so-called \emph{covariant} measurements \citep{Holevo1982,Bartlett2007,Chiribella2005}.
If we take flat prior distribution $p(\varphi)=1$, and the
natural cost function $C(\tilde{\varphi},\varphi)=4 \sin^2[(\tilde{\varphi} - \varphi)/2]$ introduced in \secref{sub:ClEstB},
symmetry conditions are fulfilled and substituting \eqref{eq:CovPOVM} into  \eqref{eq:BCquantum} we get a simple expression:
\begin{equation}
\langle C \rangle= \mbox{Tr}\left\{ \left\langle \rho_{\varphi}\right\rangle _{C}\Xi\right\} ,\label{eq:BRiskHQCovPOVM}
\end{equation}
where
$\left\langle \rho_{\varphi}\right\rangle _{C}\!\!=\!4\int\!\!\frac{d\varphi}{2\pi}\, U_{\varphi}\rho U_{\varphi}^{\dagger}\sin^{2}\!
\left(\frac{\varphi}{2}\right)$
is the final quantum state averaged with the cost function.
Looking for the optimal Bayesian strategy now amounts to minimization of the above quantity
over $\Xi$ with the POVM constraints requiring that $\Xi \geq 0$, $\int \frac{\t{d} \tilde{\varphi}}{2\pi} U_{\tilde{\varphi}} \Xi U^\dagger_{\tilde{\varphi}} =
\openone $. As shown in \secref{sec:ideal} and \secref{sec:decoherence} this minimization is indeed possible
for optical interferometric estimation models and allows to find the optimal Bayesian strategy and the corresponding minimal average cost.

\section{Quantum limits in decoherence-free interferometry \label{sec:ideal}}

\begin{figure}[h!]
\centering
\includegraphics[width=0.9\columnwidth]{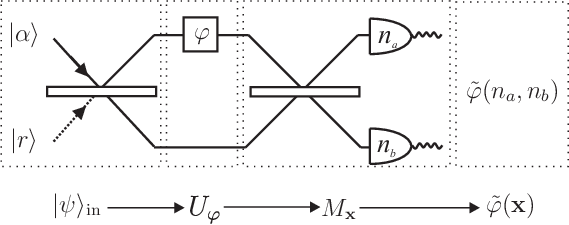}
\caption{Instead of a particular Mach-Zehnder interferometric strategy
 a general quantum interferometric scheme involves a general input probe state $\ket{\psi}_{\t{in}}$ that is subject to a unitary phase delay operation $U_\varphi$ followed by  a general quantum measurement (POVM) $M_{\mathbf{x}}$. Finally, an estimator $\tilde{\varphi}(\mathbf{x})$ is used
   to obtain the estimated value of the phase delay.} \label{fig:interchannel}
\end{figure}

In order to analyze the ultimate precision bounds in interferometry one needs to employ the quantum estimation theory introduced in \secref{sec:estimationquantum}. In this section we review the most important results of decoherence-free interferometry leaving the analysis
of the impact of decoherence for \secref{sec:decoherence}. This will provide us with precision benchmarks to which we will be able to compare
realistic estimation schemes and analyze the reasons for the departure of practically achievable precisions from idealized scenarios.

Formally, interferometry may be regarded as a channel estimation problem where a known state $\ket{\psi}_{\t{in}}$ is sent through a quantum channel
$U_\varphi=\e^{-i\varphi\hat{J}_z}$, with an unknown parameter $\varphi$, where $\hat{J}_z = \frac{1}{2}(a^\dagger a - b^\dagger b)$ is
the $z$ component of angular momentum operator introduced in \secref{sub:phase_sens_uncer}.
Quantum measurement $\hat{M}_{\mathbf{x}}$ is performed at the output state and the value of $\varphi$ is estimated based on the obtained outcome $\mathbf{x}$  through an estimator $\tilde{\varphi}(\mathbf{x})$, see \figref{fig:interchannel}.
Pursuing either QFI or Bayesian approach it is possible to derive bounds on achievable precision that
are valid irrespectively of how sophisticated is the measurement-estimation strategy employed and how exotic the input states of light are.
We start by considering definite photon-number states using both QFI and Bayesian approaches and then move on to discuss
issues that arise when discussing fundamental bounds taking into account states of light with indefinite photon number.

%%%%%%%%%%%%%%%%%%%%%%%%%%%%%%%%%%%%

\subsection{Quantum Fisher Information approach}

As discussed in detail in \secref{sec:estimation}, the QFI approach is particularly well suited to
analyze problems where one wants to estimate small deviations of $\varphi$ around a known value $\varphi_0$, as in this local regime
the QCRB, \eqref{eq:qCRB}, is known to be saturable. This is, for example, the case of gravitational-wave interferometry in which one sets the interferometer at the dark fringe and wants to estimate small changes in the interference pattern induced by the passing wave \citep{Pitkin2011,LIGO2011, LIGO2013}.

Since the state of light at the output, $\ket{\psi_\varphi}= \e^{-\mathrm{i} \varphi \hat{J}_z} \ket{\psi}_{\t{in}}$, is pure, the QFI
may be calculated using the simple formula given in \eqnref{eq:qfipure}. Realizing that $\ket{\dot{\psi}_\varphi}= -\mathrm{i} \hat{J}_z \ket{\psi_\varphi}$ we get the QFI and the corresponding QCRB on the estimation precision:
\begin{equation}
\label{eq:QFI_Jz_DecFree}
F_Q = 4\left(\bra{\psi_{\varphi}} \hat{J}_z^2 \ket{\psi_{\varphi}} -|\bra{\psi_\varphi} \hat{J}_z \ket{\psi_\varphi}|^2\right) = 4 \Delta^2 J_z,\\
\Delta {\varphi} \geq \frac{1}{2 \Delta J_z}.
\end{equation}
Note that the form of the QCRB above may be regarded as an analogue of the Heisenberg uncertainty relation for phase and angular momentum.

Clearly, according to the above bound, the optimal probe states for interferometry are the ones that maximize
$\Delta J_z$.
%Of course in principle, by choosing appropriate state of many photons, one could make $\Delta J_z$ as large as required.
%However, number of photons used is regarded as a resource in interferometric experiments and is limited, either because of
%the energy constraints, fragility of the sample sensed or by  side-effects such as recoil of the mirrors in gravitational wave detectors.
We fix the total number of photons and look for $N$-photon states  maximizing $\Delta J_z$. A General $N$-photon input state and the explicit form of the corresponding QFI read:
\begin{align}
\ket{\psi}_{\t{in}} & = \sum_{n=0}^N c_n \ket{n}\ket{N-n}, \\
F_Q &= 4\left[\sum_{n=0}^N |c_n|^2 n^2 - \left(\sum_{n=0}^N |c_n|^2 n\right)^2\right].
\end{align}

Let us first consider the situation in which
$\ket{\psi}_{\t{in}}$ is a state resulting from sending $N$ photons on the balanced beam-splitter,  as discussed in
\secref{sec:fockinterferometry}. This time, however, we do not insist on a particular measurement nor estimation scheme, but just want
to calculate the corresponding QCRB on the sensitivity. Written in the photon-number basis the state takes the form:
\begin{equation}\label{eq:PsiSep}
\ket{\psi}_{\t{in}} =\sum_{n=0}^N\sqrt{\frac{1}{2^N}{N\choose n}}\ket{n}\ket{N-n},
\end{equation}
for which $F_Q=N$ and this results in the shot noise bound on precision:
\begin{equation}\label{eq:sql}
\Delta{\varphi}^{\ket{N}\ket{0}} \geq \frac{1}{\sqrt{N}}.
\end{equation}
Recall, that this bound is saturated with the simple MZ interferometric scheme discussed in \eqnref{eq:fockmzprecision}, which
is a proof that for the considered probe state this measurement-estimation scheme is optimal.
From a particle-description point of view, see \secref{sub:par_descr}, the above considered state is a pure product state with no entanglement between the photons. More generally, the shot noise limit, sometimes referred to as the standard quantum limit, is valid for all $N$-photon separable states \citep{Pezze2009,Giovannetti2011}, and going beyond this limit requires making use of inter-photon entanglement, see \secref{sub:Id_entanglement}.

Let us now investigate general input states, which possibly may be entangled. Consider the state of the form $\ket{\t{\noon}}=\frac{1}{\sqrt{2}}(\ket{N}\ket{0}+\ket{0}\ket{N})$ which is commonly referred to as the ``NOON'' state
\citep{Lee2002,Bollinger1996}. QFI for such state is given by $F_Q=N^2$ and consequently
\begin{equation}
\label{eq:HL}
\Delta\varphi \geq \frac{1}{N},
\end{equation}
which is referred to as the \emph{Heisenberg limit}.
In fact, \noon state gives the best possible precision as it has the biggest variance of $\hat{J}_z$ among the states with a given photon number $N$ \citep{Bollinger1996, Giovannetti2006}. Still, the practical usefulness of the \noon states is doubtful. The difficulty in their preparation increases dramatically with increasing $N$, and with present technology the experiments are limited to relatively small $N$,  e.g.
$N=4$ \citep{Nagata2007} or $N=5$ \citep{Afek2010}. Moreover, even if prepared, their extreme susceptibility to decoherence with increasing $N$, see \secref{sec:decoherence}, makes them hardly useful in any realistic scenario unless $N$ is restricted to small values.
Taking into account that experimentally accessible squeezed states of light offer a comparable performance in the decoherence-free scenario, see \secref{sec:idealindefinite}, and basically optimal asymptotic performance in the presence of decoherence, see \secref{sec:decoherence}, there is not much in favor of the N00N states apart from their conceptual appeal.

%%%%%%%%%%%%%%%%%%%%%%%%%%%%%%%%%%%%

\subsection{Bayesian approach \label{sub:Bayes_DecFree}}

Let us now look for the fundamental precision bounds in the Bayesian approach  \citep{Hradil1996, Berry2000}.
We assume the flat prior distribution $p(\varphi)=1/2\pi$ reflecting our complete initial ignorance on the
true phase value, and the natural cost function $C(\tilde{\varphi},\varphi) = 4 \sin^2[(\tilde{\varphi}-\varphi)/2]$, see \secref{sub:ClEstB}.
Thanks to the phase shift symmetry of the problem, see \secref{sec:quantumbayesian}, we can restrict the class of measurements to covariant measurements $M_{\tilde{\varphi}} = U_{\tilde{\varphi}} \Xi U_{\tilde{\varphi}}^\dagger$, where $\Xi$ is the seed measurement operator, to be optimized below.
Using \eqnref{eq:BRiskHQCovPOVM}, and noting that
$\int_0^{2\pi}\frac{d\varphi}{2\pi}4 \sin^2(\varphi/2) \e^{i\varphi(n-m)}= 2\delta_{nm} -(\delta_{n,m-1} + \delta_{n,m+1})$, the averaged cost for a general $N$-photon input state $\ket{\psi}=\sum_{n=0}^{N}c_n\ket{n,N-n}$
reads:
\begin{equation}
\label{eq:costbayesianproof}
\mean{C}= 2 - 2\t{Re}\left(\sum_{n=1}^N c_n^*c_{n-1} \Xi_{n,n-1}\right).
\end{equation}
Because $\Xi$ is Hermitian, the completeness condition $\int\frac{d\varphi}{2\pi}\hat{U}_\varphi\Xi\hat{U}_\varphi^{\dagger}=\openone$
 implies that $\Xi_{nn}=1$, while due to the positive semi-definiteness condition $\Xi \geq 0$, $|\Xi_{nm}|\leq\sqrt{\Xi_{nn}\Xi_{mm}}=1$.
Therefore, The real part in the subtracted term in \eqnref{eq:costbayesianproof} can at most be $\sum_{n=1}^N |c_n| |c_{n-1}|$,
which will be the case for $\Xi_{n,m} =\e^{\mathrm{i}(\xi_{n} - \xi_{m})}$, where $\xi_{n} = \arg(c_n)$.
This is a legitimate positive semi-definite operator, as it can be written as $\Xi = \ket{e_N} \bra{e_N}$, with
$\ket{e_N} = \sum_{n=0}^N \e^{\mathrm{i}\xi_n}\ket{n,N-n}$ \citep{Holevo1982,Chiribella2005}.
Thus, for a given input state the optimal Bayesian measurement-estimation strategy yields
\begin{equation}
\mean{C}=2\left(1 - \sum_{n=1}^N |c_n c_{n-1}|\right).
\end{equation}
For the uncorrelated input state \eref{eq:PsiSep}, the average cost reads:
\begin{equation}
\label{eq:optimalbayesiancost}
\mean{C}= 2 \left(1-\frac{1}{2^N}\sum_{n=0}^{N-1}\sqrt{{N\choose n}{N\choose n+1}}\right) \overset{N\to\infty}{\approx} \frac{1}{N}.
\end{equation}
Since in the limit of small estimation uncertainty the considered cost function approaches the MSE, we
may conclude that in the limit of large $N$:
\begin{equation}
\Delta{\varphi} \overset{N\to\infty}{\approx} \frac{1}{\sqrt{N}},
\end{equation}
which coincides with the standard shot-noise limit derived within the QFI approach.

Note a subtle difference between the above solution and the solution of the optimal Bayesian transmission coefficient
estimation problem discussed in \secref{sub:ClEst_BinDistr_CoinToss} with the $\varphi$ parametrization employed.
The formulas for probabilities in  \secref{sub:ClEst_BinDistr_CoinToss}, can be regarded as arising from
measuring each of uncorrelated photons leaving the interferometer independently, while in the present considerations we have allowed
for arbitrary quantum measurements, which are in general collective. Importantly, we account for the adaptive protocols
in which a measurement on a subsequent photon depends on the results obtained previously \citep{Kolodynski2010}---practically
these are usually additional controlled phase shifts allowing to keep the setup at the optimal operation point \citep{Higgins2007}.
This approach is therefore more general and in particular, does not suffer from a $\pm \varphi$ ambiguity that
forced us to restrict the estimated region to $[0,\pi)$ in order to obtain nontrivial results given in \eqnref{eq:transmissionbayesian}.

Let us now look for the optimal input states. From \eqnref{eq:optimalbayesiancost} it is clear that we may
restrict ourselves to real $c_n$. Denoting by $\mathbf{c}$ the vector containing the state coefficients $c_n$, we rewrite the formula for $\mean{C}$ in a more appealing form
\begin{equation}
\langle C \rangle =2-\mathbf{c}^{T}\mathbf{A}\mathbf{c}, \quad
A_{n,n-1}=A_{n-1,n}=1,
%\mean{C}=2\mathbf{c}^T \left(
%\begin{array}{ccccc}
%1 & -\frac{1}{2} & 0 & \dots\\
%-\frac{1}{2} & 1 & -\frac{1}{2} & \dots\\
%0 & -\frac{1}{2} & 1 & \dots\\
%\vdots & \vdots & \vdots & \ddots
%\end{array}\right)
%\mathbf{c}
\end{equation}
from which it is clear that minimizing the cost function is equivalent to finding the eigenvector with maximal eigenvalue of
 the matrix $\mathbf{A}$, which has all its entries zero except for its first off-diagonals. This can be done analytically \citep{Summy1990,Luis1996,Berry2000} and the optimal state, which we will refer to as the \emph{sine} state, together with the resulting cost read
\begin{align}
\ket{\psi}_{\t{in}}& =\sum_{n=0}^{N}\sqrt{\frac{2}{2+N}}\sin\Big(\frac{n+1}{N+2}\pi\Big)\ket{n}\ket{N-n} \label{eq:BWstate},\\
\mean{C} & = 2\left[1-\cos\left(\frac{\pi}{N+2}\right)\right]\overset{N\to\infty}{\approx} \frac{\pi^2}{N^2}.
\end{align}
Again, in the large $N$ limit we may identify the average cost with the average MSE, so that
the asymptotic precision reads:
\begin{equation}
\label{eq:HL_Bayes}
\Delta {\varphi}  \overset{N\to\infty}{\approx} \frac{\pi}{N}.
\end{equation}
Analogously, as in the  QFI approach we arrive at the $1/N$ Heisenberg scaling of precision, but with an additional constant factor $\pi$,
reflecting the fact that Bayesian approach is more demanding as it requires the strategy to work well under complete prior ignorance of
the value of the estimated phase. Note also that the structure of the optimal states is completely different from the \noon states.
In fact, the \noon states are useless in absence of any prior knowledge on the phase, since they are invariant under $2\pi/N$ phase shift, and
hence cannot unambiguously resolve phases differing by this amount.

%%%%%%%%%%%%%%%%%%%%%%%%%%%%%%%%%%%%

\subsection{Indefinite photon-number states \label{sec:idealindefinite}}

We now consider a more general class of states with indefinite photon numbers and look for optimal probe states
treating the \emph{average} photon number $\mN$ as a fixed resource.
A state with an indefinite number of photons may posses in general
coherences between sectors with different total numbers of particles. These coherences may in principle improve
estimation precision. However, a photon number measurement performed at the output ports projects the state on one of the sectors and
necessarily all coherences between different total photon number sectors are destroyed.
In order to benefit from these coherences, one needs to make use of a more general scheme such
as e.g. homodyne detection, where  an additional phase reference beam is needed,
typically called the local oscillator, which is being mixed with the signal light at the detection stage.
Usually, the local oscillator is assumed to be strong, classical field with a well defined phase.
In other words, it provides one with reference frame with respect to which phase of the signal beams
can be measured \citep{Molmer1997, Bartlett2007}. Thus, it is crucial to explicitly state whether the reference beam is included in the
overall energy budget or is it treated as a free resource.  Otherwise one my arrive at conflicting statements on
the achievable fundamental bounds \citep{Jarzyna2012}.

\subsubsection{Role of the reference beam \label{sub:role_of_ref}}

As an illustrative example, consider an artificial one mode scheme with input in coherent state $\ket{\psi}=\ket{\alpha}$ which passes through the phase delay $\varphi$. Strictly speaking, this is not an interferometer and one may wonder how one can possibly get any information on the phase
by measuring the output state.  Still $\ket{\psi}$ evolves into a formally different state $\ket{\psi_\varphi} = U_\varphi \ket{\alpha} = \ket{\alpha \e^{-\mathrm{i} \varphi}}$, where $U_\varphi=\e^{-\mathrm{i} \varphi \hat{n}}$,  and since the corresponding QFI is non-zero
\begin{equation}
F_Q=4\left(\bra{\alpha} \hat{n}^2 \ket{\alpha}  - |\bra{\alpha} \hat{n} \ket{\alpha}|^2\right) = 4 |\alpha|^2,
\end{equation}
it is in principle possible to draw some information on the phase by measuring $\ket{\psi_\varphi}$.
Clearly, the measurement required cannot be a direct photon-number measurement, and an additional phase reference beam
needs to be mixed with the state before sending the light to the detectors.
In a fair approach one should include the reference beam into the setup and assume that whole state
of signal+reference beam is averaged over a global undefined phase. This formalizes
the notion of the relative phase delay - it is defined with respect to reference beam and there is no such thing like absolute phase delay.

More formally, let $\ket{\psi}_{ar}=\ket{\alpha}_a \ket{\beta}_r$, be the original coherent state used for sensing the phase,
accompanied by a coherent reference beam $\ket{\beta}$. The corresponding output state reads $\ket{\psi_\varphi}_{ar} =
\ket{\alpha \e^{-\mathrm{i}\varphi}}_a \ket{\beta}_r$. Now, the phase $\varphi$ plays the role of the \emph{relative} phase shift between the two modes,
with a clear physical interpretation. The combined phase shift in the two modes, i.e. an operation
$U_\theta^a U_\theta^r = \e^{-\mathrm{i} \theta (\hat{n}_a + \hat{n}_r)}$ has no physical significance as it is not detectable without an \dots additional reference beam.
Hence, before calculating the QFI or any other quantity
determining fundamental precision bounds, one should first average
the state $\ket{\psi}_{ar}$ over the combined phase shift and treat the resulting density matrix as the input probe state
\begin{equation}
\label{eq:averagingphase}
\rho=\int_{0}^{2\pi}\frac{d\theta}{2\pi}U_\theta^{a}U_\theta^{r}\ket{\psi}_{ar}\bra{\psi}
\otimes\ket{\beta}\bra{\beta}U_\theta^{a\dagger}U_\theta^{r \dagger} = \sum_{N=0}^\infty p_N \rho_N.
\end{equation}
This operation destroys all the coherences between sectors with different total photon number, and the resulting state is a mixture
of states $\rho_N$ with different total photon numbers $N$, appearing with probabilities $p_N$.

In the absence of a reference beam, i.e. when $\beta=0$, the above averaging kills all the coherences between terms with different photon numbers in the mode $a$:
\begin{multline}
\rho=\int\frac{d\theta}{2\pi}\ket{\alpha \e^{-i\theta}}\bra{\alpha \e^{-i\theta}} \otimes \ket{0}\bra{0}= \\ = \e^{-|\alpha|^2}\sum_{n=0}^{\infty}\frac{|\alpha|^{2n}}{n!}\ket{n}\bra{n} \otimes \ket{0}\bra{0}
\end{multline}
which results in a state insensitive to phase delays and gives $F=0$, restoring our physical intuition.
On the other hand, $F=4|\alpha|^2$ obtained previously is recovered in the limit $|\beta|^2\to\infty$, meaning that reference beam is classical---consists of many more photons than the signal beams.

The above averaging prescription, is valid in general also when label $a$ refers to more than one mode.
Considering the standard Mach-Zehnder interferometer fed with a state $\ket{\psi}_{ab}$ with an indefinite photon numbers and
no additional reference beam available one again needs to perform the averaging over a common phase shift. If this is not done,
one may obtain conflicting results on e.g. QFI for seemingly equivalent phase shift operations such as
$U_\varphi = \e^{-\mathrm{i} \hat{n}_a \varphi}$ or $U^\prime_\varphi  = \e^{-\mathrm{i} (\hat{n}_a -\hat{n}_b) \varphi/2}$. The reason is
that, without the common phase averaging, one implicitly assumes the existence of a strong external classical phase reference to with
respect to which the phase shifts are defined. In particular, $U_\varphi$ assumes that
 the second mode is perfectly locked with the external reference beam and is not affected by the phase shift,
  whereas $U_\varphi^\prime$ assumes that there are exactly opposite phase shifts in the two modes with respect to the reference.
 Different choices of ,,phase-shift distribution'' may lead to a factor of $2$ or even factor of $4$ discrepancies in the reported QFIs
in apparently equivalent optical phase estimation schemes---see \citep{Jarzyna2012} for further discussion and compare with some
of the results that were obtained without the averaging performed \citep{Joo2011, Spagnolo2012}.

One can also understand why ignoring the need for a reference beam may result in underestimating the required energy resources. Consider a singe mode state with an indefinite photon number $\ket{\psi}=\sum_{n=0}^N c_n \ket{n}$ evolving under $\e^{-\mathrm{i} \varphi \hat{n}}$ and note that,
 from  the phase sensing point of view, this situation is formally equivalent to a two-mode state with a definite photon number $N$:
$\ket{\psi}_{ab} = \sum_{n=0}^N c_n \ket{n}\ket{N-n}$, evolving under $\e^{-\mathrm{i}\varphi \hat{n}_a}$. Still, the
average photon number consumed in the one mode case is $\mean{N} = \sum_{n=0}^N |c_n|^2 n $ and is in general  smaller than
$N$.
%In particular for path-symmetric states $c_n = c_{N-n}$ this leads to a factor of two discrepancy in the QFI which may be
%encountered in the literature, compare e.g. \citep{Spagnolo2012} and \citep{Jarzyna2012}.

%and makes sensing of the phase delay $\varphi$ impossible with a single mode coherent state $\ket{\alpha}$.
%Still, if label $a$ denotes more than one mode, and the the phase $\varphi$ to be sensed is the relative phase between these modes, then
%in general sensing is possible without the assistance of an additional phase reference.
%
% we mode Such state, when one has more than one mode, can still be sensitive to phase delays, but only those which are relative between the arms. %Especially, for pure path symmetric input states QFIs in the absence and in presence of the reference beam are equal.

\subsubsection{Optimal indefinite photon number strategies}
\label{sec:optindefinitephotonnumber}
Looking for the optimal states with fixed \emph{average} photon number
is in general more difficult than in the definite photon-number case. Still, if we agree with the above-advocated approach
to average all the input states over a common phase-shift transformation as in \eqnref{eq:averagingphase}, then the resulting state is a probabilistic mixture of definite photon number states. Intuitively, it is then clear that, instead of sending the considered averaged state,
it is more advantageous to have information which particular component $\rho_N$ of the mixture is being sent.
This would allow to adjust the measurement-estimation procedure to a given component and improve the overall performance.
This intuition is reflected by the properties of both the QFI and the Bayesian cost, which are respectively convex and concave quantities \cite{Helstrom1976}:
\begin{align}
F_Q\left(\sum_N p_N \rho_N\right) &\leq \sum_N p_N F_Q(\rho_N), \\ \mean{C\left(\sum_N p_N \rho_N\right)} &\geq \sum_N p_N \mean{C(\rho_N)}.
\end{align}
This, however, implies that knowing the solution for the optimal \emph{definite} photon number probe states,
by adjusting the probabilities $p_N$ with which different optimal $\rho_N$ are being sent,  we may
determine the optimal strategies with the average photon-number fixed.

Taking the QFI approach for a moment, we recall that the optimal $N$-photon state, the \noon state, yields $F_Q(\rho_N)=N^2$.
Let us consider a strategy where a vacuum state and the \noon state are sent with probabilities $1-p$ and $p$ respectively, with
the constraint on the average photon number $p  N = \mN$. The corresponding QFI reads $F_Q= (1-p) \cdot 0 + p N^2$.
Substituting $ p = \mN/N$ we get $F_Q = N \mN$. Therefore, while keeping $\mN$ fixed we may increase $N$ arbitrarily and
in principle reach $F_Q = \infty$, suggesting the possibility of arbitrary good sensing precisions \citep{Rivas2012,Zhang2013}.
Note in particular, that a naive generalization of the Heisenberg limit to $\Delta \varphi \geq \frac{1}{\mN}$, does not hold, and the strategies
beating this bound are referred to as sub-Heisenberg strategies \citep{Anisimov2010}.
A universally valid bound may be written as $\Delta\varphi\geq 1/\sqrt{\langle\hat{N}^2\rangle}$ \citep{Hofmann2009}, but
the question remains, whether the bound is saturable, and in particular does quantum mechanics indeed allows for practically useful estimation protocols leading to the sub-Heisenberg scaling of precision. Closer investigations of that problem proves such hypothesis to be false \citep{Zwierz2010,Tsang2012,Giovanetti2012,Berry2012}. In principle we may achieve the sub-Heisenberg precision in the local estimation regime but in order for the local strategy to be valid, we should know the value of estimated parameter with prior precision of the same order as the one we want to obtain, what makes the utility of the procedure questionable. Actually, if no such assumption on the priori knowledge is made, the
Heisenberg scaling in the form $\Delta \varphi = \t{const}/\mN$ is the best possible scaling of precision.

This claim can also be confirmed within the Bayesian approach with flat prior phase distribution. For large $N$ the minimal Bayesian cost
behaves like $\mean{C(\rho_N)} = \pi^2/N^2$, see \eqnref{eq:BWstate}. Since this function is convex, taking convex combinations of the cost for two different total photon numbers $N_1$, $N_2$, such that $p_1 N_1 + p_2 N_2 = \mN$ will yield the cost higher than $\pi^2/\mN^2$, and the corresponding uncertainty $\Delta \varphi \geq \pi/\mN$, indicating the universal validity of
the Heisenberg scaling of precision.

\subsubsection{Gaussian states \label{sub:ideal_gaussian}}

From a practical point of view, rather than looking for the optimal indefinite photon number states for interferometry it is
more important to analyze experimentally accessible Gaussian states.  The paradigmatic example of a Gaussian state applied in quantum enhanced interferometry is the two mode state $\ket{\alpha}\ket{r}$---coherent state in mode $a$ and squeezed vacuum in mode $b$.
We have already discussed this example in \secref{sub:coh_sq_vac_inter}, and calculated the precision for a simple measurement-estimation scheme.
For such states, sent through a fifty-fifty beam splitter, quantum Fisher information can been calculated explicitly \citep{Pezze2008,Ono2010,Jarzyna2012}:
\begin{equation}\label{eq:QFICohSq}
F_Q=|\alpha|^2 \e^{2r}+\sinh^2r.
\end{equation}
For the extreme cases $|\alpha|^2=0,\,\sinh^2r=\mN$ and $|\alpha|^2=\mN,\,\sinh^2r=0$ this formula gives $F_Q=\mN$, implying the shot noise scaling. Most importantly, optimization of \eqnref{eq:QFICohSq} over $\alpha$ and $r$ with constraint $|\alpha|^2+\sinh^2r=\mN$ gives asymptotically the Heisenberg limit $\Delta\varphi\sim 1/\mN$, making this strategy as good as the \noon-one for large number of photons.
Moreover, this bound on precision can be saturated by estimation strategies based on photon-number  \citep{Pezze2008, Seshadreesan2011} or homodyne \citep{Paris1995}  measurements. This also proves that
a simple measurement-estimation strategy discussed in \secref{sub:coh_sq_vac_inter} which yielded $1/\mN^{3/4}$ scaling of precision is not optimal. Unlike the simple interferometric scheme where it was optimal to dedicate approximately $\sqrt{\mN}$ photons to the squeezed beam, from the QFI point of view it is optimal to equally divide the number of photons between the coherent and the squeezed beam.

\begin{figure}[h!]
\centering
\includegraphics[width=0.9\columnwidth]{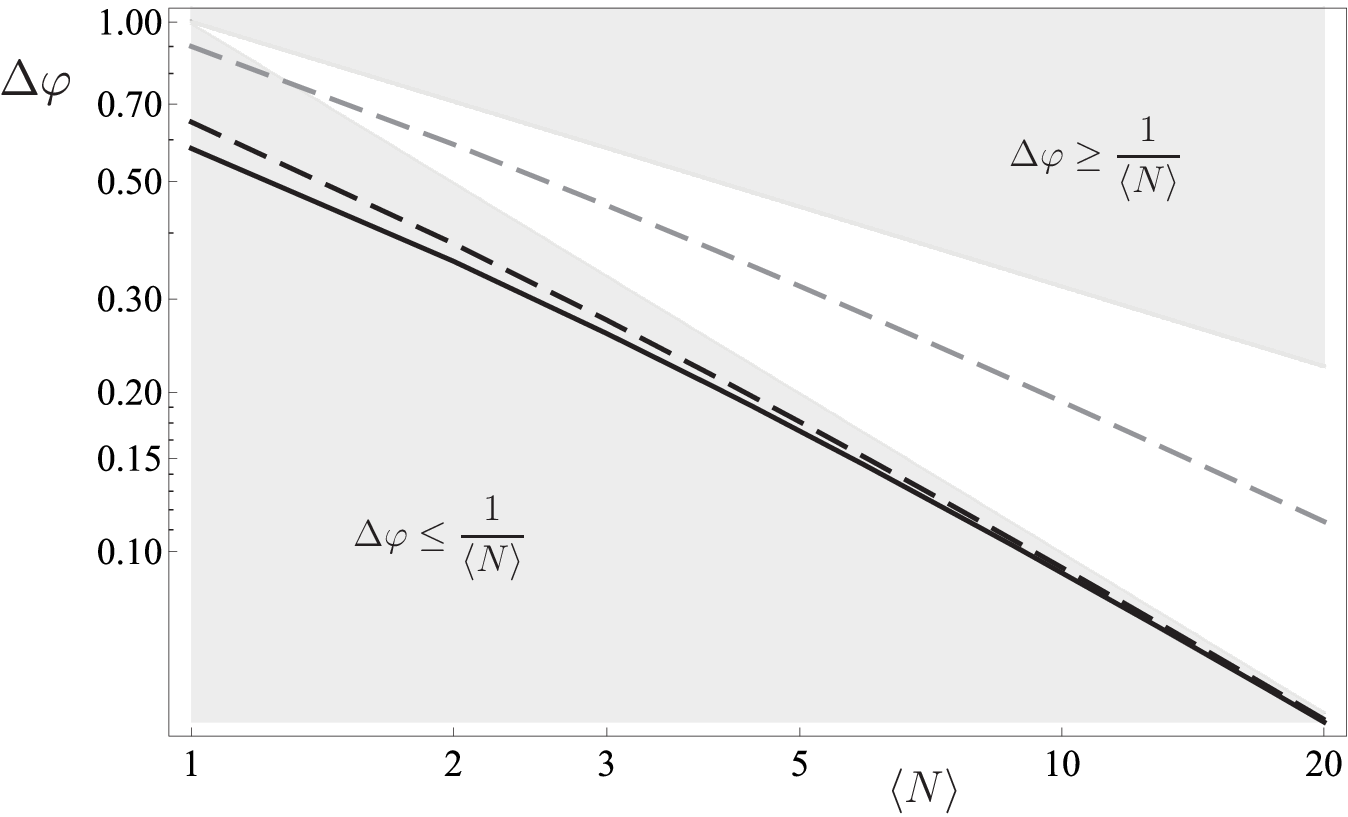}
\caption{Limits on precision obtained within QFI approach when using
two optimally squeezed states in both modes $\ket{r}\ket{r}$ (black, solid), coherent and squeezed-vacuum states $\ket{\alpha}\ket{r}$ (black, dashed). For comparison precision achievable with simple coherent and squeezed vacuum MZ interferometric scheme discussed in \secref{sub:coh_sq_vac_inter} is also depicted (gray, dashed). \label{fig:noLoss}
}
\end{figure}

More generally, finding the fundamental limit on precision achievable with general Gaussian states, requires optimization of the QFI or the Bayesian average cost function over general two-mode Gaussian input states, specified by the covariance matrix and the vector of first moments, see
\secref{sec:gaussianstates}. For the decoherence-free case this was done in \citet{Pinel2012,Pinel2013}.
Crucial observation is that for pure states, the overlap between two M-mode Gaussian states $\ket{\psi_\varphi}$ and $\ket{\psi_\varphi+d\varphi}$ is given by (up to the second order in $\t{d}\varphi$)
\begin{equation}
|\braket{\psi_{\varphi}}{\psi_{\varphi+d\varphi}}|^2=1-\frac{\t{d}\varphi^2}{4}\left(2(4\pi)^2\int \left(\frac{\t{d}W({\bf z})}{d\varphi}\right)^2
\t{d}^{2M}{\bf z}\right)
\end{equation}
where $W({\bf x})$ is the Wigner function \eref{eq:wigner} of state $\ket{\psi_\varphi}$. Thus, because $|\braket{\psi_{\varphi}}{\psi_{\varphi+d\varphi}}|^2=1-\frac{1}{4}F_Q\t{d}\varphi^2$ we may write that
\begin{equation}
\Delta\varphi\geq\left(2(4\pi)^2\int \left(\frac{\t{d}W({\bf z})}{\t{d}\varphi}\right)^2d^{2M}{\bf z}\right)^{-\frac{1}{2}}.
\end{equation}
In terms of the covariance matrix $\sigma$ and the first moments $\mean{{\bf z}}$ the formula takes an explicit form:
\begin{equation}
\Delta\varphi\geq\left(\frac{d\mean{{\bf z}}}{d\varphi}^T\sigma^{-1}\frac{d\mean{{\bf z}}}{d\varphi}+\frac{1}{4}\textrm{Tr}\left(\left(\frac{d\sigma}{d\varphi}\sigma^{-1}\right)^2\right)\right)^{-\frac{1}{2}}.
\end{equation}
Formal optimization of the above equation was done by \citet{Pinel2012}, however, the result was a one-mode squeezed-vacuum state, which in order
to carry phase information needs to be assisted by a reference beam. Unfortunately, performing a common phase-averaging procedure described in
\secref{sub:role_of_ref} in order to calculate the precision in the absence of additional phase reference destroys the Gaussian structure of the state
and makes the optimization intractable. Luckily, for path symmetric states,
i.e. the states invariant under the exchange of interferometer arms, the phase averaging procedure does not affect the QFI \citep{Jarzyna2012}.
Hence, assuming the path-symmetry the optimal Gaussian state is given by $\ket{r}\ket{r}$ with $\sinh^2r=\mN/2$---two squeezed vacuums
send into the input ports of the interferometer---and its corresponding QFI leads to the QCRB
\begin{equation}\label{eq:QFISqSq}
\Delta\varphi\geq\frac{1}{\sqrt{\mN(\mN+2)}}\approx\frac{1}{\mN}
\end{equation}
The state also achieves the Heisenberg limit for a large number of photons in the setup but it does not require any external phase reference. It is also worth noting that this state, while being mode-separable is particle-entangled and is feasible to prepare with current technology for moderate squeezing strengths. However, the enhancement over optimal squeezed-coherent strategy is rather small and vanish for large number of photons. Precisions obtained in squeezed-coherent and squeezed-squeezed scenarios are depicted in \figref{fig:noLoss}.

One can also study Gaussian states within the Bayesian framework. Optimal seed operator can be easily generalized from the definite photon number case to $\Xi=\sum_{N=0}^{\infty}\ket{e_N}\bra{e_N}$. Conceptually, the whole treatment is the same as in the definite photon number case. However, the expressions and calculations are very involved and will not be presented here.

%%%%%%%%%%%%%%%%%%%%%%%%%%%%%%%%%%%%

\subsection{Role of entanglement \label{sub:Id_entanglement}}

The issue of entanglement is crucial in quantum interferometry as it is known that only entangled states can beat the shot noise scaling \citep{Pezze2009}. This statement is sometimes questioned, pointing out the example of the squeezed+coherent light strategy, where
the interferometer is fed with seemingly unentangled $\ket{\alpha} \ket{r}$  input state. The reason of confusion is
the conflict of notions of mode and particle entanglement. As discussed in detail in \secref{sub:mode_vs_part_ent},
the two notions are not compatible, and there are states which are particle entangled, while having no mode entanglement and vice versa.
In the context of interferometry it is the \emph{particle} entanglement that is the source of quantum enhancement of precision.
In order to avoid criticism based on the ground of fundamental indistinguishability of particles and therefore
 a questionable physical content of the distinguishable particle-based entanglement picture on the fundamental level \cite{Benatti2010},
 we should stress that when considering models involving indistinguishable particles one should regard this statement as a formal (but still a meaningful and useful)  criterion where the particles are treated as formally distinguishable as described in \secref{sub:par_descr}.

%, which may
%sometimes be counter-intuitive as most of the calculations in quantum interferometry is done in the mode picture.
To see this, let us consider first a separable input state of $N$ photons of the form $\rho=\rho_1\otimes\dots\otimes\rho_N$, where $\rho_i$ denotes the state of the $i$-th photon. Since the phase shift evolution affects each of the photons independently
$\rho_\varphi = U^{\otimes N}_\varphi \rho U^{\otimes N \dagger}_\varphi$
and the QFI is additive on product states we may write:
\begin{equation}
F_Q(\rho)=\sum_{i=1}^{N}F_Q(\rho_i)\leq NF_Q(\rho_{\max})
\end{equation}
where $\rho_{\max}$ denotes state from the set $\{\rho_i\}_{i=1,\dots,N}$ for which QFI takes the largest value. But for a one photon state, the maximum value of QFI is equal to $1$, so
\begin{equation}\label{eq:entangled1}
F(\rho)\leq N, \quad \Delta \varphi \geq \frac{1}{\sqrt{N}}.
\end{equation}
For general separable states $\rho=\sum_{i}p_i\rho_{1}^{(i)}\otimes\dots\otimes\rho_{N}^{(i)}$ it is sufficient to use the convexity of QFI, together with \eqnref{eq:entangled1} to obtain the same conclusion. Above results imply that QFI, or precision, can be interpreted as a particle-entanglement witness, i.e.~all states that give precision scaling better than the shot noise must be particle-entangled \citep{Hyllus2012, Toth2012}.

The seemingly unentangled state  $\ket{\alpha} \ket{r}$  when projected on the definite total photon number sector, indeed contains particle entanglement as was demonstrated in  \secref{sub:mode_vs_part_ent}. This fact should be viewed as the fundamental source of its ability for
performing quantum-enhanced sensing.
It is also worth stressing, that unlike mode entanglement, particle entanglement is invariant under passive optical transformation like beam splitters, delay lines and mirrors, which makes it a sensible quantity to be treated as a resource for quantum enhanced interferometry.

\subsection{Multi-pass protocols}
\begin{figure}[t!]
\includegraphics[width=0.9\columnwidth]{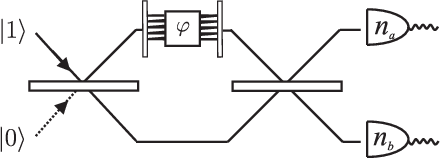}
\caption{A multi-pass interferometric protocol. A standard phase delay element is replaced by a setup  which makes the beam to pass through the phase delay multiple number of times.\label{fig:multi}}
\end{figure}

A common method, used in e.g.~gravitational wave detectors, to improve interferometric precision is to let the light bounce back and forth
through the phase delay element many times so that the phase delay signal is enhanced as shown in \figref{fig:multi}.
This method is used in GEO600 experiment \citep{LIGO2011}, where the light bounces twice in each of the sensing arms, making the detector as sensitive as the one with arms twice as long. Up to some approximation, one can also view the Fabry-Perrot cavities placed on top of the Mach-Zehnder
design as devices forcing each of the photon to pass multiple-times through the arms of the interferometer and acquire a multiple
of the phase delay \citep{Berry2009, Demkowicz2013}.

Consider a single photon in the state (after the first beam splitter)
$\ket{\psi}_{\t{in}}=\frac{1}{\sqrt{2}}(\ket{01}+\ket{10})$. After passing through the phase shift $N$ times it evolves into $\ket{\psi_\varphi}=\frac{1}{\sqrt{2}}(\ket{01}+\e^{-\mathrm{i} N\varphi}\ket{10})$. The phase is acquired $N$ times faster compared with a single pass case, mimicking the behavior of a single pass experiment with a \noon state. Hence, the precision may in principle be improved by a factor of $N$.  Treating the number of single photon passes as a resource, it has been demonstrated experimentally \citep{Higgins2007} that in the absence of noise such a device can indeed achieve the Heisenberg scaling without resorting to entanglement and efficiency of various multi-pass protocols
has been analyzed in detail in \cite{Berry2009}.
This is not to say, that all quantum strategies are formally equivalent to single-photon multi-pass strategies.
As will be discussed in the next section, the \noon states are extremely susceptible to decoherence, in particular loss, and
 this property is shared by the multi-pass strategies. Other quantum strategies prove more advantageous in this case, and they do not have a simple multi-pass equivalent \citep{Demkowicz2010,Kaftal2014}.

% Relevant decoherence: loss, dephasing (troche o tym cos pisal tez Paris)
% Bounds derrived using Bayesian and Fisher approaches - with focus on Fisher approach which allows for an easier calculation of asymptotic
% limits.  Nasza praca + Durkin.  Escher, Fujiwara  - purification method
% Classical simmulation, quantum simmulation (give explicit example of quantum simmulation for lossy interferometer).
% Definite vs indefinite photon number issues not relevant if loss is present as well as the problem with a priori knowledge.
% Application to GEO600 gravitational wave detector.

%%%%%%%%%%%%%%%%%%%%%%%%%%%%5

\section{Quantum limits in realistic interferometry\label{sec:decoherence}}

\begin{figure}[!t]
\begin{center}
\includegraphics[width=0.9\columnwidth]{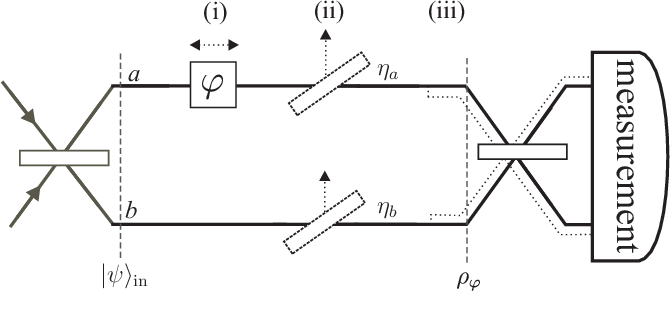}
\end{center}
\caption{\label{fig:LossyInter}
Schematic description of the decoherence processes discussed that affect the performance of an optical interferometer:
(i) \emph{phase diffusion} representing stochastic fluctuations of the estimated phase delay, (ii) \emph{losses} in the respective $a$/$b$ arms
represented by fictitious beam splitters with
$0\!\le\!\eta_{a/b}\!\le\!1$
transmission coefficients, (iii) \emph{imperfect visibility} indicated by
a mode mismatch of the beams interfering at the output beam splitter}
\end{figure}

In this section we revisit the ultimate limits on precision derived in \secref{sec:ideal}
taking into account realistic noise effects.
We study three decoherence processes
that are typically taken into account when discussing imperfections
in interferometric setups. We consider the effects of
\emph{phase diffusion}, \emph{photonic losses} and the impact of \emph{imperfect visibility}, see \figref{fig:LossyInter}.
In order to establish the ultimate limits on the estimation performance, we first analyze the above three decoherence models using the QFI perspective and secondly compare the bounds obtained with the ones derived within the Bayesian approach.

For the most part of this section, we will consider input states with definite number of photons, $N$, so that
\begin{equation}
\label{eq:nphotonprobe}
\rho_{\t{in}} = \ket{\psi}_{\t{in}}\bra{\psi}, \quad \ket{\psi}_{\t{in}} = \sum_{n=0}^N c_n \ket{n}\ket{N-n}.
\end{equation}
Similarly as in \secref{sec:ideal}, this will again be sufficient to draw conclusions also on the performance of indefinite photon number states, which will be discussed in detail in \secref{sec:decohindefinite}.

\begin{figure}
\begin{center}
\includegraphics[width=0.9\columnwidth]{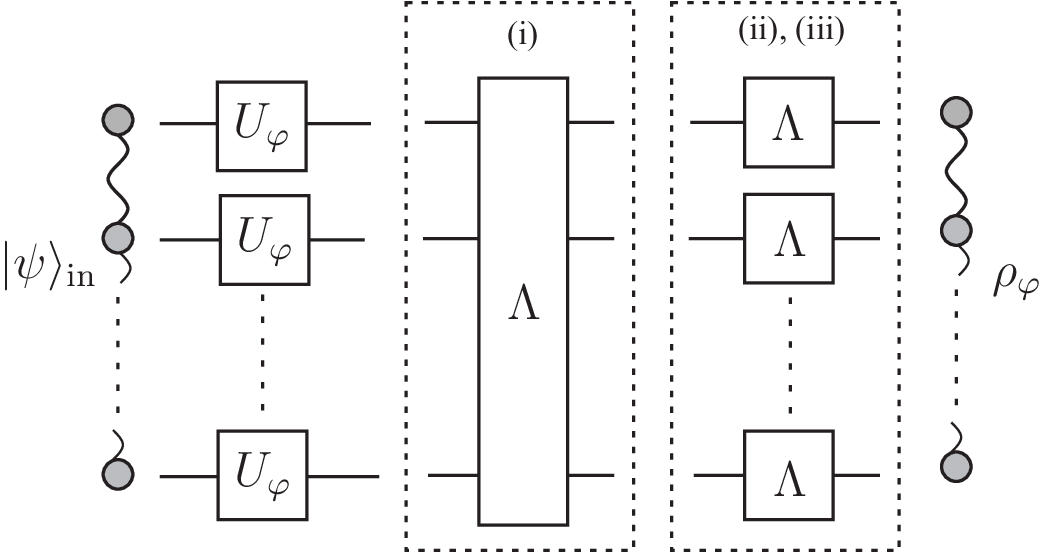}
\end{center}
\caption{General metrological scheme in case of photons being treated as formally distinguishable particles. Each photon travels through a
phase encoding transformation $U_\varphi$. Apart from that all photons are subject to either correlated (i) (phase diffusion), or uncorrelated  (ii), (iii) (loss, imperfect visibility) decoherence process.}
\label{fig:particleapproachdecoherence}
\end{figure}

In what follows it will sometimes prove useful to switch from mode to particle
description, see  \secref{sub:par_descr}, and treat photons formally as distinguishable particles but prepared in a symmetrized state.
This approach is illustrated in \figref{fig:particleapproachdecoherence} where each photon is represented by a different horizontal line,
 and travels through a phase encoding transformation $U^{(i)}_{\varphi} = \e^{-\mathrm{i} \varphi \hat{\sigma}_z^{(i)}/2}$, where
$\hat{\sigma}_z^{(i)}$ is a $z$ Pauli matrix acting on the $i$-th qubit.
The combined phase encoding operation is a simple tensor product $U_{\varphi}^{\otimes N} = \e^{-\mathrm{i} \varphi \sum_i \hat{\sigma}_z^{(i)}} =
\e^{-\mathrm{i} \varphi \hat{J}_z}$, recovering the familiar formula but with $\hat{J}_z$ being now interpreted as the $z$ component of the total angular momentum which is the sum of individual angular momenta. The photons are then subject to decoherence that acts in either correlated or uncorrelated manner. In the case of \emph{phase diffusion} (i) the decoherence has a collective character since each of the photons experiences the same fluctuation of
the phase being sensed, while in the case of \emph{loss} (ii) and \emph{imperfect visibility} (iii) the decoherence map has a tensor structure $\Lambda^{\otimes N}$
 reflecting the fact that it affects each photon independently. In the latter case of independent decoherence models
the overall state evolution is uncorrelated and may be written as:
 \begin{equation}
 \rho_\varphi = \Lambda_\varphi^{\otimes N}(\rho_\t{in}),  \quad \Lambda_\varphi(\cdot) = \Lambda(U_\varphi \cdot U_\varphi^\dagger).
 \end{equation}

\subsection{Decoherence models}
\label{sec:decoherencemodels}
In general, decoherence is a consequence of the uncontrolled interactions of a quantum system with the environment. Provided the system is initially decoupled from the environment, the general evolution of a quantum system interacting with the environment mathematically corresponds to a \emph{completely positive trace preserving} map $\Lambda$.
Every $\Lambda$ can be written using the Kraus representation \citep{Nielsen2000}:
\begin{equation}
\rho_{\t{out}} = \Lambda(\rho_{\t{in}}) = \sum_i K_i \rho_{\t{in}} K_i^\dagger, \quad \sum_{i} K_i^\dagger K_i = \openone,
\end{equation}
where $K_i$ are called the Kraus operators.

Effects of decoherence inside an interferometer are taken into account by replacing the unitary transformation $U_\varphi$ describing the action of the ideal interferometer, see \secref{sec:ideal}, with its noisy variant $\Lambda_\varphi$:
\begin{equation}
\label{eq:generaldecoherence}
\rho_\varphi = \Lambda_\varphi(\rho_{\t{in}})  = \left(\sum_i K_i U_\varphi\rho_{\t{in}}U_\varphi^\dagger K_i^\dagger\right).
\end{equation}
The formal structure of the above formula corresponds to a situation in which decoherence happens \emph{after} the unitary phase encoding. This of course might not be true in general. Still, for all the models considered in this review the decoherence part and the unitary part commute and therefore the order in which they are written is a matter of convenience.

\subsubsection{Phase diffusion\label{sec:phasediff}}
\emph{Phase diffusion}, also termed as the \emph{collective
dephasing} or the \emph{phase noise} represents the effect of
fluctuation of the estimated phase delay $\varphi$.
Such effect may be caused by any process that stochastically varies the effective optical lengths traveled by the photons, such as
thermal deformations or the micro-motions
of the optical elements.
We model the optical interferometry in the presence of phase diffusion process by the following map:
\begin{equation}
\rho_{\varphi}=\Lambda_\varphi(\rho_{\t{in}}) =
\intop_{-\infty}^{\infty}\!\!\textrm{d}\phi\; p_{\varphi}(\phi)\; U_{\phi}\,\rho_{\textrm{in}}\,U_{\phi}^{\dagger}
\label{eq:OutputNPh_PhaseDiff_Av}
\end{equation}
where the phase delay is a random variable $\phi$ distributed according to probability distribution $p_\varphi(\phi)$.
Note that the above form is actually the Kraus representation of the map $\Lambda_\varphi$ with Kraus operators $K_\phi = \sqrt{p_\varphi(\phi)} U_\phi$.
In case $p_\varphi(\phi)$ is a Gaussian distribution
with variance $\Gamma$ and the mean equal to the estimated parameter $\varphi$, $p_{\varphi}(\phi)\!=\!\frac{1}{\sqrt{2\pi\Gamma}}\textrm{e}^{\frac{-\left(\phi-\varphi\right)^{2}}{2\Gamma}}$,
the output state reads explicitly \citep{Genoni2011}:
\begin{equation}
\!\!\rho_{\varphi}=\!\sum_{n,m=0}^{N}c_{n}c_{m}^{\star}\;\textrm{e}^{-\frac{\Gamma}{2}\left(n-m\right)^{2}}\textrm{e}^{-\textrm{i}\left(n-m\right)\varphi}\left|n,N-n\right\rangle \!\left\langle m,N-m\right|\label{eq:OutputNPh_PhaseDiff},
\end{equation}
where $c_n$ are parameters of the input state given as in \eqref{eq:nphotonprobe}.
The above equation indicates that due
to the phase diffusion the off-diagonal elements of $\rho_{\varphi}$
are exponentially suppressed at a rate increasing in the anti-diagonal directions.

\subsubsection{Photonic losses}
\label{sec:photoniclosses}
In the \emph{lossy interferometer} scenario, the fictitious beam-splitters introduced in the interferometer arms with respective power transmission coefficient $\eta_{a/b}$ account for the probability of photons to leak out. Such a loss model is relatively
general, as due to the commutativity of the noise with the phase accumulation \citep{Demkowicz2009}, it accounts for the photonic losses happening at any stage of the phase sensing process. Moreover, losses at the detection as well as the preparation stages can be moved inside the interferometer provided they are equal in both arms. This makes the model applicable in typical experimental realization of quantum enhanced interferometry
 \citep{Kacprowicz2010,Vitelli2010,Spagnolo2012}, and most notably, when analyzing bounds on quantum enhancement in gravitational-wave detectors \citep{Demkowicz2013}.

Loss decoherence map $\Lambda$ may be formally described using the following set of Kraus operators \citep{Dorner2009}:
\begin{equation}
K_{l_a,l_b}=\sqrt{\frac{\left(1-\eta_{a}\right)^{l_{{a}}}}{l_{{a}}!}}
\eta_{{a}}^{\hat{a}^{\dagger}\hat{a}}\hat{a}^{l_{{a}}}\;
\sqrt{\frac{\left(1-\eta_{{b}}\right)^{l_{{b}}}}{l_{{b}}!}}
\eta_{{b}}^{\hat{b}^{\dagger}\hat{b}}\hat{b}^{l_{{b}}}
\label{eq:KrausRepOps_Loss}
\end{equation}
where the values of index $l_{a/b}$ corresponds to the number of photons lost in mode $a/b$ respectively.
For a general $N$-photon input state of the form \eref{eq:nphotonprobe}, the density
matrix representing the output state of the lossy interferometer reads
\begin{multline}
%\rho_{\varphi}=\Lambda(U_\varphi \rho_{\t{in}} U_\varphi^\dagger) = \\ =\bigoplus_{N^{\prime}=0}^{N}\,\sum_{l_{{a}}=0}^{N-N^{\prime}}\; %p_{l_{{a}},N^{\prime}-l_{{a}}}\,\left|\xi_{l_{{a}},N^{\prime}-l_{{a}}}(\varphi)\right\rangle \!\left\langle %\xi_{l_{{a}},N^{\prime}-l_{{a}}}(\varphi)\right|,
\rho_{\varphi}=\Lambda(U_\varphi \rho_{\t{in}} U_\varphi^\dagger) = \\ = \bigoplus_{N^{\prime}=0}^{N}\sum_{\underset{(l_b=N-N'-l_a)}{l_a=0}}^{N-N^{\prime}}
p_{l_a,l_b}\,\left|\xi_{l_a,l_b}(\varphi)\right\rangle \!\left\langle
\xi_{l_a,l_b}(\varphi)\right|,
\label{eq:OutputNPh_Loss}
\end{multline}
where $p_{l_{{a}},l_{{b}}}=\!=\!\sum_{n=0}^{N}\left|c_{n}\right|^{2} b_{n}^{(l_{{a}},l_{{b}})}$ is the
binomially distributed probability of losing $l_{a}$ and
$l_{b}$ photons in arms a and b respectively,
with
\begin{equation}
b_{n}^{(l_a,l_b)}=
\binom{n}{l_a}\,\eta_{a}^{n-l_a}\left(1-\eta_a\right)^{l_a}\;\binom{N-n}{l_b}\,\eta_{b}^{N-n-l_b}\left(1-\eta_{b}\right)^{l_b},
%b_{n}^{(l_{a},l_{b})}=%\begin{cases}
%\binom{n}{l_{a}}\binom{N-n}{l_{b}}\,\eta_{a}^{n}\left(1-\eta_{a}\right)^{n-l_{a}}\eta_{b}^{N-n}\left(1-\eta_{b}\right)^{N-n-l_{b}},
%,\, l_{a}\le n\le N-l_{b}\\
%0 &\!\! ,\,\textrm{otherwise}
%\end{cases},
\label{eq:b^(la,lb)_n}
\end{equation}
while the corresponding conditional pure states read:
\begin{equation}
\left|\xi_{l_{a},l_{b}}\!(\varphi)\right\rangle =\sum_{n=l_a}^{N-l_b}\, \frac{c_{n}\,\textrm{e}^{-\textrm{i}n\varphi}}{\sqrt{p_{l_{a},l_{b}}}}\,\sqrt{b_{n}^{(l_{a},l_{b})}}\left|n-l_{a},N-n-l_{b}\right\rangle.\label{eq:Xi_la_lb}
\end{equation}
The direct sum in \eqnref{eq:OutputNPh_Loss} indicates that
the output states of different total number of surviving photons,
$N^{\prime}$, belong to orthogonal subspaces, which in principle
could be distinguished by a non-demolition, photon-number counting
measurement.

In the particle-approach when photons are considered as formally distinguishable particles, the loss process acts
on each of the photons independently, see \figref{fig:particleapproachdecoherence}, so that the
overall decoherence process has a tensor product structure $\Lambda^{\otimes N}$, with $\Lambda$
being a single particle loss transformation. At the input stage, each photon occupies a two-dimensional Hilbert space spanned by vectors
$\ket{a}$, $\ket{b}$ representing the photon traveling in the mode $a/b$ respectively.
In order to describe loss, however, and formally keep the number of particles constant, one needs to introduce a
third photonic state at the output, $\ket{\t{vac}}$, representing the state of the photon being lost.
Then formally, $\Lambda$  maps states from the input two-dimensional Hilbert space to the output three-dimensional Hilbert space, and can be fully specified by means of the Kraus representation, $\Lambda(\rho)=\!\sum_{i=1}^{3} K_{i} \rho K_{i}^{\dagger}$, where $K_1$, $K_2$, $K_3$ are given respectively by the following matrices:
\begin{equation}
\left(\!\!
\begin{array}{cc}
\sqrt{\eta_{a}} & 0\\
0 & \sqrt{\eta_{b}}\\
0 & 0
\end{array}\!\right)\!,\;
\left(\!\!
\begin{array}{cc}
0 & 0\\
0 & 0\\
\sqrt{1-\eta_{a}} & 0
\end{array}\right)\!,\;
\left(
\begin{array}{cc}
0 & 0\\
0 & 0\\
0 & \sqrt{1-\eta_{b}}
\end{array}\!\right)\!.
\label{eq:LossKrauses}
\end{equation}
Intuitively, the above Kraus operators account for no photon loss, photon loss in mode $a$ and
photon loss in mode $b$ respectively.
When applied to symmetrized input states, this loss model yields output states
\begin{equation}
\rho_\varphi = \Lambda^{\otimes N}(U_\varphi^{\otimes N}\rho_{\t{in}} U_\varphi^{\dagger \otimes N}) = \Lambda_\varphi^{\otimes N}(\rho_{\t{in}}),
\end{equation}
equivalent to the ones given in \eqnref{eq:OutputNPh_Loss}, where $U_\varphi$ is a single photon phase shift operation
$U_\varphi = \e^{-\mathrm{i}\varphi\hat{\sigma}_z/2}$.

\subsubsection{Imperfect visibility}
\label{sec:impvisdecoh}
In real-life optical  interferometric experiments, it is always the
case that the light beams employed do not contribute completely to
the interference pattern. Due to spatiotemporal or polarization mode-mismatch,
caused for example by imperfect wave-packet preparation or misalignment
in the optical elements, the visibility of the interference pattern is diminished \citep{Leonhardt1997,Loudon2000}.
This effect may be formally described as an effective loss of coherence between the two arms $a$ and $b$ of an interferometer.

Consider a single photon in a superposition state of being in modes $a$ and $b$ respectively:
$\ket{\psi} = \alpha \ket{a} + \beta \ket{b}$. If other degrees of freedom such as e.g. polarization, temporal profile etc. were
identical for the two states $\ket{a}$, $\ket{b}$, we could formally write
$(\alpha \ket{a} + \beta \ket{b})\ket{0}_X$, where $\ket{0}_X$
denotes the common state of additional degrees of freedom. Loss of coherence may be formally described as the transformation of
the state $\ket{\psi}\ket{0}_X$ into
\begin{multline}
\ket{\Psi} = \alpha \left(\sqrt{\eta} \ket{a}\ket{0}_X + \sqrt{1-\eta} \ket{a} \ket{+}_X\right) + \\ +
\beta\left(\sqrt{\eta} \ket{b} \ket{0}_X + \sqrt{1-\eta} \ket{b} \ket{-}_X\right),
\end{multline}
where $\ket{+}_X$, $\ket{-}_X$ are states orthogonal to $\ket{0}_X$,
corresponding to photon traveling in e.g.~orthogonal transversal spacial modes
as depicted in \figref{fig:LossyInter} (iii), in which case parameter $\eta$ can be interpreted as transmission of fictitious beam splitters
that split the light into two orthogonal modes. Assuming we do not control the additional degrees of freedom the effective state
of the photon is obtained by tracing out the above state over $X$, obtaining the effective single-photon decoherence map:
\begin{equation}
\Lambda(\ket{\psi}\bra{\psi}) = \t{Tr}_X(\ket{\Psi}\bra{\Psi}) =
\mat{cc}{
|\alpha|^2 & \alpha \beta^* \eta \\
\alpha^* \beta \eta & |\beta|^2
},
\end{equation}
where the off-diagonal terms responsible for coherence, are attenuated by coefficient $\eta$, what corresponds to
the standard dephasing map \citep{Nielsen2000}.
Written using the Kraus representation, the above map reads:
\begin{equation}
\label{eq:dephasing}
\Lambda(\rho) = \sum_{i=1}^2 K_i \rho K_i^\dagger, \quad K_1 = \sqrt{\frac{1+\eta}{2}}\openone, \ K_{2}= \sqrt{\frac{1-\eta}{2}}\sigma_{z}.
\end{equation}

Note that similarly to the loss model we have modeled the noise with the use of fictitious beam-splitters to visualize the effects of decoherence.
Now, as we know that a beam-splitter
acts on the photons contained in its two-mode input state in an uncorrelated
manner, the effective map on the full
$N$-photon input state is $\Lambda^{\otimes N}$. In case of atomic systems, this would be a typical local dephasing model describing uncorrelated loss of coherence between the two relevant atomic levels \citep{Huelga1997}.
Still, there is a substantial difference from the loss models as the dephased photons are assumed to remain within the spatially confined beams of the interferometer arms.

We can relate the two models by a simple observation, namely that if the photons lost in the loss model with $\eta_a=\eta_b=\eta$
were incoherently injected back into the arms of the interferometer, we would recover the local dephasing model with
the corresponding parameter $\eta$.
%To see this consider the process of injecting back the photons after loss as quantum map.
%In \secref{sec:photoniclosses} we have introduced the $\ket{\t{vac}}$ state denoting the photon being lost. For the present purpose we should
%formally differentiate the states $\ket{\t{vac}_a}$, $\ket{\t{vac}_b}$ corresponding to photons being lost from mode $a$ and $b$ respectively.
%Then, the process of injecting back photons is formally described as a transformation from states living on four dimensional space, $\{ %\ket{a},\ket{b},\ket{\t{vac}_a}\ket{\t{vac}_b}\}$,
%to states living on two dimensional space, $\{\ket{a},\ket{b}\}$,
%with the following set of Kraus operators:
%\begin{equation}
%L_1 = \mat{cccc}{
%1 & 0 & 0 &0 \\
%0 &1 & 0 &0 }, \ L_2 =  \mat{cccc}{
%0 & 0 & 1 &0 \\
%0 &0 & 0 &0 },\ L_3 =  \mat{cccc}{
%0 & 0 & 0 & 0\\
%0 &0 & 0 &1 }.
%\end{equation}
%The combined effect of losses and back injection is exactly the dephasing map:
%$\sum_{ij=1}^3 L_i K^{\t{loss}}_j \rho K^{\t{loss} \dagger}_j L_j^\dagger = \sum_{i=1}^2 K_i^{\t{dephase}} \rho K_i^{\t{dephase} \dagger},$
%where  $K_j^{\t{loss}}$ are loss Kraus operators as in \eqnref{eq:LossKrauses} modified so as to distinguish between
%$\ket{\t{vac}_a}$ and $\ket{\t{vac}_b}$ states, while $K^{\t{dephase}}_i$ are given as in \eqnref{eq:dephasing}.
It should therefore come as no surprise, when we derive bounds on precisions for the two models in
\secref{sec:boundsvisibility} and \secref{sec:boundslosses}, that for the same $\eta$ the local dephasing (imperfect visibility) model
implies more stringent bounds on achievable precision than the loss model. Intuitively, it is better to get rid of the photons
that lost their coherence and do not carry information about the phase, rather than to inject them back into the setup.

The structure of the output state $\rho_\varphi$
is more complex than in the case of phase diffusion and loss models.
 This is because the local dephasing noise not
only transforms the input state into a mixed
state, but due to tracing out some degrees of freedom, the output state
\begin{multline}
\rho_{\varphi}=  \Lambda_\varphi^{\otimes N}(\rho_{\t{in}}) = \\ =\sum_{i_1,\dots,i_N=0}^1 K_{i_1} \otimes
\dots \otimes K_{i_N} U_\varphi^{\otimes N} \rho_{\t{in}} U_\varphi^{\otimes N \dagger} K_{i_1}^\dagger \otimes \dots \otimes K_{i_N}^\dagger
\label{eq:OutputDeph}
\end{multline}
is no longer supported on the bosonic space spanned by the fully symmetric states $\ket{n}\ket{N-n}$.
This makes it impossible to use the mode-description in characterization of the process.
Even though it is possible to write down the explicit form of the above state \citep{Frowis2014, Jarzyna2014} decomposing the state into SU(2) irreducible subspaces, we will not present it here for the sake of conciseness, especially that it will not be needed in derivation of the
fundamental bounds.

\subsection{Bounds in the QFI approach}
Once we have the formula for the output states $\rho_\varphi$,
given a particular decoherence model, we may use \eqnref{eq:qCRB} to calculate QFI, $F_Q(\rho_\varphi)$,
which sets the limit on practically achievable precision of estimation of $\varphi$.
In order to obtain the fundamental precision bound for a given number $N$ of photons
used, we need to maximize the resulting $F_Q$ over input states $\ket{\psi}_\t{in}$, which will in general be very different from the
 \noon states which maximize the QFI in the decoherence-free case.
 This is due to the fact that the \noon states are extremely susceptible to decoherence, as loss of e.g. a single photon makes the state completely useless for phase sensing.
Unfortunately, for mixed states, the computation of the QFI requires in general performing the eigenvalue decomposition of $\rho_\varphi$ and such a minimization
ceases to be effective for large $N$.
Therefore, while it is relatively easy to
obtain numerical bounds on precision and the form of optimal states for moderate $N$ \citep{Huelga1997, Dorner2009, Demkowicz2010}, going to the large $N$ regime poses a huge numerical challenge, making determination of the asymptotic bounds
for $N\rightarrow \infty$ with brute force optimization methods infeasible.

Over the past few years, elegant methods have been proposed
that allow to circumvent the above mentioned difficulties and obtain
explicit bounds on precision based on QFI for arbitrary $N$, and in particular
grasp the asymptotic precision scaling \citep{Escher2011, Demkowicz2012, Knysh2014}.
These methods include: the \emph{minimization over channel purifications} method \citep{Escher2011}
which is applicable in general but requires some educated guess to
obtain a useful bound, as well as \emph{classical and quantum simulation} methods
\citep{Demkowicz2012} which are applicable when the noise acts in an uncorrelated manner
on the probes, but have an advantage of being explicit and convey
some additional physical intuition on the bounds derived.
Description of the newly published method \citep{Knysh2014} which is based on continuous approximation of the probe states and the calculus of variations is beyond the scope of this review.

We will present the methods by applying them directly to
interferometry with each of the decoherence models introduced above.
We invert the order of presentation of the bounds for the decoherence
models compared with the order in \secref{sec:decoherencemodels}, as this will
allow us to discuss the methods in the order of increasing complexity.
The simplest of the methods, the classical simulation, will be
applied to the imperfect visibility model, while the quantum simulation
will be discussed
in the context of loss. Finally, minimization over channel purification method will be described in the context of the phase diffusion model, to which classical and quantum simulation methods are not applicable due to noise correlations. We should note that methods of \citep{Escher2011, Knysh2014} can also be successfully applied to uncorrelated noise models. Still, classical and quantum simulation approaches are more intuitive and that is why we present derivations based on them even though the other techniques yield equivalent bounds.

In order to appreciate the significance of the derived bounds, we
will always compare them with the precision achievable with a state of $N$ uncorrelated photons as given in \eqnref{eq:PsiSep}.
The ratio between this quantities bounds
the amount of quantum-precision enhancement that can be achieved with the help
of quantum correlations present in the input state of $N$ photons.

\subsubsection{Imperfect visibility}
\label{sec:boundsvisibility}
The fundamental QFI bound on precision in case of imperfect visibility or equivalently the local dephasing model has been derived
in \citep{Escher2011, Demkowicz2012, Knysh2014} and reads:
\begin{equation}
\Delta{\varphi} \ge \sqrt{\frac{1-\eta^{2}}{\eta^2}}\;\frac{1}{\sqrt{N}}\label{eq:PrecAsQ_Deph},
\end{equation}
where $\eta$ is the dephasing parameter, see \secref{sec:impvisdecoh}.
For the optimal uncorrelated input state, $\ket{\psi_{\t{in}}}= [(\ket{a}+\ket{b})/\sqrt{2}]^{\otimes N}$ we get $\Delta \varphi = 1/\sqrt{\eta^2 N}$, and hence the quantum precision enhancement which is the ratio of the bound on precision achievable for the optimal strategy and the precision for the uncorrelated strategy is bounded by  a constant factor of $\sqrt{1-\eta^2}$.

\paragraph{Classical simulation method}
The derivation of the formula \eref{eq:PrecAsQ_Deph} presented below makes use of the classical simulation method \citep{Demkowicz2012}, which requires viewing the quantum channel representing the action of the interferometer from a geometrical perspective.
The set of all physical quantum channels,
$\Lambda\!:\,\mathcal{L}\left(\mathcal{H}_{\textrm{in}}\right)\rightarrow\mathcal{L}\left(\mathcal{H}_{\textrm{out}}\right)$,
that map between density matrices defined on the input/ouput Hilbert
spaces ($\mathcal{H}_{\textrm{in}/\textrm{out}}$) constitutes a convex set \citep{Bengtsson2006}.
This is to say that given any two quantum channels $\Lambda_1$, $\Lambda_2$, their convex combination
$\Lambda=p \Lambda_1 + (1-p)\Lambda_2$, $0 \leq p \leq 1$ is also a legitimate quantum channel.
Physically $\Lambda$ corresponds to a quantum evolution that is equivalent to a random application of
$\Lambda_1$, $\Lambda_2$ transformations with probabilities $p$, $1-p$ respectively.

\begin{figure}[!t]
\begin{center}
\includegraphics[width=0.9\columnwidth]{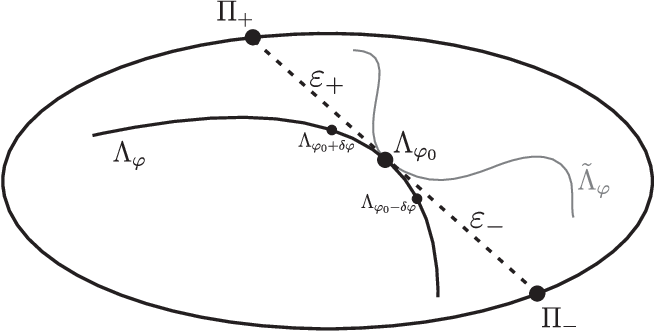}
\end{center}
\caption{
The space of all quantum channels, $\Lambda$,
which map between density matrices specified on two given Hilbert
spaces, $\Lambda:\,\mathcal{L}\left(\mathcal{H}_{\textrm{in}}\right)\rightarrow\mathcal{L}\left(\mathcal{H}_{\textrm{out}}\right)$,
represented as a \emph{convex }set. The estimated parameter $\varphi$
specifies a trajectory, $\Lambda_{\varphi}$ (\emph{black curve}),
in such a space. From the point of view of the QFI, any two channel
trajectories, e.g. $\Lambda_{\varphi}$ and $\tilde{\Lambda}_{\varphi}$
(\emph{gray curve}), are equivalent at a given $\varphi_{0}$ as
long as they and their first derivatives with respect to $\varphi$ coincide there. Moreover, they can
be optimally \emph{classically simulated} at $\varphi_{0}$ by mixing
two channels lying on the intersection of the tangent to the trajectory and the boundary of the set: $\Pi_{\pm}$.
}\label{fig:ChannelSpace}
\end{figure}

As derived in \secref{sec:impvisdecoh}, within the imperfect visibility (local dephasing) decoherence
model: $\rho_{\varphi} = \Lambda_\varphi^{\otimes N}(\rho_{\t{in}})$, and hence
the relevant quantum channel, has a simple tensor structure.
Consider a single-photon channel $\Lambda_{\varphi}$,
which $\varphi$-dependence we may depict as a trajectory within the set of all single-photon quantum maps, see \figref{fig:ChannelSpace}.
The question of sensing the parameter $\varphi$ can now be translated to the question of determining
where on the trajectory a given quantum channel $\Lambda_\varphi$ lies.

Consider a local \emph{classical simulation} (CS) of a quantum channel trajectory $\Lambda_{\varphi}$ in the vicinity of
a given  point $\varphi_0$, $\varphi=\varphi_0 + \delta\varphi$ \citep{Matsumoto2010, Demkowicz2012},
\begin{equation}
\Lambda_{\varphi}\!\left[\varrho\right]=\sum_{x}\, p_{\varphi}(x)\,\Pi_{x}\!\left[\varrho\right]+O\left(\delta\varphi^{2}\right),
\label{eq:ClassSim}
\end{equation}
which represents the variation of the channel up to the first order in $\delta \varphi$ as a classical mixture of some $\varphi$-independent
channels $\left\{ \Pi_{x}\right\} _{x}$ where the $\varphi$ dependence is present only in the mixing probabilities $p_\varphi(x)$.
Under such a construction the random variable $X$ distributed according to $p_\varphi(x)$ specifies probabilistic choice of channels $\Pi_x$ that reproduces the local action of $\Lambda_\varphi$ in the vicinity of $\varphi_0$.
Crucially, the QFI is a \emph{local} quantity---see discussions
in \secref{sub:EstTh_QFIapproach}---and at a given point $\varphi_{0}$
is a function only of the quantum state considered and
its first derivative with respect to the estimated parameter. Consequently,
when considering the parameter being encoded in a quantum channel,
all the channel trajectories at a given point $\varphi_0$ are equivalent from the point of view of QFI if
 they lead to density matrices that are identical up to the first order in $\delta\varphi$.
In other words we can replace $\Lambda_\varphi$ with any $\tilde{\Lambda}_\varphi$ and obtain the same QFI at a given
point $\varphi_0$ provided $\Lambda_{\varphi_0}=\tilde{\Lambda}_{\varphi_0}$ and
$\frac{\t{d} \Lambda_\varphi}{\t{d}\varphi} =
\left.\frac{\t{d} \tilde{\Lambda}_\varphi}{\t{d}\varphi}\right|_{\varphi=\varphi_0}$, see \figref{fig:ChannelSpace}.
This means that when constructing
a local CS of the quantum channel $\Lambda_{\varphi}$ at $\varphi_{0}$, we need only to satisfy $\sum_x p_{\varphi_0}(x)\Pi_x = \Lambda_{\varphi_0}$, as well as
$\sum_x \frac{\t{d}p_\varphi(x)}{\t{d} \varphi}|_{\varphi=\varphi_0}
\Pi_x =\frac{\t{d}\Lambda_\varphi}{\t{d}\varphi}|_{\varphi=\varphi_0}$.

Crucially, as the maps $\Lambda_{\varphi}$ in \eqnref{eq:OutputDeph}
act \emph{independently} on each photon, we can simulate the overall $\Lambda_{\varphi}^{\otimes N}$ with $N$ \emph{independent} random variables, $X^{N}$, associated
with each channel. The estimation procedure can now be described as
\begin{equation}
\varphi\rightarrow X^{N}\rightarrow\Lambda_{\varphi}^{\otimes N}\rightarrow\Lambda_{\varphi}^{\otimes N}\!\left[\left|\psi_{\textrm{in}}\right\rangle \right]\rightarrow\tilde{\varphi},\label{eq:CSMarkovChain}.
\end{equation}
where $N$ classical random variables are employed to generate the
desired quantum map $\Lambda_{\varphi}^{\otimes N}$.
It is clear that a strategy in which we could infer the parameter directly from $X^{N}$, i.e. $\varphi\!\rightarrow\! X^{N}\!\!\rightarrow\!\tilde{\varphi}$,
can perform only better than the scheme
where the information about $\varphi$ is firstly encoded into the
quantum channel which acts on the probe state and afterwards decoded from the measurement results performed on the output state. This
way, we may always construct a classically scaling lower bound on
the precision, or equivalently an upper bound on the QFI of $\rho_{\varphi}$
\eref{eq:OutputDeph}:
\begin{equation}
F_{\textrm{Q}}\!\left[\rho_{\varphi}\right]\le F_{\textrm{cl}}\!\left[p_{\varphi}^{N}\right]=N\, F_{\textrm{cl}}\!\left[p_{\varphi}\right],
\label{eq:QFICSBound}
\end{equation}
which is determined by the classical FI \eref{eq:FI} evaluated
for the probability distribution $p_{\varphi}(X)$. Importantly, \citet{Demkowicz2012}
have shown that for the estimation problems in which the parameter
is unitarily encoded, it is always optimal to choose a CS depicted
in \figref{fig:ChannelSpace}, which employs for each $\varphi$
only two channels $\Pi_{\pm}$ lying at the points of intersection
of the tangent to the trajectory with the boundary of the quantum maps set.
Such an \emph{optimal} CS leads to the tightest upper bound specified
in \eqnref{eq:QFICSBound}: $F_{\textrm{Q}}\!\left[\rho_{\varphi}\right]\!\le\!N/(\varepsilon_{+}\varepsilon_{-})$,
where $\varepsilon_{\pm}$ are the ``distances'' to the boundary marked in \figref{fig:ChannelSpace}, $\Pi_{\pm} = \Lambda_{\varphi_0} \pm \varepsilon_{\pm} \frac{\t{d} \Lambda_\varphi}{\t{d} \varphi}|_{\varphi=\varphi_0}$.

Looking for $\varepsilon_{\pm}$ parameters amounts to a search of the distances one can go along the tangent line to the trajectory of $\Lambda_\varphi$ so that the corresponding map is still a physical quantum channel, i.e. a completely positive trace preserving map. This is easiest to do making use of the Choi-Jamiolkowski isomorphism \citep{Choi1975,Jamiolkowski1972} which states that with each quantum channel, $\Lambda: \mathcal{L}(\mathcal{H}_{\t{in}}) \rightarrow \mathcal{L}(\mathcal{H}_{\t{out}})$, we can associate a positive operator $\Omega_\Lambda \in \mathcal{L}(\mathcal{H}_{\t{out}} \otimes \mathcal{H}_{\t{in}})$,
so that  $\Omega_\Lambda = (\Lambda \otimes \mathcal{I})(\ket{I}\bra{I})$, where $\ket{I} = \sum_{i=1}^{\t{dim}(\mathcal{H}_{\t{in}})} \ket{i}\ket{i}$ is a maximally entangled state on $\mathcal{H}_{\t{in}} \otimes \mathcal{H}_{\t{in}}$, while
$\mathcal{I}$ is the identity map  on $\mathcal{L}(\mathcal{H}_{\t{in}})$. Since the complete positivity of $\Lambda$ is equivalent to positivity of the $\Omega_\Lambda$ operator, one needs to analyze $\Omega_{\Lambda_{\varphi_0}} \pm \varepsilon_{\pm} \frac{\t{d} \Omega_{\Lambda_\varphi}}{\t{d} \varphi}|_{\varphi=\varphi_0}$ and find maximum $\varepsilon_\pm$ so that the above operator is still positive-semidefinite.

Taking the explicit form of the $\Lambda_\varphi$ for the case of optical interferometry with imperfect visibility, see \secref{sec:impvisdecoh},
one can show that $\varepsilon_{\pm}\!=\!\sqrt{1-\eta^{2}}/\eta$ \citep{Demkowicz2012}, which yields the ultimate quantum limit on precision given by \eqnref{eq:PrecAsQ_Deph}.

\subsubsection{Photonic losses}
\label{sec:boundslosses}
The expression for the QFI of the output state \eref{eq:OutputNPh_Loss} in  the asymptotic limit of large $N$
has been first derived by \citep{Knysh2011}. Yet, the general frameworks
proposed by \citep{Escher2011,Demkowicz2012} for generic decoherence
allowed to reconstruct this bound on precision with the following result:
\begin{equation}
\Delta{\varphi}\ge\frac{1}{2}
\left(\sqrt{\frac{1-\eta_{a}}{\eta_{a}}}+\sqrt{\frac{1-\eta_{b}}{\eta_{b}}}\right)\;\frac{1}{\sqrt{N}},\label{eq:PrecAsQ_Loss}.
\end{equation}
where $\eta_a$, $\eta_b$ are transmission in the  two arms of the interferometer respectively, see \ref{sec:photoniclosses}.
This bound simplifies to
\begin{equation}
\label{eq:equallossesbound}
\Delta \varphi \geq \sqrt{\frac{1-\eta}{\eta N}}
\end{equation}
in the case of equal losses, and since the precision achievable with uncorrelated states is given by $1/\sqrt{\eta N}$, the maximal quantum-enhancement factor is $\sqrt{1-\eta}$.
In the following, we derive the above bounds using
the quantum simulation approach of \citep{Demkowicz2012, Kolodynski2013}.

\paragraph{Quantum Simulation method \label{par:QS_Loss}}
Unfortunately, in the case of loss the simple CS method yields a trivial bound $\Delta\varphi \geq 0$, since the tangent distances to the boundary of the set of quantum channels are $\varepsilon_{\pm}=0$ in this case.
It is possible, however, to derive a useful bound
via the \emph{Quantum Simulation} (QS) \emph{method} which is a natural generalization of the CS method.
The QS method has been described in detail and developed for general metrological schemes with uncorrelated noise by \citet{Kolodynski2013} stemming from the works of \citet{Demkowicz2012} and \citet{Matsumoto2010}.

\begin{figure}[!t]
\begin{center}
\includegraphics[width=1\columnwidth]{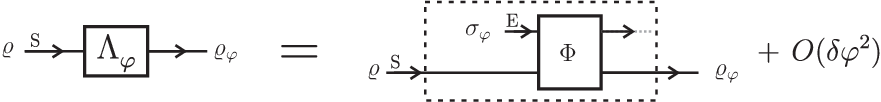}
\end{center}\caption{\label{fig:qs} The Quantum Simulation (QS) of a channel. The action of the channel $\Lambda_\varphi$ is simulated up to the first order in the vicinity of a given point $\varphi_0$ using a $\varphi$-independent $\Phi$ channel
and an auxiliary state $\sigma_\varphi$ that contains the \emph{full }information about the estimated
parameter $\varphi$.}
\end{figure}

As shown in  \figref{fig:qs}, local QS amounts to
re-expressing the action of $\Lambda_{\varphi}$ for $\varphi = \varphi_0 + \delta \varphi$ by a larger $\varphi$-\emph{independent} map $\Phi$ that also acts on the auxiliary $\varphi$-\emph{dependent} input $\sigma_\varphi$,
up to the first order in $\delta \varphi$:
\begin{equation}
\Lambda_\varphi[\varrho] = \t{Tr}_E \Phi\left[ \sigma_\varphi \otimes
\varrho\right] + O(\delta \varphi^2).
\end{equation}

Note that for $\sigma_{\varphi}\!=\!\sum_{x}
p_{\varphi}(x)\left|x\right\rangle \!\left\langle x\right|$,
and $\Phi=\ket{x}\bra{x} \otimes \Pi_x$, QS  becomes
equivalent to the CS of \eqnref{eq:ClassSim}, so that CS
is indeed a specific instance of the more general QS.
An analogous reasoning as in the case of CS leads to the conclusion that we may upper-bound
the QFI of the overall output state, here \eref{eq:OutputNPh_Loss}
for the case of losses, as
\begin{multline}
\label{eq:QFIQSbound}
%F_{\textrm{Q}}\!\left[\rho_{\varphi}\right]  =
 F_{\textrm{Q}}\!\left[\Lambda_{\varphi}^{\otimes N}\!\left[\ket{\psi}_{\textrm{in}}\bra{\psi}\right]\right]
=  F_{\textrm{Q}}\!\left[\textrm{Tr}_{\textrm{E}}\!\left\{ \Phi^{\otimes N}\!\left[\sigma_\varphi^{\otimes N} \otimes \ket{\psi}_{\textrm{in}}\bra{\psi} \right]\right\} \right] \\
  \leq  \ F_{\textrm{Q}}\!\left[ \sigma^{\otimes N}_\varphi\right]=N\,F_{\textrm{Q}}\!\left[\sigma_\varphi \right],\nonumber
\end{multline}
since $\t{Tr}_E \Phi^{\otimes N}[\cdot]$
is just a parameter independent map, under
which the overall QFI may only decrease---see \eqnref{eq:QFI_contract}.
Last equality follows from the additivity property of the QFI, which,
similarly to \eqnref{eq:QFICSBound}, constrains $F_{\textrm{Q}}\!\left[\rho_{\varphi}^{N}\right]$
to scale at most linearly for large $N$.
Similarly
to the case of CS, in order to get the tightest bound one should find QS that yields the smallest $F_Q[\sigma_\varphi]$, which
in principle is a non trivial task.

Fortunately, \citet{Kolodynski2013} have demonstrated that the search
for the optimal channel QS corresponds to the optimization over the
Kraus representation of a given channel.
Without loss of generality we may assume that $\sigma_\varphi = \ket{\varphi}\bra{\varphi}$ is a pure $\varphi$-dependent state while $\Phi[\cdot]= \mathcal{U} \cdot \mathcal{U}^\dagger$ is unitary.
For a given QS we may write the corresponding Kraus representation of the channel by choosing a particular basis $\ket{i}_\t{E}$ in the $\t{E}$ space: $K_i(\varphi) = _{\t{E}}\!\bra{i} \mathcal{U} \ket{\varphi}_{\t{E}}$. In order for the QS to be valid, these Kraus operators should correspond to a legitimate Kraus representation of the channel $\Lambda_\varphi[\cdot] =\sum_{i}K_i(\varphi) \cdot K_i(\varphi)$.
Two Kraus representation of a given quantum channel are equivalent if and only if they are related by a unitary matrix $\mathsf{u}$:
\begin{equation}
\tilde{K}_i(\varphi) = \sum_j \mathsf{u}_{ij}(\varphi) K_j(\varphi),
\end{equation}
which may in principle be also $\varphi$ dependent.
Since we require QS to be only locally valid in the vicinity of $\varphi_0$, the above equation as well as its first derivative needs to be fulfilled only at $\varphi_0$. Because of that, the search for
the optimal Kraus representation $\tilde{K}_i$ (or equivalently the optimal QS) may be restricted to the class of transformations where
$\mathsf{u}(\varphi) = \e^{\mathrm{i} (\varphi - \varphi_0) \mathsf{h}}$ with $\mathsf{h}$ being any Hermitian matrix that shifts the
relevant derivative of $K_{i}(\varphi)$ at $\varphi_{0}$, so that $\tilde{K}_{i}(\varphi_{0})\!=\! K_{i}(\varphi_{0})$
and $\dot{\tilde{K}}_{i}(\varphi_{0})\!=\!\dot{K}_{i}
(\varphi_{0})+\textrm{i}\sum_{j}\mathsf{h}_{ij}K_{j}(\varphi_{0})$.
As shown by \citep{Kolodynski2013}, the problem of finding the optimal QS i.e. the minimal
$F_{\textrm{Q}}\!\left[\ket{\varphi}\bra{\varphi}\right]$
which we term as the $F_{\textrm{QS}}$, can be formally rewritten as
\begin{gather}
F_{\textrm{QS}}=\min_{\mathscr{\mathsf{h}}}\; s\quad\textrm{s.t.} \nonumber\\
\sum_{i}\dot{\tilde{K}}_{i}(\varphi_{0})^{\dagger}\dot{\tilde{K}}_{i}(\varphi_{0})=\frac{s}{4}\,\openone_{2},\quad
\sum_{i}\dot{\tilde{K}}_{i}(\varphi_{0})^{\dagger}
\tilde{K}_{i}(\varphi_{0})=\mathbf{0},
\label{eq:F_QS}
\end{gather}
where the parameter $s$ has the interpretation of $F_{\textrm{Q}}\!\left[\ket{\varphi}\bra{\varphi}\right]$
for the particular QS at $\varphi_{0}$ and the constraints imposed in \eqnref{eq:F_QS}
are necessary and sufficient for the QS required transformation $\mathcal{U}$ and the state $\ket{\varphi}$ to exist.

The above optimization problem may not always be easy to solve. Still, its relaxed version:
\begin{equation}
\min_{\mathscr{\mathsf{h}}}\parallel \sum_i \dot{\tilde{K_i}}(\varphi_0)^\dagger
\dot{\tilde{K_i}}(\varphi_0) \parallel, \quad \sum_{i}\dot{\tilde{K}}_{i}(\varphi_{0})^{\dagger}
\tilde{K}_{i}(\varphi_{0})=\mathbf{0},
\end{equation}
where $\parallel \cdot \parallel$ is the operator norm, can always be cast in the form of an explicit semi-definite program, which
can be easily solved numerically \citep{Demkowicz2012}.
Numerical solution of the semi-definite program provides a form of the optimal $\mathsf{h}$ which may then be taken as an ansatz for further analytical optimization.

Plugging in
the Kraus operators $K_i(\varphi) = K_i U_\varphi$ representing the lossy interferometer, see  \eqnref{eq:LossKrauses}, and following the above described procedure one obtains
\begin{equation}
F_{\textrm{QS}} = \frac{4}{\left(\sqrt{\frac{1-\eta_{a}}{\eta_{a}}}+\sqrt{\frac{1-\eta_{b}}{\eta_{b}}}\right)^{2}}\label{eq:Fqs&h_opt}
\end{equation}
for the optimal $\mathsf{h}$ given by \begin{equation}
\mathsf{h}_\t{\tiny opt}=-\frac{1}{8}\,\textrm{diag}\left\{\chi,\frac{\eta_a}{1-\eta_a}\left(\frac{4}{\eta_a}-\chi\right),-\frac{\eta_b}{1-\eta_b}\left(\frac{4}{\eta_a}+\chi\right)\right\},
\end{equation}
where $\chi\!=\! F_{\textrm{QS}}\frac{\eta_{b}-\eta_{a}}{\eta_{a}\eta_{b}}$.
This indeed reproduces the bounds given in \eqref{eq:PrecAsQ_Loss}.

%%%%%%%%%%%%%%%%%%%%%%%%%%%%%%%%%%

In order to provide the reader with a simple intuition concerning the QS method, we shall present an elementary  construction of the QS for lossy interferometer in the special case of $\eta_a=\eta_b=1/2$.
In this case the bound $\eqref{eq:PrecAsQ_Loss}$ yields
$\Delta \varphi \geq 1/\sqrt{N}$ which implies that using optimal entangled probe state at the input under $50\%$ losses cannot beat the precision which can be obtained by uncorrelated probes in ideal scenario of no losses.

Consider the action of the single photon lossy channel $\Lambda_\varphi$ on the pure input state $\ket{\psi} = \alpha \ket{a} + \beta \ket{b}$:
\begin{equation}
\Lambda_\varphi(\ket{\psi}\bra{\psi}) = \frac{1}{2}
 \ket{\psi_\varphi}\bra{\psi_\varphi}+ \frac{1}{2} \ket{\t{vac}}\bra{\t{vac}}
\end{equation}
with $\ket{\psi_\varphi} = \alpha \e^{\mathrm{i}\varphi}\ket{a} + \beta \ket{b}$, which represents $1/2$ probability of photon sensing the phase undisturbed and the $1/2$ probability of the photon being lost. Let the auxiliary state
for QS be $\ket{\varphi}= (\e^{\mathrm{i}\varphi}\ket{0}+\ket{1})/\sqrt{2}$.
The joined input + auxiliary state reads:
\begin{multline}
\ket{\phi}\ket{\psi} = \frac{1}{\sqrt{2}}\left(\alpha \e^{\mathrm{i} \varphi}\ket{0}\ket{a}  + \beta \ket{1} \ket{b}\right) + \\ +
\frac{1}{\sqrt{2}}\left(\alpha \e^{\mathrm{i} \varphi}\ket{0}\ket{b}  + \beta \ket{1} \ket{a}\right).
\end{multline}
The map $\Phi$ realizing the QS consists now of two steps. First the controlled NOT operation is performed with the auxiliary system being the target qubit, this transform the above state to:
$\frac{1}{\sqrt{2}}\ket{0}(\alpha \e^{\mathrm{i}\varphi} \ket{a} + \beta \ket{b}) + \frac{1}{\sqrt{2}}
\ket{1}(\alpha \e^{\mathrm{i}\varphi} \ket{b} + \beta \ket{a})$.
The second step is the measurement of the auxiliary system.
If the result $\ket{0}$ is measured (probability $1/2$), the system is left in the correct state $\ket{\psi_\varphi}$ and the map leaves it unchanged, if the
$\ket{1}$ is measured the state of the photon is not the desired one, in which case the map returns the $\ket{\t{vac}}$ state.
This map is therefore a proper QS of the  desired lossy interferometer transformation for $\eta_a=\eta_b=1/2$. Since the auxiliary state
employed in this construction was $\ket{\varphi}=(\e^{\mathrm{i}\varphi}\ket{0} + \ket{1})/\sqrt{2}$
for which $F_Q(\ket{\varphi}\bra{\varphi}^{\otimes N})=N$ this leads to the anticipated result $\Delta\varphi \geq 1/\sqrt{N}$.

\subsubsection{Phase diffusion}
\label{sub:phasediff}
Since the phase diffusion model, see \secref{sec:phasediff}, is an example of a correlated noise model, it cannot be approached with
the CS and QS methods.
The study of the behavior of the QFI within the phase-diffusion model
was for the first time carried out by \citep{Genoni2011} considering
indefinite-photon-number Gaussian input states and
studied numerically the achievable precision and
the structure of optimal input states.  Yet, the fundamental analytical bounds on precision
have not been verified until \citet{Escher2012}, where the phase noise has been
approached using the minimization over purifications method of \citet{Escher2011} and most recently
using the calculus of variations approach of \citet{Knysh2014}.

\paragraph{Minimization over purification method}
The minimization over purification method of \citet{Escher2011} is based on the observation, already mentioned in \secref{sub:EstTh_QFIapproach}, that
QFI for a given mixed quantum state, here $\rho_{\varphi}$
\eref{eq:OutputNPh_Loss}, is not only upper bounded by the QFI of any of its purifications, but there always
exists an optimal purification, $\left|\Psi^{\textrm{opt}}_\varphi \right\rangle $,
for which $F_{\textrm{Q}}\!\left[\rho_{\varphi}\right]\!=\! F_{\textrm{Q}}\!\left[\left|\Psi^{\textrm{opt}}_\varphi\right\rangle \right]$,
where $\rho_{\varphi}\!=\!\textrm{Tr}_{\textrm{E}}\!\left\{ \left|\Psi^{\textrm{opt}}_\varphi\right\rangle \!\left\langle \Psi^{\textrm{opt}}_\varphi\right|\right\} $.
As such statement does not rely on the form of the transformation
$\ket{\psi}_{\textrm{in}} \!\rightarrow\!\rho_{\varphi}$
but rather on the properties of the output state itself, the framework
of \citet{Escher2011} in principle does not put any constraints on
the noise-model considered. Note that, even if the optimal purification itself is difficult to find, any suboptimal purification
yields a legitimate upper bound on the QFI and hence may provide a non-trivial precision bound.

In order to get a physical intuition regarding the purification method, consider a physical model of
the phase diffusion where light is being reflected from a mirror
which position fluctuations are randomly changing the effective optical length.
Formally, the model amounts to coupling the phase delay generator
$\hat{J}_{z}$ to the mirror position quadrature
$\hat{x}_{\textrm{E}}\!=\!\frac{1}{\sqrt{2}}\!\left(\hat{a}_{\textrm{E}}+\hat{a}_{\textrm{E}}^{\dagger}\right)$ \citep{Escher2012}.
Assuming the mirror, serving as the environment E, to reside in the
ground state of a quantum oscillator $\left|0\right\rangle _{\textrm{E}}$
before interaction with the light beam, the pure output state reads:
\begin{equation}
\left|\Psi_\varphi\right\rangle =\textrm{e}^{-\textrm{i}\varphi\hat{J}_{z}}\textrm{e}^{\textrm{i}\sqrt{2\Gamma}\hat{J}_{z}
\hat{x}_{\textrm{E}}}\ket{\psi}_{\textrm{in}} \ket{0}_{\textrm{E}}.
\end{equation}
Thanks to the fact that $\left|\left\langle x|0\right\rangle \right|^{2}\!=\!\textrm{e}^{-x^{2}}/\sqrt{\pi}$, the reduced state
\begin{equation}
\rho_{\varphi}=\intop_{-\infty}^{\infty} \textrm{d}x\prescript{}{\textrm{E}}{\!\left\langle x|\Psi_\varphi\right\rangle }\!\left\langle \Psi_\varphi|x\right\rangle _{\textrm{E}}
\end{equation}
indeed coincides with
the correct output phase diffused state \eref{eq:OutputNPh_PhaseDiff}. Therefore, this is a legitimate purification
of the interferometer output state in presence of phase diffusion.

Consider now another purification
$\left|\tilde{\Psi}_\varphi\right\rangle \!=\!\textrm{e}^{\textrm{i}\delta\varphi \hat{H}_{\textrm{E}}}\!\left|\Psi_\varphi\right\rangle $
generated by a local ($\varphi\!=\!\varphi_0\!+\!\delta\varphi$) rotation of the mirror modes, i.e.~ a unitary operation
on the system $\t{E}$.
We look for a transformation of the above form which hopefully erases as much information on the estimated phase as possible,
so that QFI for the purified state $\left|\tilde{\Psi}_\varphi\right\rangle$ will be minimized leading to the best bound on the QFI of $\rho_\varphi$.
Choosing $\hat{H}_{\textrm{E}}\!=\!\lambda\hat{p}_{\textrm{E}}$ we obtain the following upper bound on the QFI
\begin{multline}
F_{\textrm{Q}}\!\left[\rho_{\varphi}\right]  \le  \min_{\lambda}\left\{ F_{\textrm{Q}}\!\left[\textrm{e}^{\textrm{i}\varphi\lambda\hat{p}_{\textrm{E}}}\!\left|\Psi(\varphi)\right\rangle \right]\right\} = \\
  =  \min_{\lambda}\left\{ 2\lambda^{2}+4\left(1-\sqrt{2\Gamma}\lambda\right)^{2}\Delta^{2}J_{z}\right\} =\frac{4\,\Delta^{2}J_{z}}{1+4\Gamma\,\Delta^{2}J_{z}},\nonumber
\end{multline}
and thus a lower limit on the precision
\begin{equation}
\Delta{\varphi}\ge\sqrt{\Gamma+\frac{1}{4\,\Delta^{2}J_{z}}} = \sqrt{\Gamma+\frac{1}{N^2}}\label{eq:PrecAsQLB_PhaseDiff},
\end{equation}
where we plugged in $\Delta^2 J_z = N^2/4$ corresponding to the \noon state which maximized the variance for $N$-photon states.
See also an alternative derivation of the above result that has been proposed recently in \citep{Macieszczak2014}.
Crucially, the above result proves that the
phase diffusion constrains the error to approach a constant value $\sqrt{\Gamma}$ as $N\!\rightarrow\!\infty$,
which does not vanish in the asymptotic limit, what contrasts the $1/\sqrt{N}$ behavior characteristic for uncorrelated noise models.
Note also that due to the correlated character of the noise, the bound \eref{eq:PrecAsQLB_PhaseDiff}
predicts that it may be more beneficial to perform the estimation
procedure on a group of $k$ particles and then repeat the procedure independently
$\nu$ times obtaining $1/\sqrt{\nu}$ reduction in estimation error, rather than employing $N\!=\! k\nu$ in a single experimental
shot \citep{Knysh2014}.

Only very recently,  the exact ultimate
quantum limit for the $N$-photon input states has been derived in \citep{Knysh2014}
\begin{equation}
\Delta{\varphi}\ge\sqrt{\Gamma+\frac{\pi^{2}}{N^{2}}},
\label{eq:PrecAsQ_PhaseDiff}
\end{equation}
showing that the previous bound was not tight, with the second term following the HL-like asymptotic scaling of the
noiseless decoherence-free Bayesian scenario stated in \eqnref{eq:HL_Bayes}.
In fact, as proven by \citep{Knysh2014}, the optimal states of the
noiseless Bayesian scenario, i.e. the sine states \eref{eq:BWstate},
attain the above correct quantum limit.
In \secref{sub:Bayes_Phase_Diff}, we show that within the Bayesian approach,
with the phase-diffusion effects incorporated, the sine states are always the optimal inputs.
%The convergence of the input state distributions obtained by means of, in principle competing,
%QFI and Bayesian approaches is a consequence of the general connection between the
%two when applied to estimation problems possessing a Gaussian prior distribution
%\citep{Macieszczak2013}. As shown in \secref{sub:Bayes_Phase_Diff},
%the Gaussian phase-noise model falls into such category, as even when no prior knowledge about the parameter
%is assumed the phase-noise process \eref{eq:OutputNPh_PhaseDiff_Av} leads to an
%effective Gaussian prior.

\subsection{Bayesian approach}
%comment that imperfect visibility is under preparation, but gives asymptotically the same as in QFI approach
Minimizing the average Bayesian cost, as given by \eqnref{eq:BRiskHQCovPOVM}, over input probe states $\ket{\psi}_{\t{in}}$
is in general more demanding than minimization of the QFI due to the fact that
it is not sufficient to work within the local regime and analyze only the action of a channel and its first derivative at a given estimation point
as in the QFI approach. For this reason we do not apply the Bayesian approach to the imperfect visibility model
as obtaining the bounds requires a significant numerical and analytical effort \citep{Jarzyna2014}, and constrain ourselves
to loss and phase diffusion models.

\subsubsection{Photonic losses}
The optimal Bayesian performance of $N$-photon
states has been studied by \citet{Kolodynski2010}.
Assuming the natural cost function \eref{eq:CostFunH}
and the flat prior phase distribution, the average cost \eref{eq:BRiskHQCovPOVM} reads:
\begin{equation}
\left\langle C \right\rangle =\textrm{Tr}\!\left\{ \left\langle \rho_{\varphi}\right\rangle _{C}\Xi \right\}
\label{eq:AvCost_Loss}
\end{equation}
where $\left\langle \rho_{\varphi}\right\rangle _{C}\!\!=\!4\int\!\!\frac{d\varphi}{2\pi}\, \rho_{\varphi}\sin^{2}\!
\left(\frac{\varphi}{2}\right)$ and $\rho_\varphi$ is given by \eref{eq:OutputNPh_Loss}.
The optimal measurement seed  operator $\Xi$ can be found analogously as in the decoherence-free case.
The block diagonal form of $\rho_{\varphi}$, implies that without losing optimality one can assume  $\Xi\!=\!\bigoplus_{N^{\prime}=0}^{N}\left|e_{N^{\prime}}\right\rangle \!\left\langle e_{N^{\prime}}\right|$
with $\left|e_{N^{\prime}}\right\rangle \!=\!\sum_{n=0}^{N^{\prime}}\left|n,N^{\prime}\!-\! n\right\rangle$.
Physically, the block-diagonal structure of $\Xi$
indicates that the optimal covariant measurement requires a non-demolition
photon number measurement to be performed before carrying out any
phase measurements, so that the orthogonal subspaces, labeled by the
number of surviving photons $N^{\prime}$, may be firstly distinguished, and subsequently the measurement which is optimal
in the lossless case is performed \citep{Kolodynski2010}.
Plugging in the explicit form of $\Xi$ together with the explicit form of the output
state $\rho_\varphi$, we arrive at
\begin{equation}
\langle C \rangle =2-\mathbf{c}^{T}\mathbf{A}\mathbf{c}, \quad
A_{n,n-1}=A_{n-1,n}=\sum_{l_a,l_b=0}^{n,N-n} \sqrt{b_{n}^{(l_a,l_b)}\,b_{n-1}^{(l_a,l_b)}},
\label{eq:OffDiagEls}
\end{equation}
where  $\mathbf{A}$ is a symmetric $\left(N\!+\!1\right)\!\!\times\!\!\left(N\!+\!1\right)$
matrix that is non-zero only on its first off-diagonals, $b_{n}^{(l_{a},l_{b})}$ are the binomial
coefficients previously defined in \eqnref{eq:b^(la,lb)_n},
while $\mathbf{c}$ is a state vector containing coefficients $c_{n}$ of
the $N$-photon input state \eref{eq:nphotonprobe}.

The \emph{minimal} average cost \eref{eq:AvCost_Loss} for the lossy
interferometer then equals $\left\langle C\right\rangle _{\min}\!=\!2-\lambda_{\textrm{max}}$,
where $\lambda_{\textrm{max}}$ is the maximal eigenvalue of the matrix
$\mathbf{A}$ and the corresponding eigenvector $\mathbf{c}_{\textrm{max}}$
provides the optimal input state coefficients. $\left\langle C\right\rangle _{\min}$
quantifies the maximal achievable precision and
in the $N\!\rightarrow\!\infty$ limit may be interpreted as the average
MSE \eref{eq:BMSE} due to the convergence of the cost function
\eref{eq:CostFunH} to the squared distance as $\tilde{\varphi}\!\rightarrow\!\varphi$.

The procedure described above allows only to obtain numerical values of the achievable precision, and
ceases to be feasible for $N\rightarrow \infty$. The main result of \citep{Kolodynski2010} was to construct a valid
analytical lower bound on the minimal average cost \eref{eq:AvCost_Loss}:
\begin{equation}
\left\langle C\right\rangle _{\textrm{min}}\ge2\left[1-A_{\textrm{max}}\,\cos\!\left(\frac{\pi}{N+2}\right)\right],
\label{eq:MinAvCostLB_Loss}
\end{equation}
where $A_{\textrm{max}}\!=\!\max_{1\le n\le N}\!\left\{ A_{n,n-1}\right\} $
is the largest element of the matrix $\mathbf{A}$, contained within
its off-diagonal entries \eref{eq:OffDiagEls}. The bound yields exactly the same formula
as the QFI bound \eref{eq:PrecAsQ_Loss}, proving that in this case the Bayesian and QFI approaches are equivalent:
\begin{equation}
\!\!\Delta \varphi \approx \sqrt{\left\langle C\right\rangle} \geq
\sqrt{\left\langle C\right\rangle _{\textrm{min}}}\geq \frac{1}{2}\left(\sqrt{\frac{1-\eta_{a}}{\eta_{a}}}+
\sqrt{\frac{1-\eta_{b}}{\eta_{b}}}\right)\;\frac{1}{\sqrt{N}},
\label{eq:PrecAsQ_Bayes_Loss}
\end{equation}
where $\approx$ represents the fact that Bayesian cost approximated the variance only in the limit of large $N$.
The fact that both approaches
lead to the same ultimate bounds on precision suggests that
 the optimal input states may be approximated for $N\rightarrow \infty$ up to an arbitrary good precision
with states manifesting only local finite-number of particle correlations and may in particular be
efficiently simulated with the concept of matrix-product states \citep{Jarzyna2013, Jarzyna2014}.

%Lastly, we should note that the
%above argument makes us conjecture that the \emph{information theoretic
%}approaches of \citep{Hall2012,Nair2012}, which also allow to quantify
%the ultimate attainable precision incorporating the prior information
%about the parameter, should similarly lead to the asymptotic results
%obtained via the QFI methods whenever an uncorrelated noise is present
%Therefore the bound on the asymptotic quantum enhancement is the same in both approa, which is an
%example of a more general result on equivalence between Bayesian on QFI approaches in case of estimation on uncorrelated probes \citep{Gill2013}.

\subsubsection{Phase diffusion}\label{sub:Bayes_Phase_Diff}
Similarly to the case of losses discussed in the previous section,
we study the estimation precision achieved within the Bayesian approach
but in the presence of \emph{phase diffusion}. The analysis follows exactly
in the same way, so that likewise assuming \emph{no prior knowledge}
and the natural cost function introduced in \eqnref{eq:BRiskHQCovPOVM}
the formula for the average cost reads
\begin{equation}
\left\langle C\right\rangle =\textrm{Tr}\!\left\{ \left\langle
\rho_\varphi \right\rangle _{C}\Xi\right\} =2-\mathbf{c}^{T}\mathbf{A}\mathbf{c},
\label{eq:AvCost_PhaseDiff}
\end{equation}
where this time one may think of the effective state,
as of the input state $\rho_{\textrm{in}}$ which is firstly averaged
over the Gaussian distribution dictated by the evolution \eref{eq:OutputNPh_PhaseDiff_Av} and
then over the cost function in accordance with \eqnref{eq:BRiskHQCovPOVM}.
The optimal seed element of the covariant POVM is identical as in the decoherence-free case
  $\Xi = \left|e_{N}\right\rangle \!\left\langle e_{N}\right|$ and
the matrix $\mathbf{A}$ possesses again only non-zero
entries on its first off-diagonals, but this time all of them are equal to $\textrm{e}^{-\frac{\Gamma}{2}}$.
 As a result, the \emph{minimal
}average cost \eref{eq:AvCost_PhaseDiff} may be evaluated analytically
following exactly the calculation of \citep{Berry2000} for the noiseless
scenario, which leads then to $\lambda_{\textrm{max}}\!=\!2\textrm{e}^{-\frac{\Gamma}{2}}\cos\left(\frac{\pi}{N+2}\right)$
and hence
\begin{multline}
\left\langle C\right\rangle _{\min}=2\left[1-\textrm{e}^{-\frac{\Gamma}{2}}\cos\left(\frac{\pi}{N+2}\right)\right]\,\overset{N\rightarrow\infty}{\approx}\, \\ 2\left(1-\textrm{e}^{-\frac{\Gamma}{2}}\right)+\textrm{e}^{-\frac{\Gamma}{2}}\frac{\pi^{2}}{N^{2}}.
\label{eq:MinAvCost_PhaseDiff}
\end{multline}
The optimal input states are the same as in the decoherence-free
case, i.e.~they are the $N$-photon sine states of \eqnref{eq:BWstate}.
%Hence, in these regimes
%the complementary Bayesian and local approaches lead to the same form
%of the optimal input states \eref{eq:InputNPh}.
Note that in contrast to the photonic loss which is an example of an uncorrelated noise,
the minimal average cost \eref{eq:MinAvCost_PhaseDiff} does not asymptotically
coincide with the QFI-based precision limit \eref{eq:PrecAsQ_PhaseDiff} unless $\Gamma \ll 1$.

\subsection{Practical schemes saturating the bounds}
Deriving the fundamental bounds on quantum enhanced precision in presence of decoherence is
interesting in itself from a theoretical a point of view. Still, a practical question remains whether the bounds derived
are saturable in practice. Note that \noon states and the sine states that are optimal in case of QFI and Bayesian approaches
in the decoherence-free case are notoriously hard to prepare apart from regime of very small $N$.
For large photon numbers, the only practically accessible states of light are squeezed Gaussian states
and one of the most popular strategies in performing quantum-enhanced interferometry
amounts to mixing a coherent beam with a squeezed vacuum
state on the input beam splitter of the Mach-Zehnder interferometer, see \secref{sub:coh_sq_vac_inter}.
We demonstrate below that in presence of uncorrelated decoherence, such as loss or imperfect visibility,
this strategy is indeed optimal in the asymptotic regime of large $N$ and allows to saturate the fundamental bounds derived
above. We will not discuss the phase-diffusion noise, since the estimation uncertainty is finite in the asymptotic limit, and
the issue of saturating the asymptotic bound becomes trivial as practically all states lead to the same asymptotic precision value,
while saturating the bound for finite $N$ requires the use of experimentally inaccessible sine states.

\subsubsection{Bounds for indefinite photon number states}
\label{sec:decohindefinite}
Derivation of the bounds presented in this section both in the QFI and Bayesian approaches assumed \emph{definite}-photon number states
at the input. We have already discussed the issue of translating the bounds from a definite photon number input state case to a general
indefinite-photon number state case in \secref{sec:idealindefinite} in the case of decoherence-free metrology,
where we have observed that due to quadratic dependence of QFI on number of photons used,
maximization of QFI over states with fixed \emph{averaged} photon number $\mN$ is ill defined and arbitrary high QFI
are in principle achievable. Controversies related to this observation, discussed in \secref{sec:optindefinitephotonnumber},
are fortunately not present in the noisy metrology scenario.

For the decoherence models, analyzed in this paper, the QFI scales at most linearly with $N$. Following the reasoning presented
in \secref{sec:idealindefinite}, consider a mixture of different
photon number states $\sum_N p_N \rho_N$. Since in the presence of decoherence $F_Q(\rho_N) \leq c N$, where $c$ is a
constant coefficient that depends on the type and strength of the noise considered, thanks to the convexity of the QFI we can write:
\begin{equation}
F_Q\left(\sum_N p_N \rho_N\right) \leq \sum_N p_N F_Q(\rho_N) \leq \sum p_N  c N  = c \mN.
\end{equation}
Hence the bounds on precision derived in \secref{sec:boundslosses} (losses) and  \secref{sec:boundsvisibility} (imperfect visibility) are
valid also under replacement of $N$ by $\mN$. Still, one may come across claims of precisions going beyond the
above mentioned bounds typically by a factor of two \citep{Aspachs2009, Joo2011}. This is only possible, however,
if classical reference beam required to perform e.g. the homodyne detection is not
treated as a resource. As discussed in detail in \secref{sub:role_of_ref}, we take the position that
such reference beams should be treated in the same way as the light traveling through the interferometer and as such
also counted as a resource.

\subsubsection{Coherent + squeezed vacuum strategy}
In section \secref{sub:phase_sens_uncer}, we have derived
an error-propagation formula for the phase-estimation
uncertainty \eqref{eq:deltaphi}
for the standard Mach-Zehnder interferometry in
absence of decoherence. For this purpose we have adopted the Heisenberg picture
and expressed the precision in terms of expectation values, variances
and covariances of the respective angular momentum observables calculated
for the input state.
Here, we follow the same procedure but take additionally into account the effect
of imperfect visibility (local dephasing) and loss. For simplicity, in the case of loss
we restrict ourselves to equal losses in both arms.
The Heiseberg picture transformation of
an observable $\hat{O}$ corresponding to a general map $\Lambda_\varphi$ \eqref{eq:generaldecoherence}
reads
\begin{equation}
\sum_i U_\varphi^\dagger K^\dagger_i \hat{O} K_i U_\varphi = \Lambda_\varphi^*(\hat{O}),
\end{equation}
where $\Lambda^*$ is called the conjugated map.

For a more direct comparison with the
decoherence-free formulas of \secref{sub:phase_sens_uncer}, we explicitly
include the action of the Mach-Zehnder
input and output balanced beam splitters in the description of
the state transformation---in terms of \figref{fig:LossyInter}
this corresponds to moving $\ket{\psi}_\t{in}$ to the left and
$\rho_\varphi$ to the right of the figure.
In case of loss the decoherence map has the same form as given in
\secref{sec:photoniclosses}, but with $U_\varphi = \e^{-\mathrm{i}\frac{\varphi}{2} \sigma_y}$,
while in the case of imperfect visibility the Kraus operators \eqref{eq:dephasing}
will be modified
to $K_1=\sqrt{\frac{1+\eta}{2}}\openone$, $K_2 = \sqrt{\frac{1-\eta}{2}}\sigma_y$,
so that the local dephasing is defined
with respect to the $y$ rather than the $z$ axis.
For the two decoherence models,
the resulting Heisenberg picture transformation of the $J_z$ observable
yields \citep{Ma2011}:
\begin{eqnarray}
\label{eq:E[J]Var[J]_Deph}
\frac{\langle \hat{J}_z \rangle}{\eta} &=&  \cos \varphi \langle \hat{J}_z \rangle_{\t{in}} - \sin \varphi \langle \hat{J}_x \rangle_{\t{in}},\\
\frac{\Delta^2 J_z}{\eta^2} & = &
f(\eta) \frac{\langle N \rangle}{4} + \cos^2\varphi\, \Delta^2J_z|_{\t{in}}+ \sin^2\varphi\, \Delta^2 J_x|_{\t{in}} + \nonumber\\
&& -2 \sin\varphi\cos\varphi\,\t{cov}(J_x,J_z)|_{\t{in}}.\nonumber
\end{eqnarray}
where $f(\eta) = (1-\eta)/\eta$ for the loss model and
$f(\eta) = (1-\eta^2)/\eta^2$ in the case of local dephasing model.
The above expressions have a clear intuitive interpretation.
The signal $\langle \hat{J}_z\rangle$ is rescaled by a factor $\eta$
compared with the decoherence-free case,
while the variance apart from the
analogous rescaling is enlarged by an
 additional noise contribution $f(\eta)\mN/4$ due to
 lost or dephased photons.

In order to calculate the precision achievable with
coherent+squeezed-vacuum strategy, we may use the already
obtained quantities presented in \eqnref{eq:coh+sq_vac_Jrels}.
After substituting the input
variances and averages into \eqnref{eq:E[J]Var[J]_Deph}
and optimally setting
$\alpha\!=\!\textrm{Re}(\alpha)$ as before, we arrive at a modified
version of the formula \eref{eq:coh+sq_vac_prec} for the phase estimation
precision:
\begin{multline}
\Delta{\varphi}^{\left|\alpha\right\rangle
\left|r\right\rangle }= \\ = \tfrac{\sqrt{\cot^{2}\varphi
\left(\left|\alpha\right|^{2}+\frac{1}{2}\sinh^{2}2r\right)+
\left|\alpha\right|^{2}\textrm{e}^{-2r}+\sinh^{2}r
+ f(\eta)\frac{\left|\alpha\right|^{2}+\sinh^{2}r}{\sin^{2}\varphi}}}{\left|\left|\alpha\right|^{2}-\sinh^{2}r\right|}.
\end{multline}
The optimal operation points are again $\varphi =\pi/2, 3 \pi/2$.
Considering the asymptotic limit $\langle N\rangle
= |\alpha|^2 + \sinh^2r  \rightarrow \infty$ and assuming
the coherent beam to carry the dominant part of the energy
$|\alpha| \gg \sinh^2 r$, the
 formula for precision at the optimal operation point reads:
\begin{equation}
\Delta{\varphi}^{\left|\alpha\right\rangle \left|r\right\rangle }
\approx \frac{\sqrt{\langle N \rangle  \e^{-2r} + f(\eta)
\langle N \rangle }}{\langle N \rangle } =
\frac{\sqrt{\e^{-2r} + f(\eta)}}{\sqrt{\langle N \rangle}}.
\label{eq:PrecAsSqVac_Deph}
\end{equation}
Clearly, even for relatively small
squeezing strength $r$ the $\e^{-2r}$ term becomes negligible,
and hence we can effectively approach arbitrary close precision given by:
\begin{equation}
\Delta{\varphi}^{\left|\alpha\right\rangle \left|r\right\rangle }
\approx \frac{\sqrt{f(\eta)}}{\sqrt{\langle N \rangle}},
\end{equation}
which recalling the definition of $f(\eta)$ for the two decoherence models
considered coincides exactly with the fundamental bounds
\eqref{eq:PrecAsQ_Deph}, \eqref{eq:equallossesbound} derived before.
This proves that the fundamental bounds can be asymptotically saturated with
a practical interferometric scheme.
 One should note
that this contrasts the noiseless case and the suboptimal performance
of simple estimation scheme based on the photon-number difference measurements,
see \eqnref{eq:precAs_coh_sq_noDec}.

\begin{figure}[!t]
\begin{center}
\includegraphics[width=0.9\columnwidth]{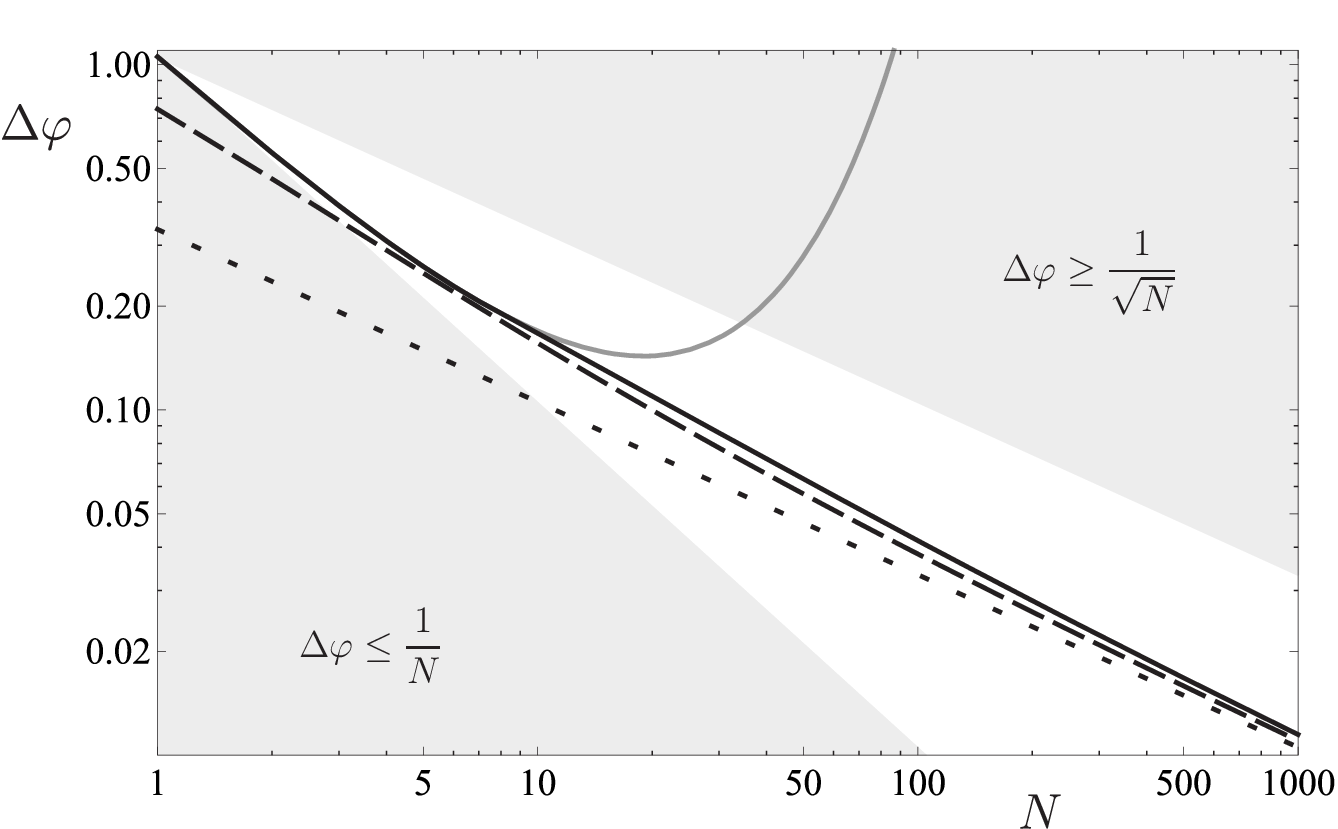}
\end{center}
\caption{\label{fig:PrecPlotLossInter}
The phase estimation precision of an
interferometer with equal losses in both arms ($\eta\!=\!0.9$). The
performance of the optimal $N$-photon input states \eref{eq:nphotonprobe}
is shown (\emph{solid black}) that indeed saturate the asymptotic
quantum limit \eref{eq:PrecAsQ_Loss} (\emph{dotted}): $\sqrt{\left(1-\eta\right)/\left(\eta N\right)}$.
The \noon states (\emph{solid grey}) achieve nearly optimal precision
only for low $N$ ($\le\!10$) and rapidly diverge becoming
out-performed by classical strategies. For comparison, the precision
attained for an indefinite photon number scheme is presented, i.e.
a coherent state and squeezed vacuum optimally mixed on a beam-splitter
\citep{Caves1981} (\emph{dashed}), which in the presence of loss
also saturates the asymptotic quantum limit \eref{eq:PrecAsQ_Loss}.}
\end{figure}
To summarize the results obtained in this section,
in \figref{fig:PrecPlotLossInter}, we present a plot of the maximal
achievable precision for the lossy interferometer in the equal-losses
scenario with $\eta\!=\!0.9$, i.e.
$\Delta{\varphi}\!=\!1/\sqrt{\bar{F}_{\textrm{Q}}[\rho_{\varphi}]}$
as a function of $N$ compared
with the \noon state--based strategy as well as the asymptotic bound
\eref{eq:PrecAsQ_Loss}. On the one hand, the \noon states remain
optimal for relatively small $N$($\le\!10$), for which
the effects of losses may be disregarded. This fact supports the choice
of \noon-like states in the quantum-enhanced experiments
with small number of particles \citep{Mitchell2004,Nagata2007,Resch2007,Okamoto2008,Xiang2010, Krischek2011}.
However, one should note that
in the presence of even infinitesimal losses, the precision achieved
by the \noon states quickly diverges with $N$, because their corresponding
output state QFI, $F_{\textrm{Q}}^{\t{\noon}}\!=\!\eta^{N}N^{2}$,
decays exponentially for any $\eta\!<\!1$.

Most importantly, it should be stressed that the coherent+squeezed vacuum strategy discussed above has been implemented in
recent gravitational-wave interferometry experiments
\citep{LIGO2011,LIGO2013}. The main factor limiting the quantum enhancement of precision in this experiments
is loss, which taking into account detection efficiency, optical instruments imperfections and
imperfect coupling  was estimated at the level of  $38\%$ \citep{LIGO2011}.
In \citep{Demkowicz2013} it has been demonstrated that the sensing precision achieved in \citep{LIGO2011}
using the $10\t{dB}$ squeezed vacuum (corresponding to the squeezing factor  $\e^{-2r} \approx 0.1$), was
strikingly close to the fundamental bound, and only $8\%$ further reduction in estimation uncertainty would be possible
if more advanced input states of light were used. \section{Conclusions \label{sec:conclusions}}
In this review we have showed how the tools of quantum estimation theory can
be applied in order to derive fundamental bounds on achievable precision
in quantum-enhanced optical interferometric experiments.
The main message to be conveyed is the fact that while the power of quantum enhancement
is seriously reduced by the presence of decoherence, and in general the Heisenberg scaling
cannot be reached, non-classical states of light offer a noticeable improvement in interferometric precision
and simple experimental schemes may approach arbitrary close the fundamental quantum bounds.
It is also worth noting that in the presence of uncorrelated decoherence the Bayesian approaches coincide asymptotically with the QFI approaches
easing the tension between this two often competing ways of statistical analysis.

We would also like to mention an inspiring alternative approach to the derivation of limits on precision of phase estimation, where
the results are derived making use of information theoretic concepts such as rate-distortion theory \citep{Nair2012} or entropic uncertainty relations \citep{Hall2012}. Even though the bounds derived in this way are weaker than the bounds presented in this review and obtained via Bayesian or QFI approaches, they carry a conceptual appeal encouraging to look for deeper connections between quantum estimation and communication theories.

Let us also point out, that while we have focused our discussion on optical interferometry using the paradigmatic Mach-Zehnder model,
the same methods can be applied to address the problems of fundamental precision bounds in atomic interferometry \citep{Cronin2009},
magnetometry \citep{Budker2007}, frequency stabilization in atomic clocks \citep{Diddams2004} as well as the limits on resolution of quantum enhanced lithographic protocols \citep{Boto2000}.
All these setups can be cast into a common mathematical framework, see \secref{sec:otherinterferometers},
but the resulting bounds will depend strongly on the nature of dominant decoherence effects and the
relevant resource limitations such as: total experimental time, light power, number of atoms etc.,
as well as on the chosen figure of merit. In particular, it is not excluded that in some atomic metrological scenarios
 one may still obtain a better than $1/\sqrt{N}$ of precision if decoherence is of a special form
 allowing for use of the decoherence-free subspaces \citep{Dorner2012, Jeske2013}
 or when its impact may be significantly reduced by
considering short evolution times,
in which the SQL-like scaling bounds may in principle be circumvented by:
adjusting decoherence geometry \citep{Chaves2013, Kessler2013, Dur2014, Arrad2014}
or by considering Non-Markovian short-time behaviour \citep{Matsuzaki2011, Chin2012}.

We should also note that application of the tools presented in this review to a proper analysis of fundamental limits to
the operation of quantum enhanced atomic clocks  \citep{Leibfried2004, Andre2004} is not that direct as it requires
taking into account precise frequency noise characteristic of the local oscillator, allowing to determine
the optimal stationary operation regime of the clock \citep{Macieszczak2013} ideally
in terms of the Allan variance taken as a figure of merit \citep{Fraas2013}. A deeper theoretical insight into this problem
is still required to yield computable fundamental bounds.

The applicability of the tools presented has also been restricted to single phase parameter estimation.
A more general approach may be taken, were multiple-phases \citep{Humphreys2013} or
 the phase as well as the decoherence strength itself are the quantities to be estimated \citep{Knysh2013, Crowley2014}. This poses an additional theoretical challenge
as then the multi-parameter quantum estimation theory needs to be applied, while most of the tools discussed in this review
are applicable only to single-parameter estimation. Developing non-trivial multi-parameter fundamental bounds for quantum metrology is
therefore still an open field for research.

This research work supported by the FP7 IP project SIQS co-financed by the
Polish Ministry of Science and Higher Education, Polish NCBiR under the ERA-NET CHIST-ERA project QUASAR,
and Foundation for Polish Science TEAM project.

%bibliography
%\bibliographystyle{elsarticle-harv}
%\bibliography{limits}
%merlin.mbs apsrmp4-1.bst 2010-07-25 4.21a (PWD, AO, DPC) hacked
%Control: author (11) reversed first initials

\end{document}